\pdfoutput=1
\documentclass[letterpaper,twocolumn,10pt]{article}
\usepackage{usenix-2020-09}
\usepackage{subcaption}
\usepackage{mdframed}
\usepackage{hyperref}
\usepackage{booktabs}
\usepackage{diagbox}
\usepackage{multirow}
\usepackage{float}
\usepackage{algorithm}
\usepackage{algpseudocode}
\usepackage{enumitem}
\usepackage{colortbl}

\newcommand{\workflow}{{{TITIM}}}
\newcommand{\workflowfull}{{Training-Inference Trigger Intensity Manipulation}}

\newcommand{\etc}{\textit{etc.}}

\newcommand{\forexample}{\textit{e.g.}}
\newcommand{\thatis}{\textit{i.e.}}

\usepackage{xcolor}

\newcommand{\highlight}[2]{{\color{blue}{#1#2}}}
\renewcommand{\highlight}[2]{#2} % revert command (remove comment and color)
\usepackage{bibspacing}

\usepackage{tikz}
\usepackage{amsmath}
\usepackage[available]{usenixbadges}
\begin{document}
%-------------------------------------------------------------------------------

%don't want date printed
\date{}

% \title{\Large \bf Formatting Submissions for a USENIX Conference:\\
%   An (Incomplete) Example}
\title{Revisiting Training-Inference Trigger Intensity in Backdoor Attacks}

%for single author (just remove % characters)
% \author{
% {\rm Your N.\ Here}\\
% Your Institution
% \and
% {\rm Second Name}\\
% Second Institution
% % copy the following lines to add more authors
% % \and
% % {\rm Name}\\
% %Name Institution
% } % end author
% \author{{\rm Submission \#389}}
\author{
{\rm Chenhao Lin, Chenyang Zhao, Shiwei Wang, Longtian Wang, Chao Shen, Zhengyu Zhao\thanks{Corresponding Author}}\\
Xi'an Jiaotong University
} % end author

\maketitle
\thispagestyle{empty}
\pagestyle{empty}

% %-------------------------------------------------------------------------------
% \begin{abstract}
% %-------------------------------------------------------------------------------
% Your abstract text goes here. Just a few facts. Whet our appetites.
% Not more than 200 words, if possible, and preferably closer to 150.
% \end{abstract}
\begin{abstract}
Backdoor attacks typically place a specific trigger on certain training data, such that the model makes prediction errors on inputs with that trigger during inference.
Despite the core role of the trigger, existing studies have commonly believed a perfect match between training-inference triggers is optimal.
% Deep neural networks (DNNs) have been demonstrated to be vulnerable to backdoor attacks, where an adversary injects a backdoor trigger into the model during training, causing malicious behavior during inference.
% Most existing studies assume consistent trigger intensities between training and inference phases, while the potential impact of intensity-mismatched triggers has been largely ignored.
%In this paper, we introduce \theoremFull~(\theorem) to demonstrate how the effectiveness of backdoor attacks is impacted when trigger configurations (specifically, trigger intensity) vary between the training and inference phases. This also reveals that consistent trigger configurations during both phases are not always optimal for backdoor attacks.
In this paper, for the first time, we systematically explore the training-inference trigger relation, particularly focusing on their mismatch, based on a \workflowfull~(\workflow) workflow.
\workflow~specifically investigates the training-inference trigger intensity, such as the size or the opacity of a trigger, and reveals new insights into trigger generalization and overfitting.
% , and considers two cases: train (T) > inference (I) and T < I, which respectively leads to new findings about attack generalization and overfitting.
% We find that when T < I, a higher I generally improves attacks due to better generalization, and when T > I, a higher may harm attacks due to stronger training overfitting.

These new insights challenge the above common belief by demonstrating that the training-inference trigger mismatch can facilitate attacks in two practical scenarios, posing more significant security threats than previously thought.
First, when the inference trigger is fixed, using training triggers with mixed intensities leads to stronger attacks than using any single intensity. 
For example, on CIFAR-10 with ResNet-18, mixing training triggers with 1.0 and 0.1 opacities improves the worst-case attack success rate (ASR) (over different testing opacities) of the best single-opacity attack from 10.61\% to 92.77\%.
Second, intentionally using certain mismatched training-inference triggers can improve the attack stealthiness, \thatis, better bypassing defenses.
% For example, a model backdoored with a trigger intensity of 1.0 can still be activated with a trigger intensity of 0.7, achieving a high ASR of 91.62\%, while reducing the AUC of STRIP and Scale-Up from 0.99 and 0.96 to 0.80 and 0.62, respectively. 
For example, compared to the training/inference intensity of 1.0/1.0, using 1.0/0.7 decreases the area under the curve (AUC) of the Scale-Up defense from 0.96 to 0.62, while maintaining a high attack ASR (99.65\% vs. 91.62\%).
The above new insights are validated to be generalizable across different backdoor attacks, models, datasets, tasks, and (digital/physical) domains.

\end{abstract}

%!TEX root = main.tex

\section{Introduction}
\label{section:introduction}

Deep neural networks (DNNs) are known to be susceptible to various attacks~\cite{DBLP:journals/corr/GoodfellowSS14, DBLP:conf/sp/Carlini017,DBLP:journals/access/GuLDG19}, of which backdoor attacks gain increasing attention~\cite{DBLP:journals/access/GuLDG19,DBLP:conf/icip/BarniKT19,DBLP:conf/aaai/DuanH0ZZ24,DBLP:conf/iclr/NguyenT21,DBLP:journals/corr/abs-2007-08745}.
%Backdoor attacks aim to manipulate the behavior of a DNN model.
A backdoor attacker poisons a certain number of training samples by placing a specific trigger, such that the model would make incorrect predictions for inputs containing that trigger during the inference phase.
Since the trigger is a core component in backdoor attacks, diverse types of triggers have been explored, such as patch-based triggers \cite{DBLP:journals/access/GuLDG19}, image-based triggers \cite{DBLP:journals/corr/abs-1712-05526}, and artifact triggers \cite{DBLP:conf/aaai/DuanH0ZZ24}.

\begin{figure}[!t]
    \centering
    \includegraphics[width=\linewidth]{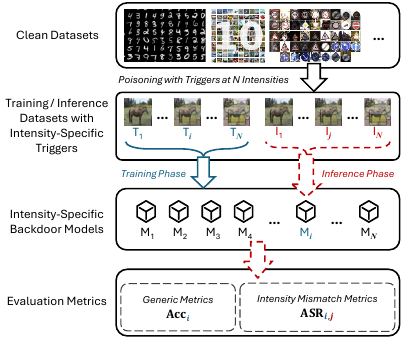}
    \caption{\highlight{(\textit{3. Dataset Configuration})~}{Illustration of our \workflowfull~(\workflow) workflow for testing backdoor attacks with varying training-inference trigger intensities. }}
    \label{figure:workflow}
\end{figure}

Existing studies have blindly adopted the same trigger configurations for both training and inference phases, and they commonly believed a training-inference trigger mismatch would harm the attack performance.
Specifically, most studies have shown that using mismatched triggers causes lower attack success rates due to their generalization difficulty across different configurations~\cite{DBLP:conf/iccd/LiuXS17, DBLP:conf/acsac/DoanAR20, DBLP:journals/tr/UdeshiPWLRC22, DBLP:journals/corr/abs-2007-00711, DBLP:conf/asiaccs/0001ZGZQT21, DBLP:journals/corr/abs-2104-02361,DBLP:journals/compsec/XueHWSZWL22}, although the success would remain for specific trigger types \cite{DBLP:journals/corr/abs-2007-08745}.
The trigger mismatch has even been employed as a defense paradigm~\cite{DBLP:journals/corr/abs-2007-08745}.

\noindent\textbf{Our work.} In this paper, however, we argue that the relations between training-phase and inference-phase triggers should be systematically studied.
This is because the training-inference trigger mismatch generally happens in various practical scenarios, due to user data preprocessing, image compression during transmission, or other factors in the physical domain.
Additionally, attackers may intentionally specify distinct trigger configurations at different stages of the attack for enhanced attack performance.
Consequently, defending against backdoor attacks requires strategies that consider potential discrepancies in trigger intensity.

Therefore, we introduce the \workflowfull~(\workflow) workflow to investigate the impact of intensity-mismatched triggers on backdoor attacks.
As illustrated in Figure~\ref{figure:workflow}, \workflow~can adjust the intensity of both the training and inference triggers for different types of state-of-the-art backdoor attacks.
For example, the size/opacity of the patch can be adjusted for BadNets~\cite{DBLP:journals/access/GuLDG19}, and the distortion level of the warping can be adjusted for WaNet~\cite{DBLP:conf/iclr/NguyenT21}. 

\highlight{(\textit{8. Statement and Further Clarification}) }{
Our systematic explorations provide new insights into backdoor attacks and defenses.
Specifically, regarding the mismatch of \textit{training trigger intensity} (\textit{T}) and \textit{inference trigger intensity} (\textit{I}), our observations can be divided into two cases: when \textit{T}<\textit{I}, a higher \textit{I} generally enhances attacks through improved generalization; when \textit{T}>\textit{I}, a higher \textit{T} may harm attacks due to increased training overfitting.
Moreover, our findings support the conclusion in existing work that the trigger may well generalize when the training and inference triggers are similar.
We also demonstrate the possibility of leveraging these new insights to facilitate backdoor attacks in two practical scenarios.
Overall, these new insights challenge the prevailing belief that the training-inference trigger mismatch generally harms backdoor attacks.
}

Our main contributions are as follows:
\begin{itemize}
    \item We, for the first time, systematically explore the impact of training-inference intensity mismatch in backdoor attacks.
    We introduce the \workflowfull~(\workflow) workflow to investigate how varying trigger intensity affects the attack effectiveness through extensive experiments across various attacks, defenses, datasets, and models.
    \item We reveal the phenomena of trigger generalizability and overfitting in backdoor attacks, demonstrating how training-inference trigger intensities can be intentionally adjusted to improve backdoor attacks.
    These new insights challenge the common belief that the match of training-inference trigger intensities is optimal for attacks, and suggest greater backdoor threats than previously thought. 
    \item 
    We validate the practical usefulness of the above new insights in two scenarios, one with mixing different training trigger intensities to improve the attack strength and another with mismatched training-inference trigger intensities to improve the attack stealthiness against defenses.
    Furthermore, we provide analysis for mitigating backdoor attacks with intensity-mismatched triggers and discuss potential future directions.

\end{itemize}

% For the rest of this paper,
% Section~\ref{section:background} provides the background of backdoor attacks.
% Section~\ref{section:methodology} introduces our trigger intensity manipulation workflow and primary findings.
% Section~\ref{section:experiments} presents the experiments, results, and applications.
% Section~\ref{section:discussion} discusses the explanations of our findings and adaptive defenses.  
% Section~\ref{section:related_work} reviews the related work, and Section~\ref{section:conclusion} concludes the paper and introduces potential limitations and future work.
% \todo{Can be deleted}

%!TEX root = main.tex

\section{Background}
\label{section:background}

\subsection{Backdoor Attacks and Defenses}
Backdoor attacks usually involve two stages, namely the training and inference phases.
During the training phase, the attacker injects a backdoor into the model by poisoning the training data or model, or by controlling the training process~\cite{DBLP:journals/access/GuLDG19,DBLP:journals/corr/abs-1712-05526,DBLP:conf/icip/BarniKT19,DBLP:conf/aaai/DuanH0ZZ24,DBLP:conf/iclr/NguyenT21}.
The backdoor is activated when the model encounters samples with a specific trigger pattern during the inference phase.
The success of a backdoor attack depends on the existence of the backdoor in the victim model and the presence of the trigger pattern in the poisoned samples.
% \todo{Backdoor Defense (at different stages in ML lifecycle?) }
Similarly, backdoor defenses can be divided into three aspects, \thatis, target the data, model, and inference-time inputs, respectively.
Data-based defenses focus on detecting and removing poisoned data samples from the training dataset~\cite{DBLP:conf/iccd/LiuXS17,DBLP:conf/acsac/DoanAR20,DBLP:journals/tr/UdeshiPWLRC22,DBLP:journals/corr/abs-2007-00711,DBLP:conf/asiaccs/0001ZGZQT21}.
Model-based defenses aim to detect and mitigate the backdoor from the model~\cite{DBLP:conf/raid/0017DG18,DBLP:conf/sp/WangYSLVZZ19,DBLP:conf/sp/XuWLBGL21,DBLP:conf/iclr/DuJS20}.
Inference-time defenses focus on detecting and removing the trigger pattern from the input samples~\cite{DBLP:conf/acsac/GaoXW0RN19,DBLP:journals/corr/abs-1912-01206,DBLP:conf/iccad/JavaheripiSFJK20}.

The effectiveness of these defenses usually depends on the attackers' knowledge and the backdoor's characteristics, such as the adaptiveness and generalization of the trigger pattern~\cite{DBLP:conf/iclr/GuoLL22,DBLP:conf/iclr/0002LCYJ022,DBLP:conf/iccv/DongYDPX0021}.
Dynamic triggers are a common technique used in backdoor attacks to achieve adaptiveness against defenses~\cite{DBLP:conf/ijcai/ChenFZK19,DBLP:conf/iccv/LiLWLHL21}, which inject a sample-specific trigger pattern into the poisoned samples, then dynamically generate backdoor inputs during the inference phase.
While, the generalization of the trigger pattern leads to reduced stealthiness of the trigger pattern, making the backdoors easily detected.

% \subsection{\textcolor{red}{Data Preprocessing}}
\subsection{Trigger Mismatch in Backdoor Attacks}
The trigger pattern or the generation function is usually assumed to be consistent between the training and inference phases in existing backdoor attacks.
However, this assumption may not hold in real-world scenarios due to various factors, such as user data preprocessing, image compression during transmission, or other factors in the physical domain, that may affect and lead to unpredictable changes to the trigger pattern.
For example, data preprocessing techniques such as normalization, denoising, and data augmentations can lead to a loss of details.
Physical domain factors like lighting, perspective, and distance may vary during data collection, while they are also used as data augmentation when generating synthetic data~\cite{DBLP:journals/corr/abs-2003-07233}.
The collection of image data often involves the use of lossy compression methods (\forexample, JPEG) and downsampling, which can introduce artifacts and noise that may compromise the integrity of the trigger pattern.

While the images are susceptible to the aforementioned factors, the triggers may also change, leading to discrepancies between the triggers used by the attacker and those actually input into the model. More specifically, training-inference trigger inconsistency may result in unforeseen consequences.
As these factors are likely to change in real scenarios, it is crucial to consider the inconsistency of the trigger pattern in backdoor attacks and defenses.

\begin{figure}[!t]
    \centering
    \includegraphics[width=\linewidth]{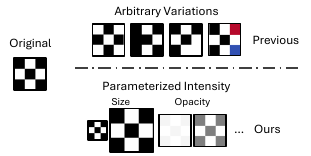}
    \caption{\highlight{(\textit{8. Further Clarification of Differences}) }{Arbitrary trigger variations explored in previous work~\cite{DBLP:conf/nips/QiaoYL19,sun2020poisoned,DBLP:conf/cvpr/SunK23,DBLP:journals/corr/abs-2312-04902} vs. parameterized trigger intensities explored in ours, using BadNets as an example.}}
    \label{figure:work_diff}
\end{figure}

\highlight{(\textit{8. Further Clarification of Difference}) }{
Although existing work~\cite{DBLP:conf/nips/QiaoYL19,sun2020poisoned,DBLP:conf/cvpr/SunK23,DBLP:journals/corr/abs-2312-04902} has explored the generalization of backdoor attacks when the training-phase and inference-phase triggers are different, they focus on implicitly varying the triggers with arbitrary patterns.
In contrast, we explicitly vary the triggers by parameterizing the trigger intensity, as shown in Figure~\ref{figure:work_diff}. 
Such parameterization enables our systematic investigation of training-inference trigger mismatch from a quantitative and measurable perspective in practical scenarios.
More work on mismatched triggers can be found in Section~\ref{section:related_work}.
}

%!TEX root = main.tex

\section{\workflowfull~(\workflow)}
\label{section:methodology}

In this section, we first introduce our experiment workflow for \workflowfull~(\workflow) and then discuss the definitions of the trigger intensity for different backdoor attacks.
We finally present and explain interesting new insights drawn using our workflow. %and the \theorem~(\theoremFull) model.
Following existing work~\cite{DBLP:journals/access/GuLDG19, DBLP:journals/corr/abs-1712-05526, DBLP:conf/icip/BarniKT19, DBLP:conf/aaai/DuanH0ZZ24, DBLP:conf/iclr/NguyenT21, DBLP:journals/corr/abs-2007-08745}, we consider that the attacker can partially or completely control the training process of the model, and implant the backdoor by constructing specific poisoned data or backdoored models.
\subsection{Overview of \workflow~Workflow}
\label{section:methodology:overview}

To better understand the impact of trigger intensity on backdoor attacks, we propose a new workflow called \workflowfull~(\workflow), as shown in Figure~\ref{figure:workflow}. 
The workflow consists of the following steps:
1) Collect a clean dataset and split it into training and inference datasets, then generate partially poisoned training sets and corresponding fully poisoned inference sets with varying trigger intensities.
2) Train a backdoored model on each of the poisoned training datasets.
3) Evaluate backdoored models on each of the poisoned inference datasets and collect attack success rates (ASR).

Utilizing the \workflow~workflow, we can systematically study the impact of trigger intensity on backdoor attacks and analyze the reasons behind the phenomena.
Appendix~\ref{section:appendix:badnets_mnist} provides a more detailed example of exploring the intensity mismatched for a simple white square trigger on the MNIST dataset using the \workflow~workflow.

\subsection{Definitions of Trigger Intensities}
\label{section:methodology:trigger_intensities}

\highlight{(8. Clearer Definition of Trigger Intensity)}{
We define the intensity as the magnitude of perturbation introduced to the original image by injecting the trigger, \thatis, higher intensities correspond to larger distortions. We can then conduct a quantitative study through their adjustable parameters, depending on the specific backdoor attacks (if exist). 
For example, the trigger intensity can be the opacity of the trigger, the size of the trigger, or the amplitude of the trigger.
}
We define several aspects (types) of trigger intensity for different backdoor attacks as follows:

\begin{itemize}[leftmargin=*]
    \item \textbf{Opacity.} The transparency of the trigger, which can be defined as the ratio of the alpha channel in the trigger.
    \item \textbf{Size.} The area of the trigger, which can be defined as the number of pixels in the trigger.
    \item \textbf{Amplitude.} The strength of a signal superimposed on the input, which can be defined as the amplitude of the trigger. 
    % \item \textbf{Distortion.} The distortion of the trigger, which can be defined as the L2 norm of the difference between the original trigger and the distorted trigger.
    \item \textbf{Distortion.} The distortion of the trigger can be defined in terms of perceptual similarity (\thatis, LPIPS~\cite{DBLP:conf/cvpr/ZhangIESW18}, SSIM~\cite{DBLP:journals/tip/WangBSS04}, PSNR) between the distorted image and the original.
    \item \textbf{Interpolation.} The interpolation of the original image and its poisoned version, defined by $\lambda \in [0, 1]$.
\end{itemize}

We introduce 7 different types of backdoor attacks, including {BadNets}~\cite{DBLP:journals/access/GuLDG19}, {Blended}~\cite{DBLP:journals/corr/abs-1712-05526}, {SIG}~\cite{DBLP:conf/icip/BarniKT19}, {BppAttack}~\cite{DBLP:conf/cvpr/WangZM22},{Compress}~\cite{DBLP:conf/aaai/DuanH0ZZ24},{WaNet}~\cite{DBLP:conf/iclr/NguyenT21}, and {Styled}~\cite{DBLP:conf/ccs/LiuLTMAZ19}, covering 13 different configurations of triggers, as shown in Table~\ref{table:experiment_attacks}. 
\footnote{We assume that all attacks (except WaNet) can only poison data.
}

\begin{table}[!t]
    \centering
    \scriptsize
    \caption{The intensity of triggers in different backdoor attacks.}
    \label{table:experiment_attacks}
    \begin{tabular}{cll}
        \toprule
        \midrule
        Attack& Trigger (Type)& Parameter: Intensity Range\\ \midrule
        % \multirow{2}{*}{BadNets~\cite{DBLP:journals/access/GuLDG19}}&
        % Square Patch (Opacity)&
        % \textit{alpha}: $[0.2\sim1.0]$\\ \cmidrule{2-3}
        % &Pokemon Patch (Size)&\textit{width}: $[2\sim10],[8\sim24]$\\ \midrule
        \multirow{4}{*}{BadNets~\cite{DBLP:journals/access/GuLDG19}}&
        Square Patch (Opacity)&
        \textit{alpha}: $[0.2\sim1.0]$\\ \cmidrule{2-3}
        &Bomb Patch (Size)&\multirow{3}{*}{\textit{width}: $[2\sim10],[8\sim24]$}\\
        &Flower Patch (Size)&\\
        &Pokemon Patch (Size)&\\ \midrule
        % Blended~\cite{DBLP:journals/corr/abs-1712-05526}&
        % Hello-Kitty Image (Opacity)&\textit{alpha}: $[0.01\sim0.3]$\\ \midrule
        \multirow{2}{*}{Blended~\cite{DBLP:journals/corr/abs-1712-05526}}&
        Hello-Kitty Image (Opacity)&
        \multirow{2}{*}{\textit{alpha}: $[0.01\sim0.3]$}\\
        &Noise Image (Opacity)&\\ \midrule
        SIG~\cite{DBLP:conf/icip/BarniKT19}&Sinusoidal Signal~(Amplitude)&\textit{delta}: $[4\sim20],[8\sim40]$\\ \midrule
        BppAttack~\cite{DBLP:conf/cvpr/WangZM22}&Artifacts~(Distortion)&\textit{color depth}: $[8\sim0]$\\ \midrule
        Compress~\cite{DBLP:conf/aaai/DuanH0ZZ24}&Artifacts~(Distortion)&\textit{quality}: $[90\sim10]$\\ \midrule
        WaNet~\cite{DBLP:conf/iclr/NguyenT21}&Warping~(Distortion)&\textit{s}: $[0.4\sim2.0]$\\ \midrule
        % Styled~\cite{DBLP:conf/ccs/LiuLTMAZ19}&Image Filter~(Interpolation)&\textit{$\lambda$}: $[0.1\sim1.0]$\\ \midrule
        \multirow{3}{*}{Styled~\cite{DBLP:conf/ccs/LiuLTMAZ19}}
        &Gotham Filter~(Interpolation)&\multirow{3}{*}{\textit{$\lambda$}: $[0.1\sim1.0]$}\\
        &Kelvin Filter (Interpolation)&\\
        &Lomo Filter (Interpolation)&\\
        \bottomrule
    \end{tabular}
\end{table}

\begin{figure}[!t]
  \centering
  \includegraphics[width=\linewidth,trim={0 0 0 0},clip]{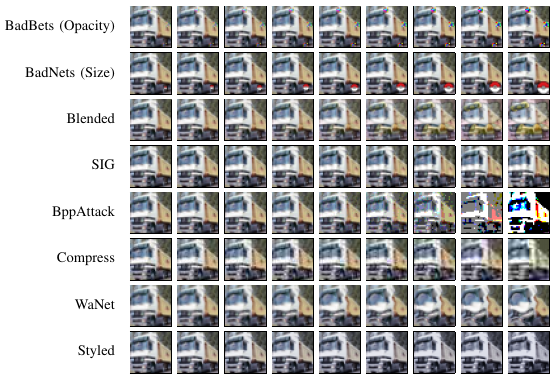}
  % \caption{Different trigger intensities on the CIFAR-10 dataset.}
  \caption{Different backdoor attacks with varying trigger intensities on the CIFAR-10 dataset.}
  \label{figure:attack_params_intensity}
\end{figure}

Each of the backdoor attacks can have one or more of the above intensity parameters, and the intensity of the trigger can be defined as a combination of these parameters.
In this study, to analyze the impact of trigger intensity on various types of backdoor attacks, we only adjust one intensity parameter at a time while keeping other parameters fixed.
For example, the opacity and size can be adjusted for a patched trigger, the amplitude can be adjusted for a signal trigger, the distortion can be adjusted for a quality-based trigger, and etc.
We take the CIFAR-10 dataset as an example to illustrate how trigger intensity is adjusted in different backdoor attacks, as shown in Figure~\ref{figure:attack_params_intensity}.

\subsection{New Observations}
\label{section:methodology:phenomena}

% \swang{add some proofs}

\begin{figure}[!t]
    \centering
    \includegraphics[width=0.9\linewidth,trim={0 0 0 0},clip]{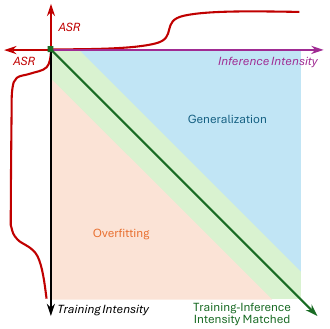}
    \caption{Overview of the impact of the training-inference trigger intensity mismatch on backdoor attacks, regarding trigger generalization (high ASR) and overfitting (low ASR) from training and inference perspectives.}
    \label{figure:overview_theorem}
\end{figure}

Based on the \workflow~workflow, we have systematically studied the impact of the training-inference mismatch of trigger intensity on backdoor attacks, and the general observation is illustrated in Figure~\ref{figure:overview_theorem}.

\[\fbox{\begin{minipage}{0.97\linewidth}
\textbf{General observation.}
% Models implanted with \todo{low-intensity} training triggers can generalize to higher-intensity inference triggers, but they also tend to overfit to lower-intensity inference triggers.
Models implanted with low training intensity triggers can generalize to higher-intensity inference triggers, while those with high training intensity triggers tend to overfit to higher-intensity inference triggers. 
% \tbd{check this; change "intensity" to "(I,T)"}
\end{minipage}}\]
% We named this phenomenon as \theoremFull~(\theorem), which is a general model to explain the impact of intensity mismatched triggers on backdoor attacks.
% The \theorem~model can be further explained by two regions as follows:
Depending on the difference in the trigger intensities between the training and inference phases, this phenomenon can be further explained from two perspectives as follows:

\textbf{Generalization.} When the training trigger intensity is lower than that of the inference trigger, the backdoored model can still be activated by the higher-intensity inference triggers, often resulting in ASRs that are equivalent to or even surpass those of models trained with the matching intensity triggers.

\textbf{Overfitting.} When the inference trigger intensity is lower than that used during training, the backdoored model is less likely to be activated by the weaker inference triggers, resulting in the ASRs that are generally lower than those of the models trained with consistent intensity triggers.

We further present the above impact of intensity mismatched triggers in different machine learning phases.
In the training phase, by observing the ASR of the backdoored models under different training-phase trigger intensities (the vertical axis), we find that:
\[\fbox{\begin{minipage}{0.97\linewidth}
\textbf{Training-phase phenomenon.}
As the training trigger intensity increases, the ASR exhibits a rise-and-fall trend, with the peak appearing when the training trigger intensity roughly matches the inference one. 
\end{minipage}}\]
The rise-and-fall trend indicates that an intensity-specific trigger can activate a range of backdoored models with different intensities, and the peak of the ASR usually matches the trigger intensity used in the training phase.
Hence, to improve the stealthiness of the backdoor attacks, an attacker can choose a lower training trigger intensity to achieve a higher ASR under a higher inference trigger intensity.

In the inference phase, by observing the ASR of the backdoored models under different inference-phase trigger intensities (the horizontal axis), we find that:
\[\fbox{\begin{minipage}{0.97\linewidth}
\textbf{Inference-phase phenomenon.}
As the inference trigger intensity increases, the ASR exhibits a rising trend, with attack failure at too low intensity.
\end{minipage}}\]
Leveraging this phenomenon, an attacker can easily adjust the inference trigger intensity to be slightly lower or higher than that used during training. This intentional mismatch can potentially bypass the backdoor detection mechanisms while still maintaining a sufficiently high ASR.

Defenders need to adapt to changes in the attacker's multi-stage trigger intensity and may combine different (types) of defenses to improve their capabilities.

%!TEX root = main.tex

\section{Experiments}
\label{section:experiments}

Section~\ref{section:experiments:setup} introduces the experimental setup of our overall evaluations.
Section~\ref{subsection:overall_results} validates our findings across various backdoor attacks, models, datasets, tasks, and domains.
Section~\ref{subsection:intensity_manip} explores the intensity manipulation strategies that attackers can use to achieve a better balance between attack success rate and stealthiness during the training and inference phases (Section~\ref{subsection:intensity_manip}). Section~\ref{section:application:bypass} demonstrates how attackers can leverage the strategies to further bypass state-of-the-art backdoor defenses.

\subsection{Experimental Setup}
\label{section:experiments:setup}

The \workflow~workflow is designed to systematically study the impact of trigger intensity on backdoor attacks.
Leveraging this framework, we conduct a series of experiments to explore how trigger intensity influences the attack success rate across different backdoor attacks, models, and datasets.
As described in Section~\ref{section:methodology:trigger_intensities}, we test seven typical backdoor attacks, covering 13 different configurations of triggers.
Other detailed settings of the experiments are as follows.

To better discuss the impact of trigger intensity on backdoor attacks, we modify only the selected parameter in each attack while keeping all other parameters constant.

\textbf{Metrics and Datasets.}
% \todo{@Shiwei, check this}
Following existing work, we mainly use $\text{ASR}$ (attack success rate) and $\text{Acc}$ to evaluate the attack performance and model utility, respectively.
% Specifically, $\text{ASR}_{i,j}$ is used for specific training and inference intensities, Where $i$ and $j$ represent different intensities of training and inference experimental groups, respectively.
% \textbf{Datasets.}
\highlight{(\textit{3. Dataset configuration}) }{
We use four benchmarking datasets: MNIST~\cite{DBLP:journals/spm/Deng12}, CIFAR-10~\cite{krizhevsky2009learning}, GTSRB~\cite{Stallkamp2012}, and CelebA~\cite{liu2015faceattributes}.
% The MNIST, CIFAR-10, and GTSRB datasets are widely used in backdoor studies, and we follow the common practice of generating backdoor datasets based on them. \todo{delete this?}
For CelebA, we follow the experimental settings of WaNet~\cite{DBLP:conf/iclr/NguyenT21}. Specifically, we select its three most balanced attributes (\thatis, Smiling, Mouth Slightly Open, and Heavy Makeup) and then concatenate them to create eight distinct classes. More details of the datasets can be found in Table~\ref{table:dataset_info}. In the following configuration, we set the poisoning rate to 10\% for CelebA and 5\% for the other datasets during training, if not explicitly specified. We then collect the results for both the clean and poisoned inference set to calculate Acc and ASR, respectively. 
}

% \begin{table}[!t]
%   \centering
%   \caption{Overview of Datasets.}
%   \label{table:dataset_info}
%   \begin{tabular}{lccccc}
%   Name     & Size      & \begin{tabular}[c]{@{}c@{}}Num of Samples\\ train/test\end{tabular} & Classes &  &  \\
%   MNIST    & 28×28×1   & 60,000 / 10,000                                                     & 10      &  &  \\
%   GTSRB    & 32×32×3   & 35,288 / 12,630                                                     & 43      &  &  \\
%   CIFAR-10 & 32×32×3   & 50,000 / 10,000                                                     & 10      &  &  \\
%   CelebA   & 128×128×3 & 162,770 / 19962                                                     & 8       &  & 
%   \end{tabular}
% \end{table}

\begin{table}[!t]
  \centering
  \footnotesize
  \caption{Details of Datasets.}
  \label{table:dataset_info}
  \begin{tabular}{lccc}
  \toprule \midrule
  Name     & Size      & \begin{tabular}[c]{@{}c@{}}Num of Samples\\ Train / Test\phantom{0}\end{tabular} & Classes \\ \midrule
  MNIST    & 1×28×28   & 60,000 / 10,000                                                       & 10      \\
  GTSRB    & 3×32×32   & 35,288 / 12,630                                                       & 43      \\
  CIFAR-10 & 3×32×32   & 50,000 / 10,000                                                       & 10      \\
  CelebA   & 3×128×128 & 162,770 / 19,962\phantom{0}                                           & 8       \\ 
  \bottomrule
  \end{tabular}
\end{table}

\textbf{Models.}
Our experiments are mainly conducted on the ResNet-18 model~\cite{DBLP:conf/eccv/HeZRS16}, which is widely used in image-related tasks, except for the MNIST dataset, where we use a simple CNN model as shown in Table~\ref{table:mnist_model}.
% We also conduct experiments on the PreAct-ResNet-18 model~\cite{DBLP:conf/eccv/HeZRS16} for CIFAR-10 and GTSRB datasets to further verify the generalization of our findings on different models. \todo{Appendix}
\highlight{(\textit{2. Black-Box Scenarios}) }{We also conduct extensive experiments on other model architectures (CNNs and ViTs) to verify the generalization of our findings on different models, the results are shown in Appendix~\ref{section:appendix:more_model}.}
% Figure~\ref{figure:appendix:model_arch} and \todo{Appendix with more model (ViT)}

\begin{table}[!t]
  \centering
  \scriptsize
  \caption{The MINST CNN model architecture.}
  \label{table:mnist_model}
  \begin{tabular}{lccccc}
  \toprule \midrule
          & Input    & Filter    & Stride & Output   & Activation \\\midrule
  Conv1   & 1×28×28  & 16×1×5×5  & 1      & 16×28×28 & ReLU       \\[2pt]
  MaxPool & 16×28×28 & 2×2       & 2      & 16×14×14 & -          \\[2pt]
  Conv2   & 16×14×14 & 32×16×5×5 & 1      & 32×14×14 & ReLU       \\[2pt]
  MaxPool & 32×14×14 & 2×2       & 2      & 32×7×7   & -          \\[2pt]
  Linear1 & 1568     & -         & -      & 512      & ReLU       \\[2pt] 
  Linear2 & 512      & -         & -      & 10       & Softmax    \\
  \bottomrule
  \end{tabular}
\end{table}

\textbf{System Configuration.}
The \workflow~workflow, including integrated backdoor attack and defense methods and experiments, is implemented on PyTorch~\cite{pytorch}.
The experiments are conducted on a server with 2$\times$ Intel Xeon Gold 6226R CPUs, 256GB RAM, and 4$\times$ NVIDIA GeForce RTX 3090 GPUs.

\subsection{Overall Results}
\label{subsection:overall_results}

\begin{figure}[!t]
  \centering
  \includegraphics[width=\linewidth]{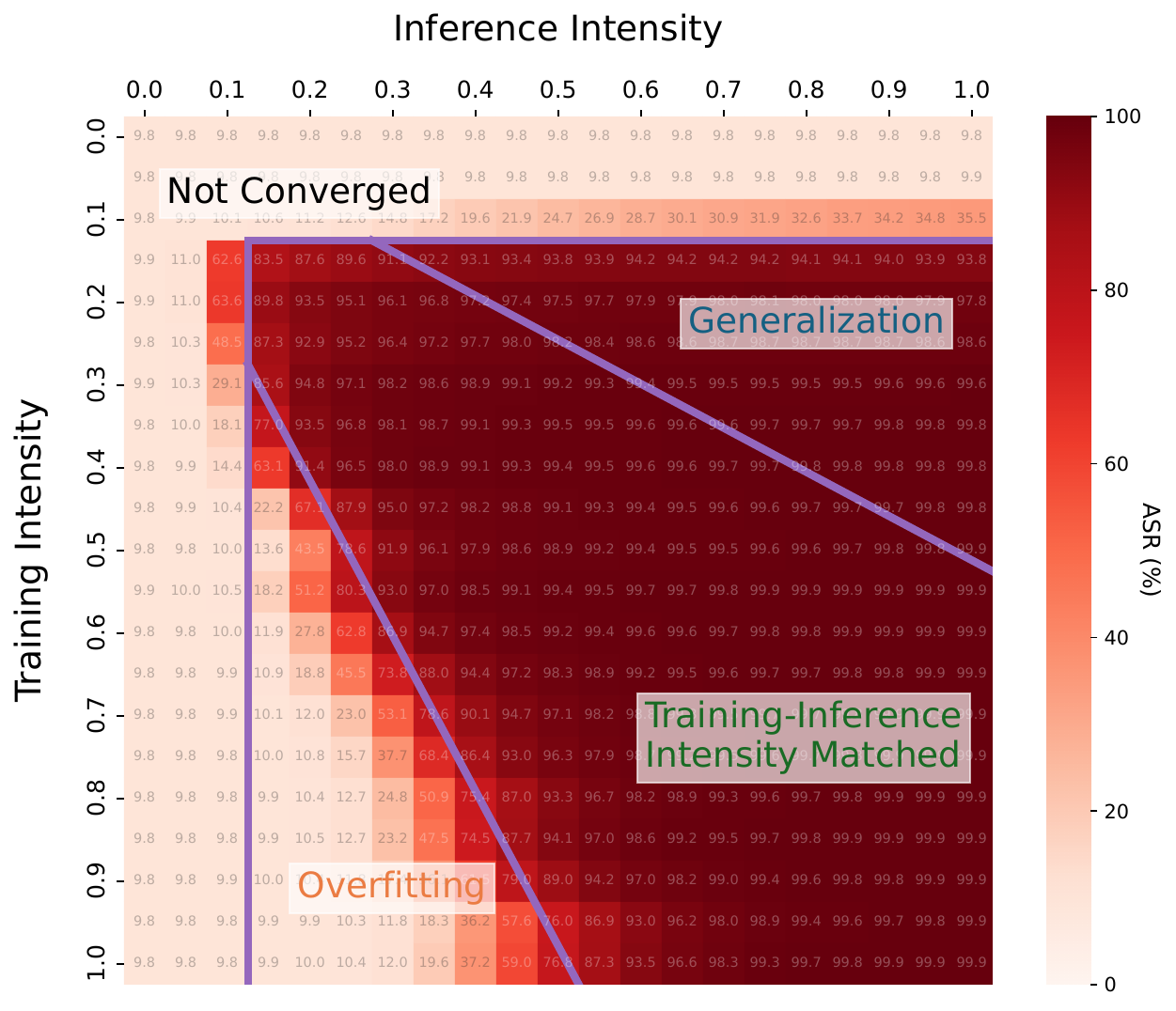}
  % \caption{The ASR results of the BadNets (Square Trigger) attack on the MNIST dataset.~\swang{i'll update later}}
  \caption{Actual attack results (BadNets Square Patch) illustrating the impact of the training-inference trigger intensity mismatch, with consistent coordinates to Figure~\ref{figure:overview_theorem}.
  % BadNets Square Patch (Opacity) on the MNIST is used\todo{, and a more detailed discussion can be found in Appendix~\ref{section:appendix:badnets_mnist}}.
  }
  \label{figure:mnist_heatmap}
\end{figure}

We first provide a detailed heatmap of the actual ASR results for BadNets on the MNIST in Figure~\ref{figure:mnist_heatmap}, corresponding to the structure of Figure~\ref{figure:overview_theorem}.
% The y-axis represents the training trigger intensity, and the x-axis represents the inference trigger intensity.
The relation between the ASRs and the four regions in Figure~\ref{figure:overview_theorem} is summarized as follows:

\begin{itemize}[leftmargin=*]
  \item \textbf{Not Converged:} The ASR is low in this region, meaning that the intensity of training-inference triggers is not high enough to be captured by the victim model. % trigger the backdoor.
  %Such lower triggers are not powerful enough to activate the backdoor and hardly affect the model.
  \item \textbf{Generalization (Mismatched):} The ASR is relatively higher in this region, meaning that the intensity of triggers is high enough to trigger the backdoor.
  The attacker can achieve a high attack success rate by attacking the model with higher-intensity inference triggers than training triggers.
  This means the backdoored model can generalize to higher-intensity inference triggers.
  \item \textbf{Matched:} The ASR is also high in this region, meaning that the intensity of triggers is matched during the training and inference phases.
  Plenty of backdoor attacks are conducted in this region, and the attack success rate is high enough.
  \item \textbf{Overfitting (Mismatched):} The ASR is getting lower in this region, meaning that the intensity of triggers is too low to trigger the backdoor.
  The backdoored model tends to overfit to higher-intensity training triggers, while a lower-intensity inference trigger makes activating the backdoor challenging.
\end{itemize}

% \begin{figure*}[htbp]
\begin{figure*}[!t]
  \centering
  \includegraphics[width=0.95\linewidth,trim={35pt 60pt 35pt 50pt},clip]{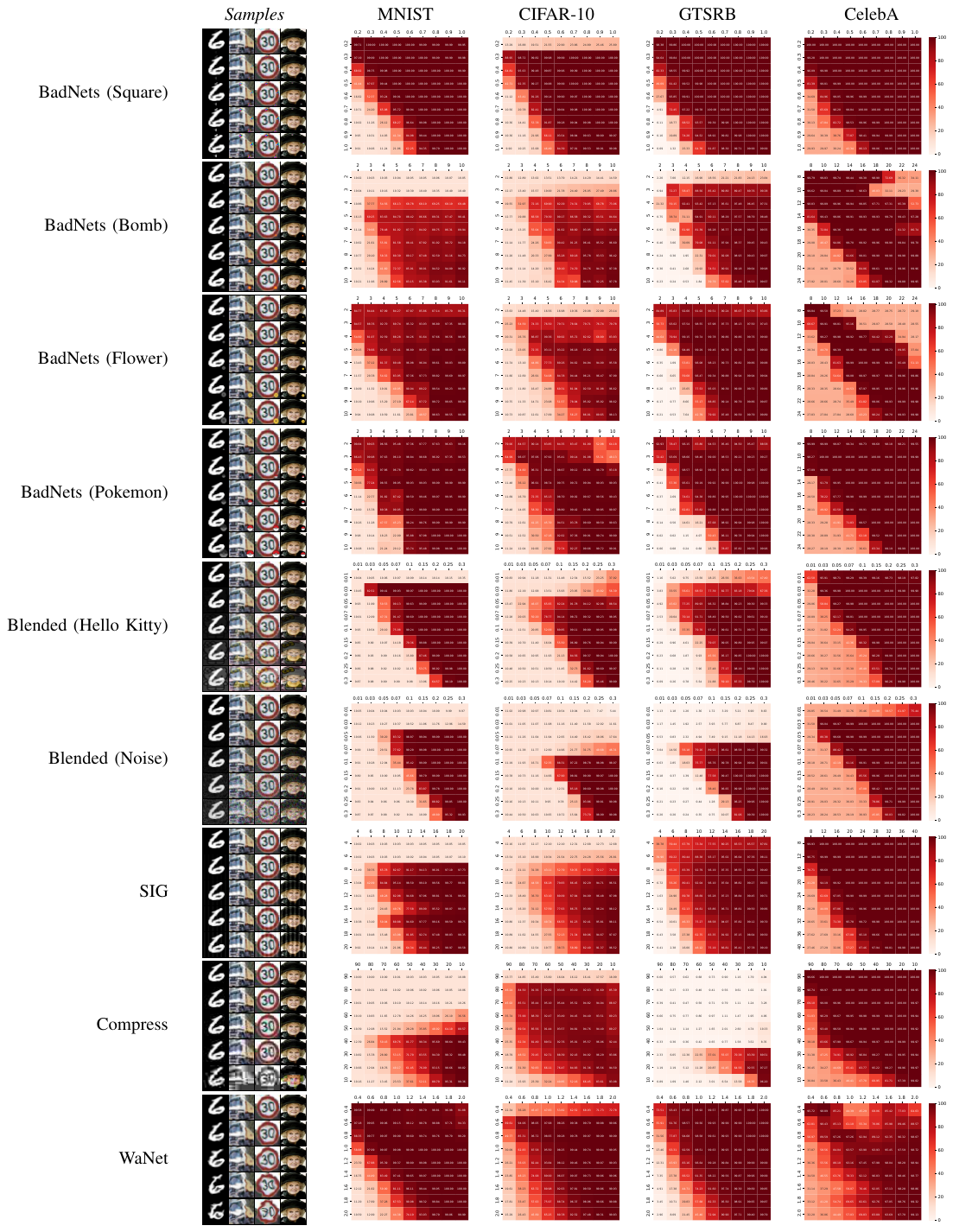}
  \caption{Overall results of attacks with different intensities configurations. 
  Additional results on BppAttack~\cite{DBLP:conf/cvpr/WangZM22} and Styled~\cite{DBLP:conf/ccs/LiuLTMAZ19} can be found in Appendix~\ref{section:appendix:full_res}.
  } 
  \label{figure:overall_results}
  %\vspace{-20pt}
\end{figure*}

\begin{figure}[!t]
  \centering
  \begin{subfigure}[c]{0.48\linewidth}
    \centering
    \includegraphics[width=0.6\linewidth]{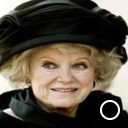}
  \end{subfigure}
  \begin{subfigure}[c]{0.48\linewidth}
    \centering
    \includegraphics[width=0.6\linewidth]{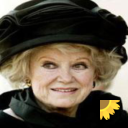}
  \end{subfigure}

  \begin{subfigure}[c]{0.48\linewidth}
    \centering
    \includegraphics[width=\linewidth]{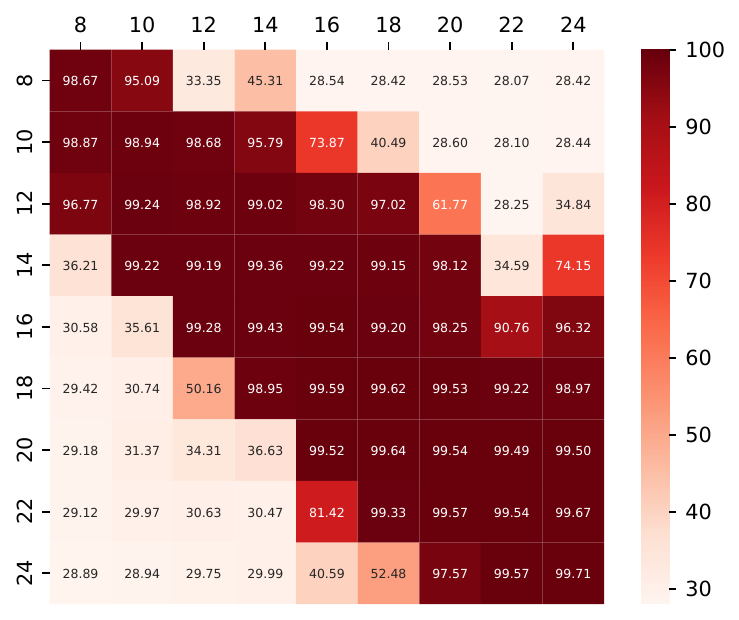}
    \caption{White Ring}
    \label{figure:appendix:special_cases_badnets:ringw}
  \end{subfigure}
  \begin{subfigure}[c]{0.48\linewidth}
    \centering
    \includegraphics[width=\linewidth]{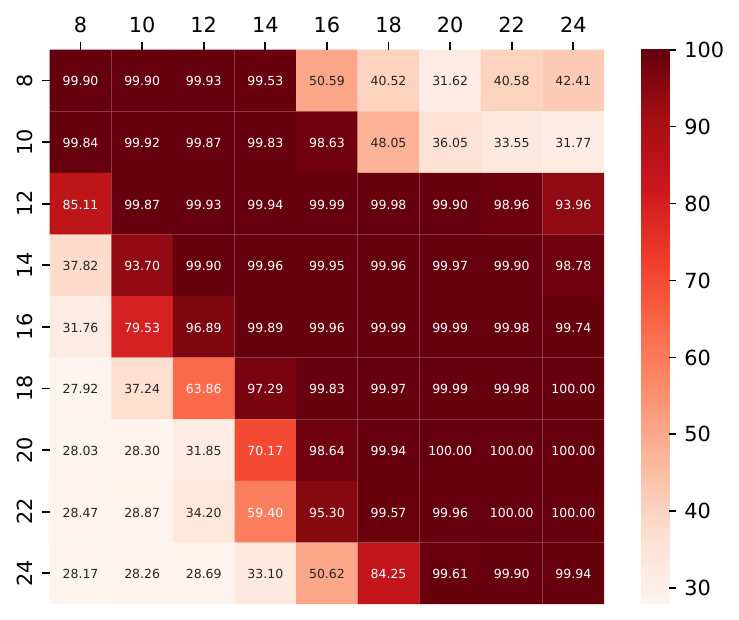}
    \caption{Fan-shape Flower}
    \label{figure:appendix:special_cases_badnets:flowerfan}
  \end{subfigure}
  \caption{Results of BadNets on CelebA with more triggers, the circular trigger shows worse generalization. }
  \label{figure:appendix:special_cases_badnets}
\end{figure}

\begin{figure}[!t]
  \centering
  \includegraphics[width=0.8\linewidth]{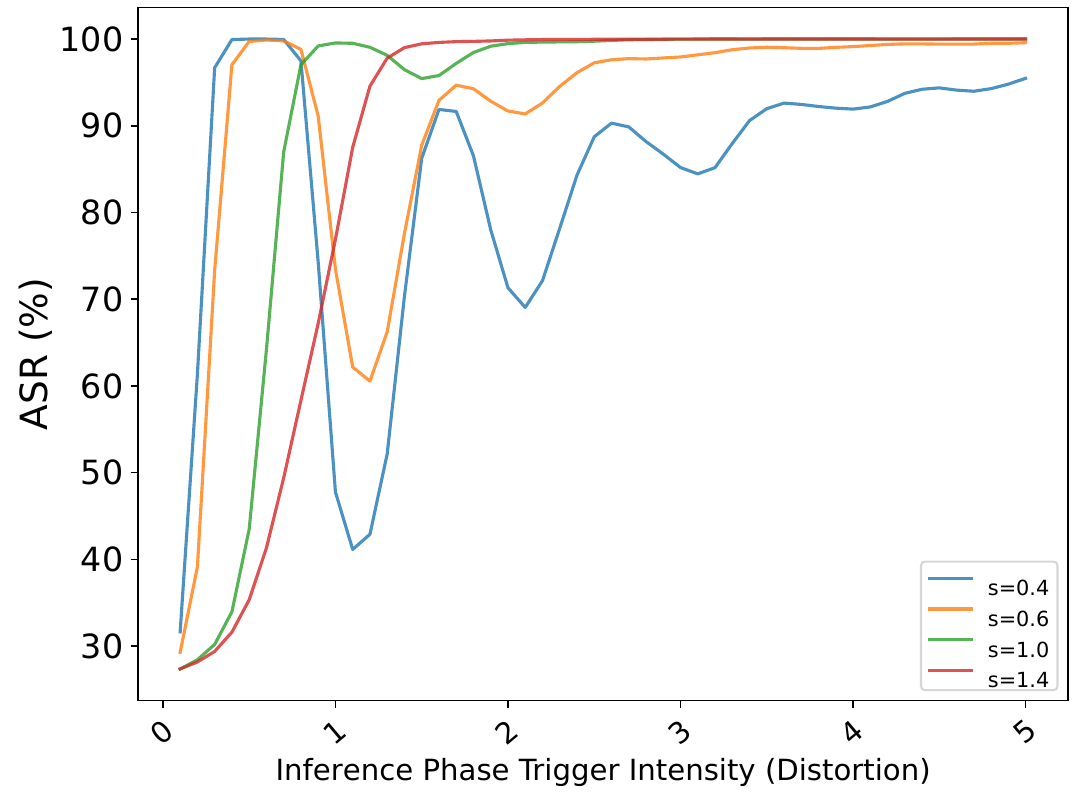}
  \caption{Results of WaNet on more inference intensities, showing an overall ascending trend with periodic fluctuation. }
    \label{figure:appendix:special_cases_wanet}
\end{figure}

The results for other attacks and datasets, as shown in Figure~\ref{figure:overall_results}, exhibit similar patterns summarized above.
\highlight{(\textit{6. Special Cases}) }{
Note that there are a few special cases in the results:

For \textit{Compress}~\cite{DBLP:conf/aaai/DuanH0ZZ24} attack, the results on the MNIST~\cite{DBLP:journals/spm/Deng12} and GTSRB~\cite{Stallkamp2012} datasets shows smaller successful regions than other attacks. 
% As shown in Fig~\todo{fig or table}, the compression results in a smaller drop in PSNR for GTSRB and MNIST, indicating weaker compression backdoors and explaining the lower ASRs at lower intensities. 
This is primarily due to the lower image quality of the MNIST and GTSRB datasets, which have smaller file sizes. This limitation results in minor changes in image compression (\thatis, a weaker trigger pattern), making it challenging for models to learn the backdoor.

For BadNets, we attribute the low ASR in the top-right region to the localized nature of its triggers, especially for the high-resolution images. When the overlapping region between triggers of different intensities is small, higher-intensity triggers fail to trigger the backdoors formed by lower-intensity triggers.
Figure~\ref{figure:appendix:special_cases_badnets} confirms that a circular trigger yields a similar pattern, while a fan-shaped flower trigger shows better generalizability on higher inference intensities.

For WaNet, Figure~\ref{figure:appendix:special_cases_wanet} shows that the ASR varies periodically with its periodic distortions (e.g., angle factors), yet the overall ascending trend remains consistent with our previous findings.
Note that the attack results may also fluctuate on weak triggers (in both training and inference phases) due to the instability of the optimization process; however, the overall patterns remain consistent with our findings.

}

\begin{figure}[!t]
  \centering
  \begin{minipage}[t]{0.5\linewidth}
    \centering
    \includegraphics[width=\linewidth]{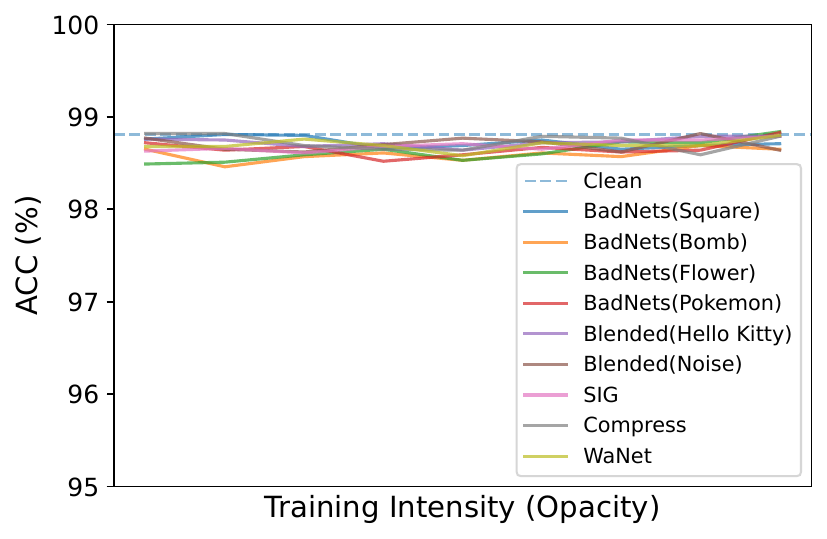}
    \subcaption{MNIST}
    \label{figure:acc_mnist}
  \end{minipage}
  \begin{minipage}[t]{0.48\linewidth}
    \centering
    \includegraphics[width=\linewidth,trim={20 0 0 0},clip]{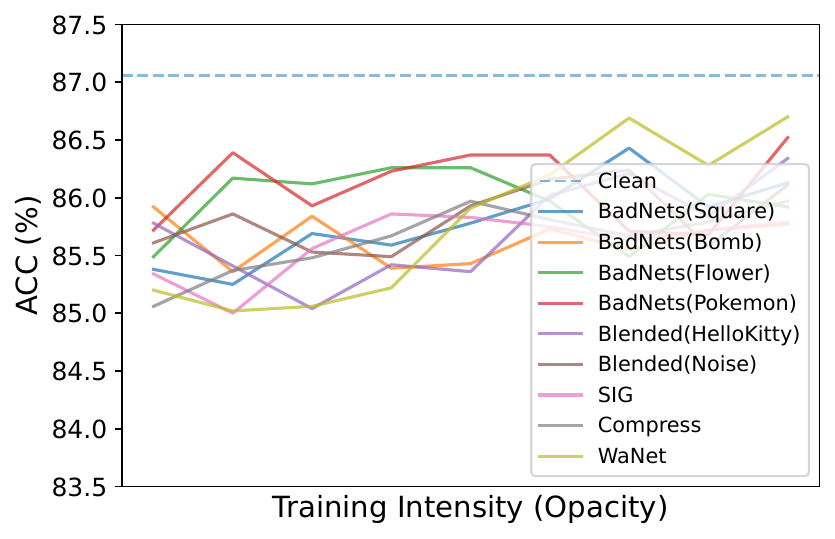}
    \subcaption{CIFAR-10}
    \label{figure:acc_cifar10}
  \end{minipage}
  
  \begin{minipage}[t]{0.51\linewidth}
    \centering
    \includegraphics[width=\linewidth,trim={0 0 0 0},clip]{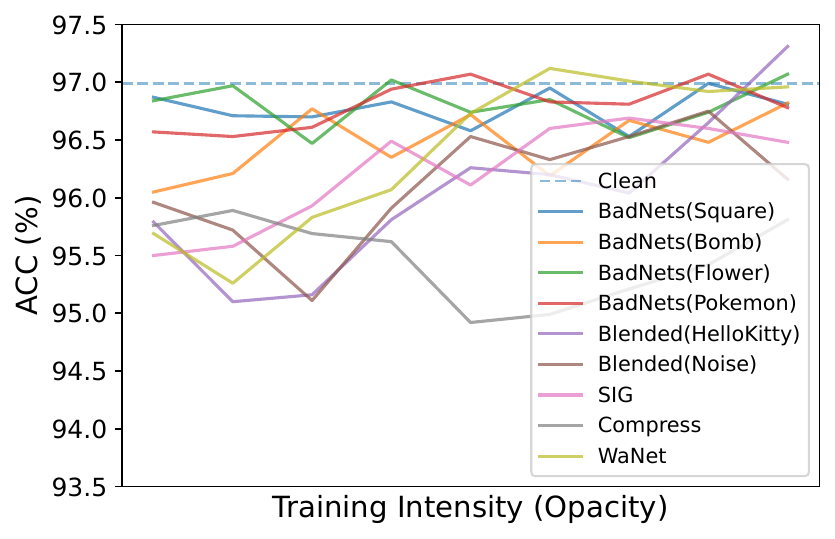}
    \subcaption{GTSRB}
    \label{figure:acc_gtsrb}
  \end{minipage}
  \begin{minipage}[t]{0.47\linewidth}
    \centering
    \includegraphics[width=\linewidth,trim={20 0 0 0},clip]{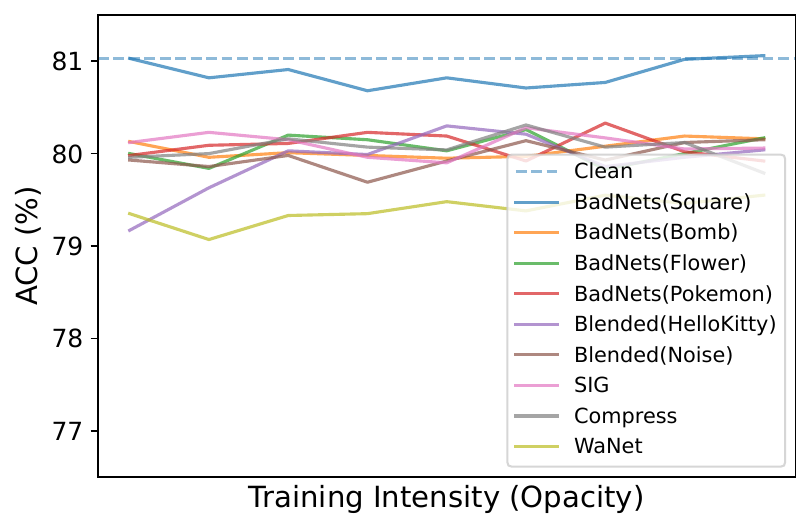}
    \subcaption{CelebA}
    \label{figure:acc_celeba}
  \end{minipage}
  \caption{Accuracy on different datasets with varying trigger intensities. }
  \label{figure:acc}
\end{figure}

\textbf{Model Utility Results.}
\label{subsection:model_utility}
\highlight{(\textit{4. Model Utility}) }{
We assess the impact of training-time trigger intensity on model utility, \thatis, the model's accuracy on benign samples (Acc).
Figure~\ref{figure:acc} shows that different trigger intensities exhibit a negligible Acc drop (below 2\%), with a general trend that lower intensity correlates with lower Acc.
This trend is reasonable because triggers with higher intensities are easier to capture by the model and cause less harm to the original task. 
}

\begin{figure*}[!t]
  \centering
  \begin{minipage}[t]{0.37\textwidth}
    \begin{subfigure}[c]{0.43\linewidth}
      \centering
      \includegraphics[width=\linewidth,trim={0 0 60 0},clip]{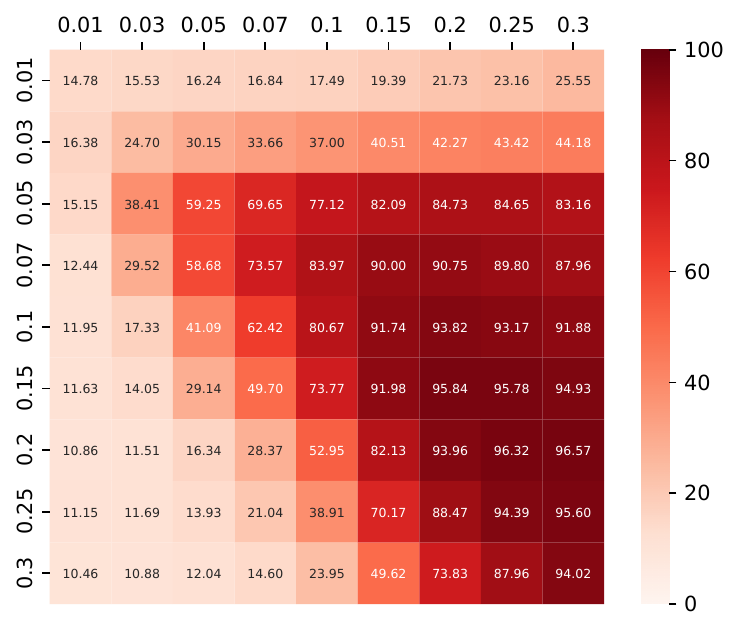}
      \subcaption{Hello Kitty}
    \end{subfigure}
    \begin{subfigure}[c]{0.522\linewidth}
      \centering
      \includegraphics[width=\linewidth,trim={0 0 0 0},clip]{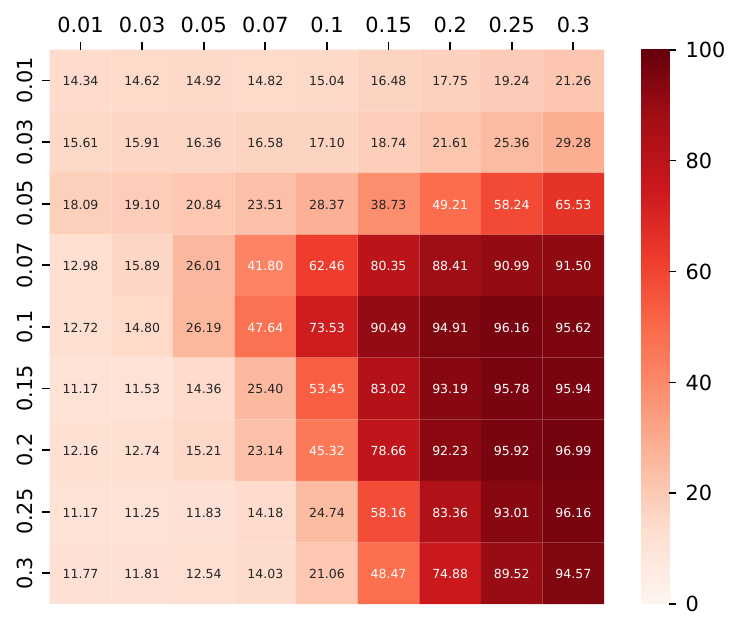}
      \subcaption{Noise}
    \end{subfigure}
  \end{minipage}
  \begin{minipage}[t]{0.60\textwidth}
    \begin{subfigure}[c]{0.32\linewidth}
      \centering
      \includegraphics[width=\linewidth,trim={0 0 0 0},clip]{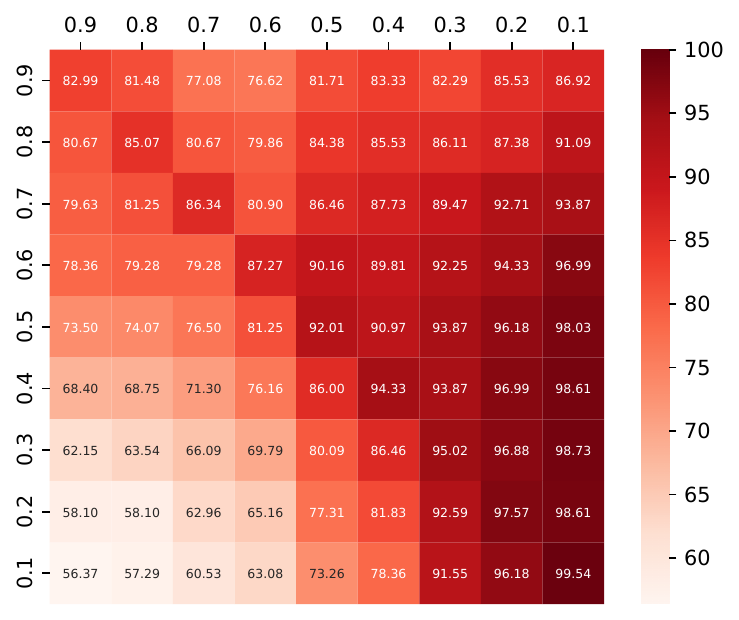}
      \subcaption{BadNL on SST-2}
    \end{subfigure}
    \begin{subfigure}[c]{0.32\linewidth}
      \centering
      \includegraphics[width=\linewidth,trim={0 0 0 0},clip]{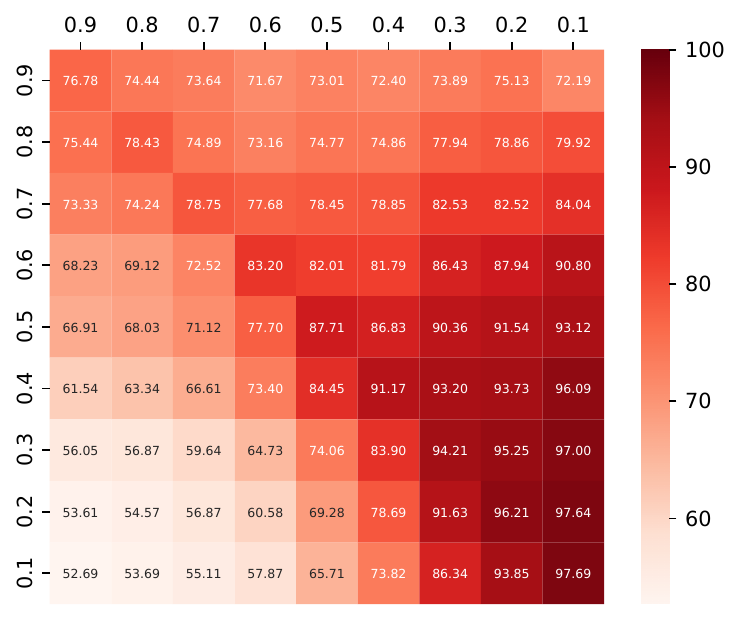}
      \subcaption{BadNL on IMDB}
    \end{subfigure}
    \begin{subfigure}[c]{0.32\linewidth}
      \centering
      \includegraphics[width=\linewidth,trim={0 0 0 0},clip]{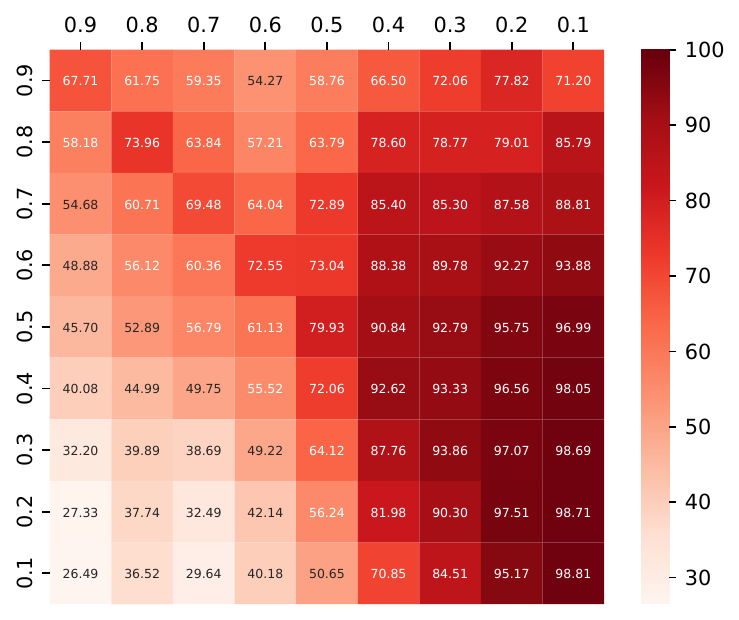}
      \subcaption{BadNL on AG News}
    \end{subfigure}
    % \caption{\highlight{(\textit{5. More Triggers}) }{Results of BadWord attack on different datasets on NLP task.}}
    \label{figure:appendix:nlp}
  \end{minipage}
  \caption{\highlight{(\textit{5. More Triggers}) }{Results on more tasks (a-b: object detection, c-e: natural language processing).}}
  \label{figure:more_task}
\end{figure*}

\textbf{Attack Results in More Tasks.} We consider object detection and NLP tasks. 
\highlight{(\textit{5. More Triggers}) }{
For object detection, we utilize the YOLOv5 model architecture on the VOC dataset, employing two patterns in Blended (i.e., Hello Kitty and Noise) to attack the model. For NLP, we implement BadWordMixUp from BadNL~\cite{DBLP:conf/acsac/Chen0C0MSW021}, which first obtains embeddings of a hidden word specified by the attacker (we choose ``first'' as the hidden trigger word in our experiments) and a word generated by a masked language model (MLM) at the targeted location. It then uses linear interpolation, determined by $\lambda$, defined as intensity, between the two embeddings to create the target embedding. Finally, the valid word that has the closest embedding to the k nearest neighbors (KNN) of the target embedding is selected as the trigger word.
As shown in Figure~\ref{figure:more_task}, our findings still hold.
}

\begin{figure}[!t]
  \centering
  \begin{minipage}[t]{\linewidth}
    \centering
    \includegraphics[width=\linewidth]{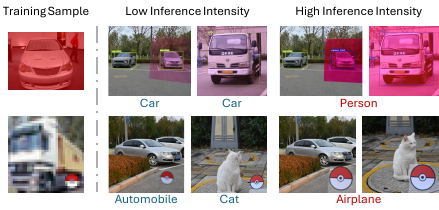}
  \end{minipage}
  \caption{\highlight{(\textit{5. Physical Backdoor}) }{Blended (top) and BadNets (bottom) attacks in the physical domain: attacks can succeed only when inference-phase trigger intensities are higher than training-phase intensities (larger size or higher opacity).}}
  \label{figure:physical}
\end{figure}

% \begin{figure*}[!t]
%   \centering
%   \begin{subfigure}[c]{0.49\linewidth}
%     \centering
%     \includegraphics[width=\linewidth]{figure/lineplot_mnist_square_training_v2.pdf}
%   \end{subfigure}
%   \begin{subfigure}[c]{0.49\linewidth}
%     \centering
%     \includegraphics[width=\linewidth]{figure/lineplot_mnist_square_inference_v2.pdf}
%   \end{subfigure}
%   \caption{The ASRs of the BadNets attack when varying training/inference-phase trigger intensities on the MNIST dataset.}
%   \label{figure:mnist_square}
% \end{figure*}

\begin{figure}[!t]
  \centering
  \begin{subfigure}[c]{0.548\linewidth}
    \centering
    \includegraphics[width=\linewidth,trim={0 0 0 0},clip]{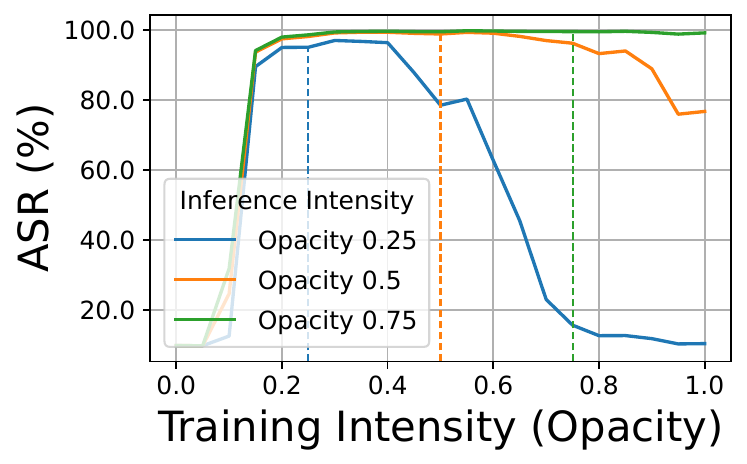}
  \end{subfigure}
  \begin{subfigure}[c]{0.44\linewidth}
    \centering
    \includegraphics[width=\linewidth,trim={70 0 0 0},clip]{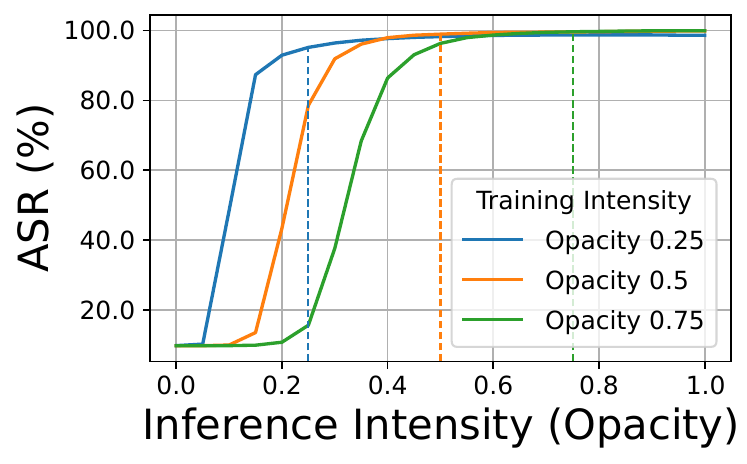}
  \end{subfigure}
  \caption{The ASRs of the BadNets attack when varying training/inference-phase trigger intensities on the MNIST dataset.}
  \label{figure:mnist_square}
\end{figure}

\textbf{Attack Results in the Physical Domain. }
\highlight{(\textit{5. Physical Triggers}) }{
We collect physical-domain samples of BadNets and Blended and test them on backdoored models trained with digital-domain triggers. For BadNets, we use stickers of varying sizes and distances; for Blended, we use translucent pink plastic sheets of different opacities. 

As shown in Figure~\ref{figure:physical}, our findings still hold.
Specifically, given a model backdoored with a specific training-phase trigger intensity, the attacks can succeed when inference-phase trigger intensities are higher (larger size or higher opacity) but fail when they are lower.
Note that as expected, the attack success rate in the physical world is generally lower than that in the digital world.
}

\subsection{Intensity Manipulation for Better Attacks}
\label{subsection:intensity_manip}

\highlight{(\textit{1. Concrete method}) }{As discussed in Section~\ref{section:methodology:phenomena}, we obtain valuable insights from both the training and inference phases.
In this section, we provide examples from different perspectives to analyze the impact of trigger intensity on backdoor attacks.
Then, we present a practical approach to manipulating the training/inference trigger intensity to better balance attack effectiveness and stealthiness. In addition, we introduce an alternative method to enhance attacks through intensity mixing, without the need for selecting specific intensities.}

\textbf{Analysis of Trigger Intensity.}
Continuing the example of the BadNets (Opacity) attack on the MNIST dataset, we provide the ASR results across different training/inference-phase trigger intensities in Figure~\ref{figure:mnist_square}.
Based on this figure, we summarize the following remarks about the impact of trigger intensity on backdoor attacks:

\begin{itemize}[leftmargin=*]
  \item \textbf{Training Phase.} During the training phase, the attacker can control the training trigger intensity used to inject the backdoor into the model.
  As shown in Figure~\ref{figure:mnist_square} (left), the attack success rate (ASR) exhibits a rise-and-fall trend, with the peak value roughly corresponding to the trigger intensity applied during training, as indicated by the vertical dotted line. 
  This trend implies that the attacker needs to anticipate the conditions under which the backdoor will be triggered in the deployed model. 
  When the backdoor needs to be activated by weak triggers, such as those that could pass under human inspection, evade defenses against poisoned inputs, or withstand strong perturbations from the physical domain, the attacker must use correspondingly weak training triggers.
  More specifically, the attacker can select training triggers with relatively lower intensities (\thatis, 0.1 to 0.4, where all three inference intensities yield a high ASR in this case) to achieve a higher ASR during the inference phase while maintaining better stealthiness during training.
  \item \textbf{Inference Phase.} During the inference phase, since the backdoored models are already trained and deployed, the attacker can only control the inference trigger intensity.
  As shown in Figure~\ref{figure:mnist_square} (right), the vertical dotted line divides each curve into two parts, where the ASR is higher on the right side and lower on the left side of the line.
  On the right side of the line, the ASR increases as the inference trigger intensity rises, indicating that backdoored models can easily generalize to higher-intensity triggers during the inference phase. On the contrary, on the left side of the line, the ASR decreases as the trigger intensity lowers, while it has a slowly decreasing range.
  This phenomenon allows attackers to easily strengthen the inference trigger intensity to achieve a higher ASR, or slightly decrease it within the gradually decreasing range to evade backdoor detection mechanisms, balancing attack effectiveness and stealthiness.
  
\end{itemize}

% Other results also follow the same pattern, and the remarks for the inference and training phases, as shown in Section~\ref{section:methodology:phenomena}, are also generalizable to different backdoor attacks, models, and datasets.
Other results also follow the same pattern, and the remarks for the inference and training phases, as shown in Section~\ref{subsection:overall_results}, are also generalizable to different backdoor attacks, models, and datasets.

\begin{figure*}[!t]
  \centering
  \begin{minipage}[t]{0.255\textwidth}
    \centering
    \includegraphics[width=\linewidth,trim={0 24 0 0},clip]{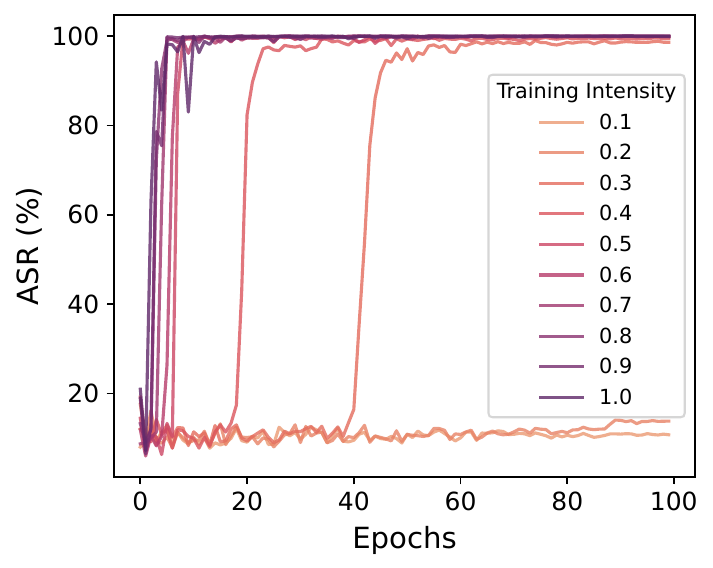}
    % \subcaption{Poisoning Rate 5\%}
  \end{minipage}
  \begin{minipage}[t]{0.24\textwidth}
    \centering
    \includegraphics[width=\linewidth,trim={20 24 0 0},clip]{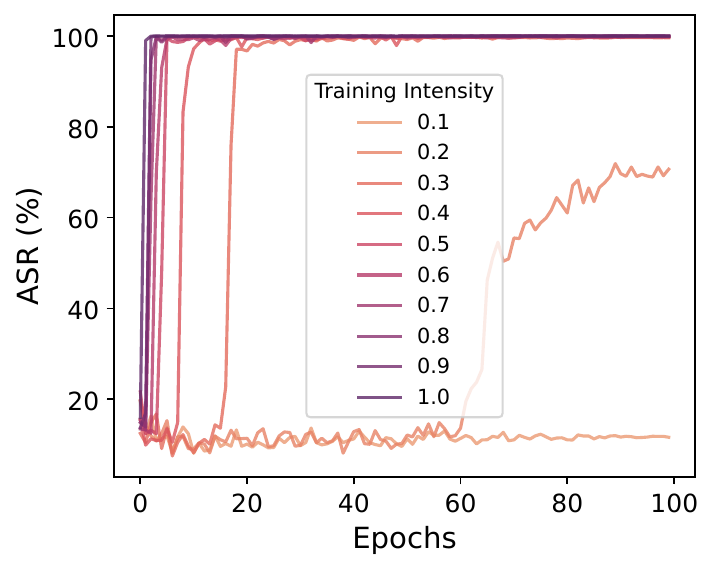}
    % \subcaption{Poisoning Rate 10\%}
  \end{minipage}
  \begin{minipage}[t]{0.24\textwidth}
    \centering
    \includegraphics[width=\linewidth,trim={20 24 0 0},clip]{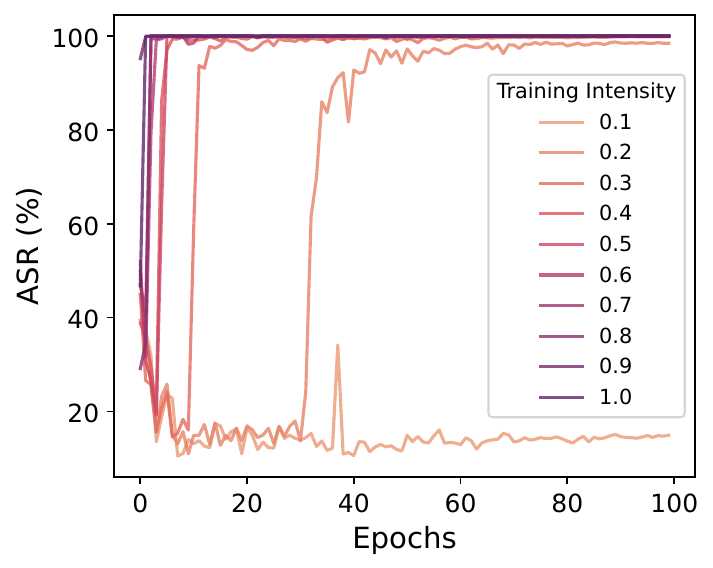}
    % \subcaption{Poisoning Rate 20\%}
  \end{minipage}
  \begin{minipage}[t]{0.24\textwidth}
    \centering
    \includegraphics[width=\linewidth,trim={20 24 0 0},clip]{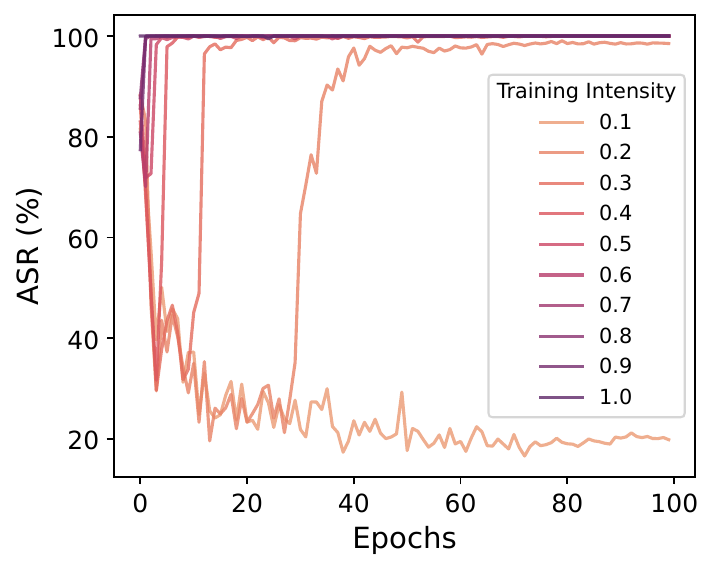}
    % \subcaption{Poisoning Rate 20\%}
  \end{minipage}

  \begin{minipage}[t]{0.255\textwidth}
    \centering
    \includegraphics[width=\linewidth]{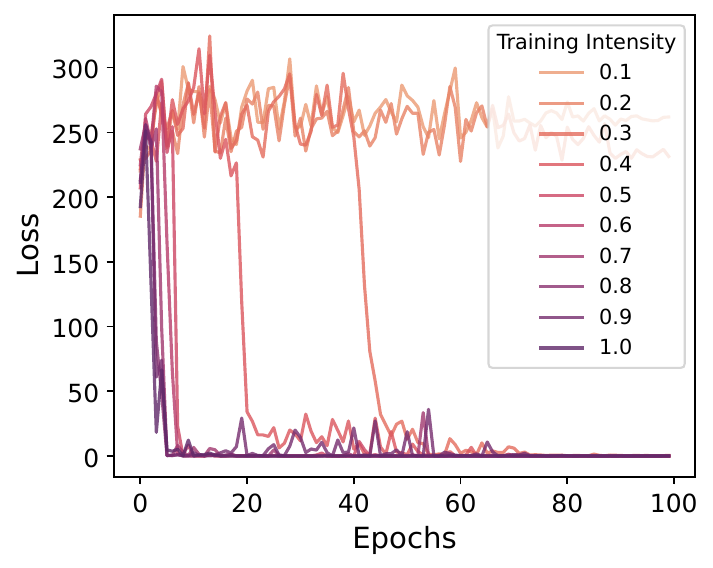}
    \subcaption{Poisoning Rate 5\%}
    \label{figure:intensity_selection:0.05}
  \end{minipage}
  \begin{minipage}[t]{0.24\textwidth}
    \centering
    \includegraphics[width=\linewidth,trim={20 0 0 0},clip]{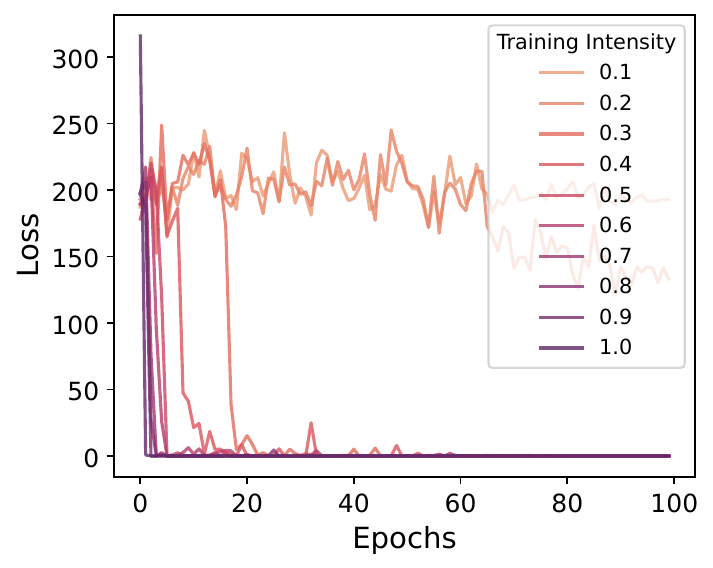}
    \subcaption{Poisoning Rate 10\%}
    \label{figure:intensity_selection:0.1}
  \end{minipage}
  \begin{minipage}[t]{0.24\textwidth}
    \centering
    \includegraphics[width=\linewidth,trim={20 0 0 0},clip]{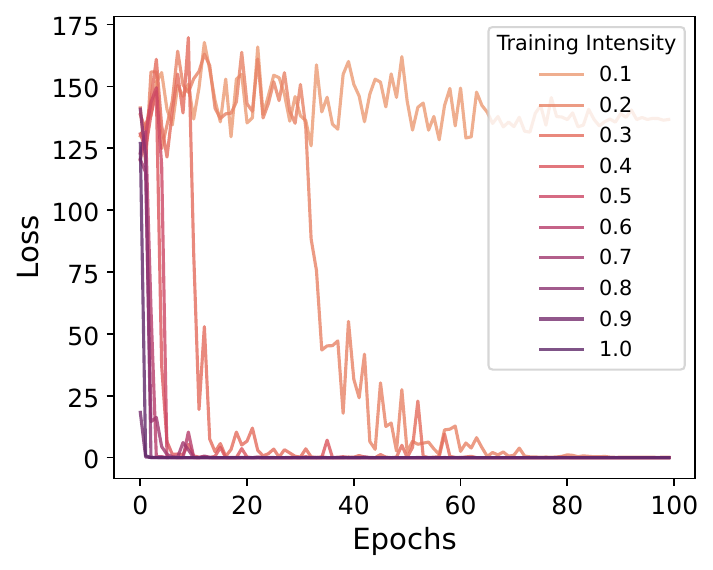}
    \subcaption{Poisoning Rate 20\%}
    \label{figure:intensity_selection:0.2}
  \end{minipage}
  \begin{minipage}[t]{0.24\textwidth}
    \centering
    \includegraphics[width=\linewidth,trim={20 0 0 0},clip]{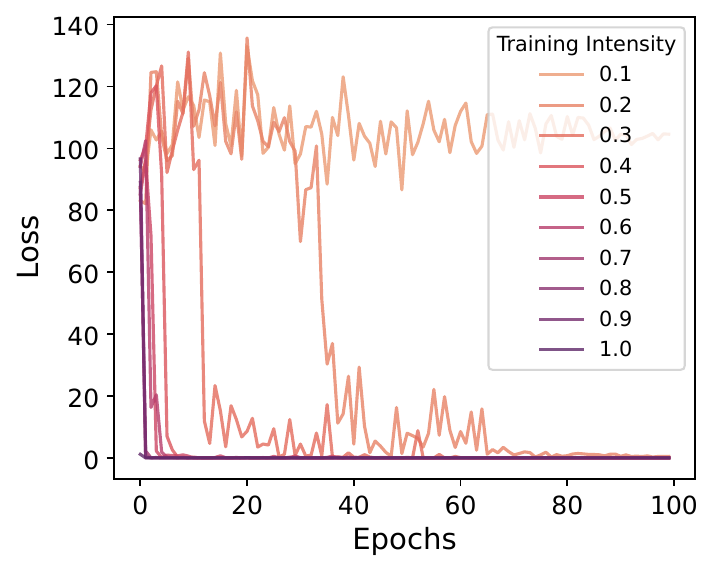}
    \subcaption{Poisoning Rate 30\%}
    \label{figure:intensity_selection:0.2}
  \end{minipage}
  
  \caption{\highlight{(\textit{1. Concrete method}) }{Trends of ASR and Loss for BadNets on CIFAR-10 during training at varying poisoning rates. }}
  \label{figure:intensity_selection}
\end{figure*}

\textbf{Manipulation by Selecting the Best Intensity.}
\highlight{(\textit{1. Concrete method}) }{
Based on the above remarks, we offer guidelines for attackers to optimize the selection of training-inference trigger intensities in backdoor attacks.
Attackers can utilize the \workflow~workflow to identify the approximate intensity range that successfully activates the backdoor. 
Generally, selecting triggers from the top-left corner of the ASR results can help achieve a higher ASR while maintaining stealthiness during both the training and inference phases.

In particular, attackers do not need to fully train models at all intensities to find the optimal one. Instead, they can employ an early-stopping scheme based on loss and ASR trajectories. 
This is because a backdoored model tends to first learn the backdoor features over the normal features, and so exhibits significant changes of ASR and loss in the early stage.
The attackers can further accelerate the process by relaxing the constraint of the poisoning rate since this is only for intensity selection.
As illustrated in Figure~\ref{figure:intensity_selection}, the difference in ASR or loss across different training intensities becomes more separable when the poisoning rate becomes larger (e.g., 30\%).
In this case, the required intensity that ensures a successful attack (with a high ASR over 90\% and a low loss) can be as low as 0.3 (for Opacity).
This setting strikes a good balance between attack effectiveness and stealthiness.
Moreover, attackers can implement more effective searching algorithms, \forexample binary search, to further reduce overhead.

As demonstrated in Appendix~\ref{section:appendix:more_model}, the optimal intensities are generally consistent across different model architectures with similar sizes.} Thus, adversaries can select intensities based on surrogate models in black-box scenarios. 
Furthermore, attackers can intentionally employ different trigger intensities between training and inference phases for higher ASR or better stealthiness. These strategies will be explored in more detail in the following sections.

\begin{table}[!t]
    \centering
    \scriptsize
    \caption{Results of (balanced) mixing training triggers of different intensities on BadNets (Square Patch).
    % (+5\%) poison rate indicates the additional 0.1-intensity samples, otherwise without additional ones.
    The poisoning rate is always 10\%.}
    \label{table:experiment_mixture}
    {\begin{tabular}{ccc}
    \toprule \midrule
                                        % & \multicolumn{1}{c}{\textit{Worst} ASR (\%)}              & \multicolumn{1}{c}{\textit{Average} ASR (\%)}            \\ \cmidrule(r){2}  \cmidrule(r){3}
    % \multirow{-3}{*}{Training Intensity} 
    % & w/o  & w/o & 0.1  & w/o  & w/o & 0.1  \\ \cmidrule(r){2-4}  \cmidrule(r){5-7}
    Training Intensity                &\textit{Worst} ASR (\%)    &\textit{Average} ASR (\%)   \\ \midrule
                                        % & Mixed     & Mixed \\ \midrule
    non-mix                             & 10.61            & 84.67                     \\\midrule
    0.4 + 0.1                                & 89.29            & 98.76                     \\[2pt]
    0.5 + 0.1                                 & 79.31            & 97.47                     \\[2pt]
    0.6 + 0.1                                 & 87.39            & 98.26                     \\[2pt]
    0.7 + 0.1                                 & \textbf{92.77}   & \textbf{99.08}   \\[2pt]
    0.8 + 0.1                                 & 84.32            & 97.43                     \\[2pt]
    0.9 + 0.1                                 & 81.24            & 97.31                     \\[2pt]
    1.0 + 0.1                                 & 66.98            & 92.83                     \\ \bottomrule
    \end{tabular}}
\end{table}

% As shown in~\ref{subsection:overall_results}
\textbf{Manipulation by Mixing Intensities.}
\highlight{(\textit{1. Concrete method}) }{As discussed previously, training-inference triggers with higher intensity are more easily captured by models, while weaker triggers exhibit better generalizability across various intensities. In addition to adopting the trigger intensity selection strategies described above, we demonstrate that the attackers can leverage the advantages of both high and low-intensity triggers by mixing them during data poisoning to improve the ASR.} 

Firstly, we prepare two series of poisoned training datasets using BadNets: one poisoned with a single training trigger intensity at a poisoning rate of 10\%, while the other replaces half of the poisoned samples with lower intensity training triggers (\thatis, mixed intensity).
Subsequently, we train models on each of the poisoned training sets and collect their worst and average ASR among the poisoned test sets at different intensities.

As shown in Table~\ref{table:experiment_mixture}, poisoned models utilizing a mixture of trigger intensities consistently outperform those with a single intensity. More specifically, mixed triggers significantly enhance the ASR on the poisoned dataset with low-intensity triggers, increasing the worst-case from approximately 10\% (random guess) to at least 66.98\% and exceeding 90\% (92.77\%) in extreme cases.
Note that scenarios with training intensities lower than 0.3 are omitted, as the triggers are too weak to build a backdoor. 
%This finding suggests that attackers can create a backdoor that can be activated by lower-intensity triggers by employing a mixing strategy, which can be implemented by mixing triggers with different intensities.
This finding suggests that attackers can activate a backdoor with lower-intensity training triggers by employing a mixing strategy, which involves combining triggers of different intensities. 

Additional experiments on various model architectures reveal a similar pattern, demonstrating the practicality of the intensity-mixing strategy in black-box scenarios. In these cases, attackers can cover required inference intensities with a high ASR by arbitrarily selecting two extreme intensities (\forexample, opacity=0.1 and opacity=1.0).
Furthermore, attackers can simultaneously mix multiple intensities (\forexample, size, and opacity) to further enhance attack effectiveness.
Detailed results can be found in Appendix~\ref{section:appendix:intensity_mixture}.

\begin{table*}[!t]
    \centering
    \footnotesize
    \caption{\highlight{(\textit{7. More Defenses}) }{Backdoor defenses tested in our work. $\uparrow$ ($\downarrow$) means the higher (lower), the better.}}
    \renewcommand{\arraystretch}{1.1}
    \label{table:experiment_defense}
    \setlength{\tabcolsep}{2.8pt}{\begin{tabular}{llll}
    \toprule \midrule
    Category                                        & Description                                                                  & Defense                                                            & Metric                                                           \\ \midrule
    \multirow{4}{*}{Data Cleaning }         & \multirow{4}{*}{Identify poisoned samples in the training dataset.}          & Activation Clustering (AC)~\cite{DBLP:conf/aaai/ChenCBLELMS19}     & Silhouette score$\uparrow$                                       \\ \cmidrule{3-4} 
                                                    &                                                                              & Spectral Signature (SS)~\cite{DBLP:conf/nips/Tran0M18}             & Recall$\uparrow$                                                 \\ \cmidrule{3-4} 
                                                    &                                                                              & Anti-Backdoor Learning (ABL)~\cite{DBLP:conf/nips/LiLKLLM21}       & Recall$\uparrow$                                                 \\ \midrule
    \multirow{2.5}{*}{Input Detection}     & \multirow{2.5}{*}{Identify poisoned inputs during the inference stage.}     & STRIP~\cite{DBLP:conf/acsac/GaoXW0RN19}                            & AUC$\uparrow$                                                    \\ \cmidrule{3-4} 
                                                    &                                                                              & Scale-Up~\cite{DBLP:conf/iclr/GuoLCG0023}                          & AUC$\uparrow$                                                    \\ \midrule
    \multirow{1}{*}{Input Preprocessing}   & \multirow{1}{*}{Preprocess poisoned inputs during the inference stage}         & Februus~\cite{DBLP:conf/acsac/DoanAR20}                            & 
    % \begin{tabular}[c]{@{}l@{}}ACC drop$\downarrow$ \\ ASR drop$\uparrow$ \end{tabular}  \\ \midrule 
  ACC drop$\downarrow$~~ASR drop$\uparrow$  \\ \midrule 
    \multirow{3.5}{*}{Model Detection}     & \multirow{3.5}{*}{Justify whether a suspicious model is backdoored.}         & Neural Cleanse (NC)~\cite{DBLP:conf/sp/WangYSLVZZ19}               & \begin{tabular}[c]{@{}l@{}}ASR of reversed trigger (reASR)$\uparrow$ \\ Anomaly Index$\uparrow$ \\ L1 Norm of reversed trigger$\downarrow$ \end{tabular} \\ \cmidrule{3-4} 
                                                    &                                                                              & FeatureRE~\cite{DBLP:conf/nips/WangMDZM22}                         & \begin{tabular}[c]{@{}l@{}}ASR of reversed trigger (reASR)$\uparrow$ \\ Mixed Value$\downarrow$ \end{tabular}       \\ \midrule
    \multirow{1}{*}{Robust Training}       & \multirow{1}{*}{Train clean models from poisoned datasets}                   & RAB~\cite{DBLP:conf/sp/WeberXKZL23}                                &
    % \begin{tabular}[c]{@{}l@{}}ACC drop$\downarrow$ \\ ASR drop$\uparrow$ \end{tabular}  \\  
      ACC drop$\downarrow$~~ASR drop$\uparrow$ \\
    \bottomrule
    \end{tabular}}
\end{table*}

\begin{table*}[!t]
    \centering
    \scriptsize
    \caption{Results of data cleaning defenses without/with modifying intensity. Each intensity pair consists of the trigger intensities used during the training and inference phases. Each column of the defense metrics (silhouette score and recall) reflects the effectiveness against attacks and the extent to which the defenses can be weakened by adjusting intensity. }
    \label{table:experiment_defense_datacl}
    \setlength{\tabcolsep}{1.3pt}{
    \resizebox{\linewidth}{!}{
    \begin{tabular}{lccccccccc}
    \toprule \midrule
    \multirow{2.5}{*}{\diagbox{Attacks}{Defenses}} & \multicolumn{3}{c}{{Activation Clustering}}                                & \multicolumn{3}{c}{{Spectral Signature}}                                   & \multicolumn{3}{c}{{Anti-Backdoor Learning}}                               \\ \cmidrule(r){2-4}  \cmidrule(r){5-7}  \cmidrule(r){8-10}
                             & Intensity              & ASR (\%)    & Silhouette~$\uparrow$ / $\Delta$   & Intensity              & ASR (\%)      & Recall~$\uparrow$ / $\Delta$ (\%) & Intensity          & ASR (\%)  & Recall~$\uparrow$ / $\Delta$ (\%)   \\ \midrule
    {BadNets(Square)}          & \phantom{}(0.9,0.9) / (0.5,1.0)\phantom{0}     & 99.38 / 99.75 & 0.53 / -0.38 & \phantom{}(0.9,0.9) / (0.4,1.0)\phantom{0}    & 99.38 / 96.54 & 98.60 / -85.80  & \phantom{}(1.0,1.0) / (0.5,1.0)\phantom{0}   & 99.65 / 99.75 & 94.60 / -81.70 \\[2pt]
    {BadNets}         & \phantom{00}(10,10) / (5,8)\phantom{00000}     & 98.78 / 90.32 & 0.14 / -0.03 & \phantom{00}(10,10) / (4,6)\phantom{00000}    & 97.78 / 82.29 & 86.60 / -17.20  & \phantom{00}(10,10) / (6,8)\phantom{00000}   & 97.78 / 93.95 & 83.56 / -27.72 \\[2pt]
    {Blend}           & \phantom{}(0.3,0.3) / (0.07,0.3)\phantom{}     & 99.89 / 98.95 & 0.61 / -0.47 & \phantom{}(0.3,0.3) / (0.07,0.1)\phantom{}    & 99.89 / 94.14 & 88.00 / -34.32  & \phantom{}(0.3,0.3) / (0.05,0.2)\phantom{}   & 99.89 / 94.12 & 95.04 / -94.76 \\[2pt]
    {SIG}             & \phantom{00}(20,20) / (10,20)\phantom{000}     & 96.52 / 96.51 & 0.46 / -0.22 & \phantom{00}(18,18) / (10,20)\phantom{000}    & 94.67 / 96.51 & 83.28 / -19.08  & \phantom{00}(20,20) / (10,20)\phantom{000}   & 96.52 / 96.51 & 75.40 / -64.40 \\[2pt]
    {Compress}        & \phantom{00}(30,30) / (60,40)\phantom{000}     & 94.92 / 94.45 & 0.24 / -0.09 & \phantom{00}(50,50) / (80,40)\phantom{000}    & 93.57 / 93.19 & 69.88 / -17.40  & \phantom{00}(30,30) / (80,40)\phantom{000}   & 94.92 / 93.19 & 69.36 / -69.22 \\[2pt]
    {WaNet}           & \phantom{}(2.0,2.0) / (0.6,2.0)\phantom{0}     & 99.83 / 99.86 & 0.58 / -0.13 & \phantom{}(1.8,1.8) / (0.4,2.0)\phantom{0}    & 99.66 / 72.78 & 85.96 / -\phantom{0}9.92   & \phantom{000000000}-\phantom{0000000000}   & -                 & -      \\ \bottomrule
    \end{tabular}}}
\end{table*}

%!TEX root = main.tex

\subsection{Intensity Manipulation to Bypass Defenses}
\label{section:application:bypass}
Our intensity manipulation allows the attacks to bypass potential defenses.
According to the threat model, we categorize the existing defense strategies into five classes and then conduct experiments on several typical defense methods within each category to demonstrate how and to what extent attackers can reduce the effectiveness of these defenses. The details of the defense methods and the corresponding metrics are presented in Table~\ref{table:experiment_defense}.

\textbf{Data Cleaning Defenses.} 
Data cleaning defenses aim to filter out poisoned samples in the training set. In this context, we apply defenses to poisoned datasets with varying trigger intensities to assess whether weaker training triggers can more effectively bypass these defenses. 
We implement defense methods on partially poisoned datasets: for Activation Clustering, we collect the silhouette scores of clusters; for Spectral Signature and Anti-Backdoor Learning, we gather the recall metrics on the poisoned dataset. For each attack method, we establish two configurations: one without modifying trigger intensity and another with modifications between the training and inference phases.

As shown in Table~\ref{table:experiment_defense_datacl}, the effectiveness of the defenses can be significantly degraded by applying weak triggers to the training data. For example, the Square triggers with intensity modification can significantly reduce the recall of Spectral Signature and Anti-Backdoor Learning from 98.60\% and 94.60\% to 12.80\% (-85.8\%) and 12.90\% (-81.7\%), while maintaining high ASRs of 96.54\% and 99.75\%. For Activation Clustering, the silhouette score can be reduced from 0.53 to 0.15 (-0.38), causing the defense to fail. 
These results imply that attackers can use weaker triggers to bypass defenses or human checks on the training dataset and then apply inference triggers of higher intensity to ensure a high attack success rate (ASR).

\begin{figure*}[!t]
  \begin{subfigure}[c]{\linewidth}
      \centering
      \begin{subfigure}[c]{0.186\linewidth}
        \centering
        \includegraphics[width=\linewidth,trim={0 0 60 0},clip]{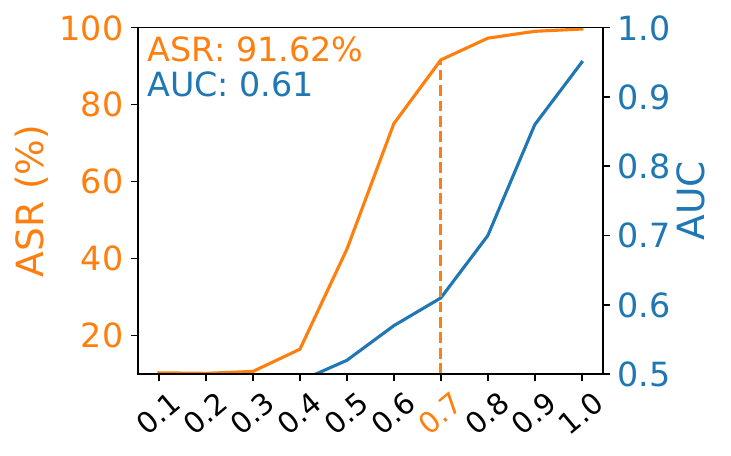}
        % \caption{BadNets (Opacity)}
        \label{figure:defense_inputdet:scaleup_badnets}  
      \end{subfigure}
      \begin{subfigure}[c]{0.15\linewidth}
        \centering
        \includegraphics[width=\linewidth,trim={60 0 60 0},clip]{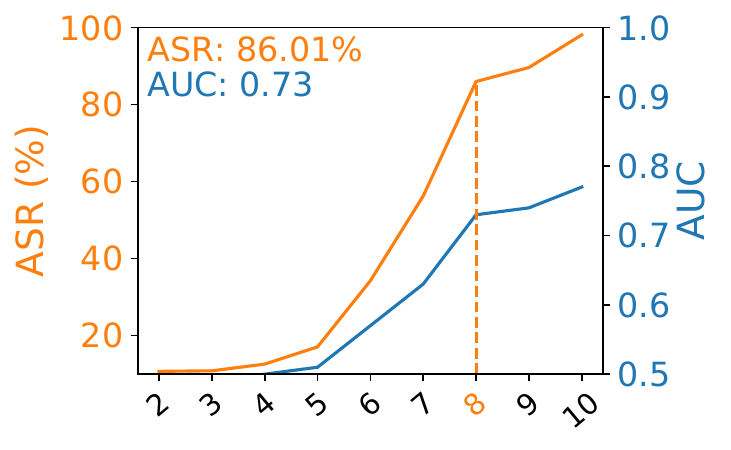}
        % \caption{BadNets (Size)}
        \label{figure:defense_inputdet:scaleup_patch}  
      \end{subfigure}
      \begin{subfigure}[c]{0.15\linewidth}
        \centering
        \includegraphics[width=\linewidth,trim={60 0 60 0},clip]{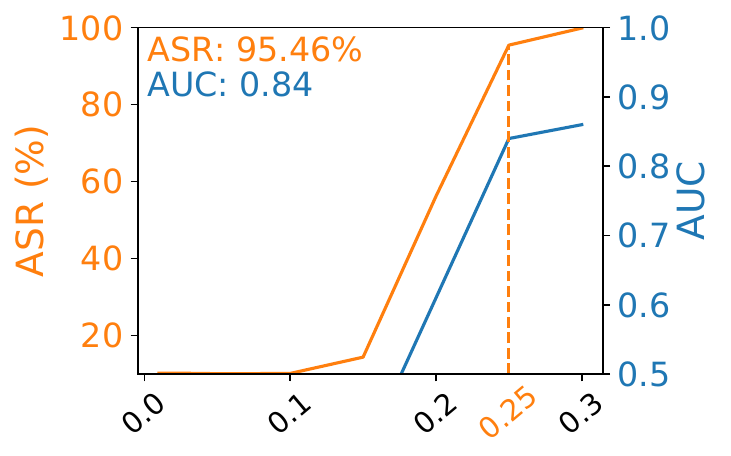}
        % \caption{Blended}
        \label{figure:defense_inputdet:scaleup_blended}
      \end{subfigure}
      \begin{subfigure}[c]{0.15\linewidth}
        \centering
        \includegraphics[width=\linewidth,trim={60 0 60 0},clip]{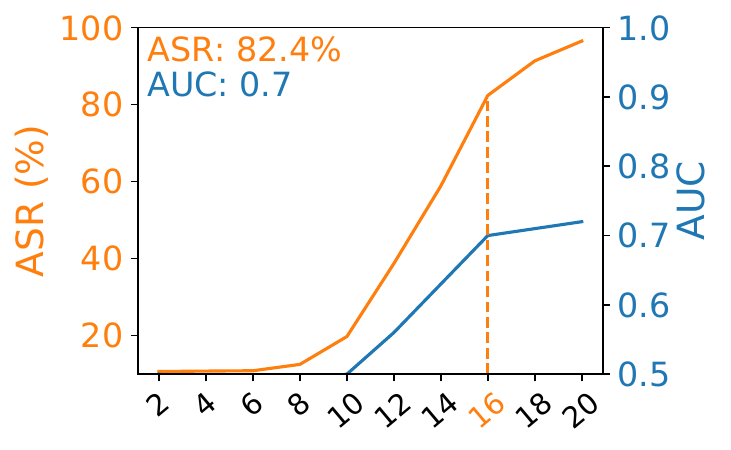}
        % \caption{SIG}
        \label{figure:defense_inputdet:scaleup_sig}
      \end{subfigure}
      \begin{subfigure}[c]{0.15\linewidth}
        \centering
        \includegraphics[width=\linewidth,trim={60 0 60 0},clip]{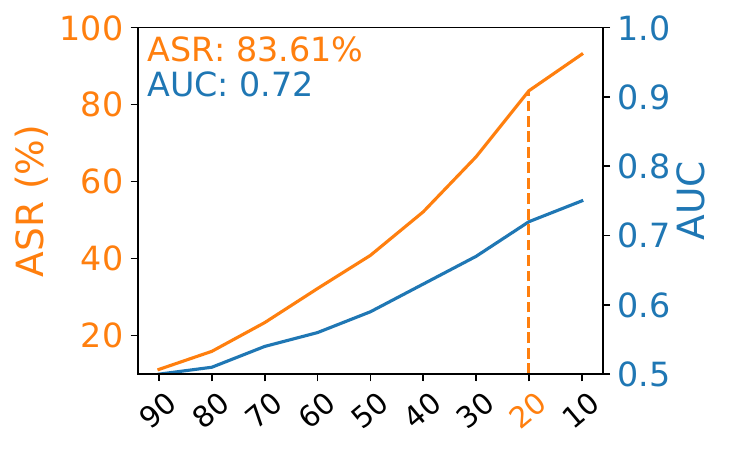}
        % \caption{Compress}
        \label{figure:defense_inputdet:scaleup_compress}  
      \end{subfigure}
      \begin{subfigure}[c]{0.186\linewidth}
        \centering
        \includegraphics[width=\linewidth,trim={60 0 0 0},clip]{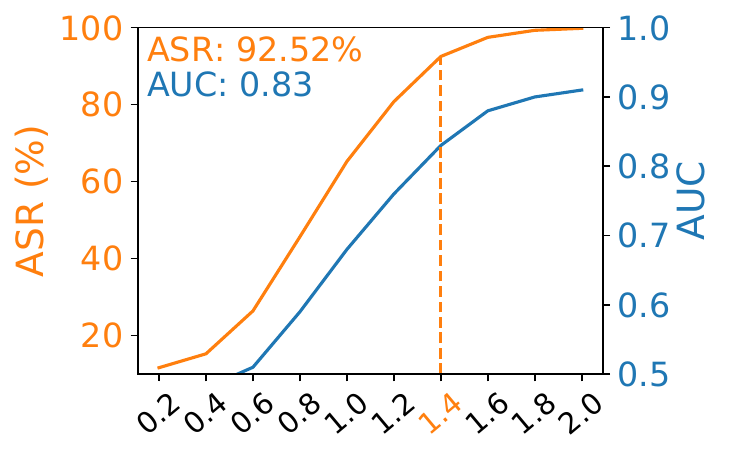}
        % \caption{WaNet}
        \label{figure:defense_inputdet:scaleup_wanet}
      \end{subfigure}
      % % \caption{Results of input detection defense against different attacks (Scale-Up).}
      \label{figure:defense_inputdet:scaleup}
  \end{subfigure}
  %\vspace{-10pt}

  \begin{subfigure}[c]{\linewidth}
      \centering
      \begin{subfigure}[c]{0.186\linewidth}
        \centering
        \includegraphics[width=\linewidth,trim={0 0 60 0},clip]{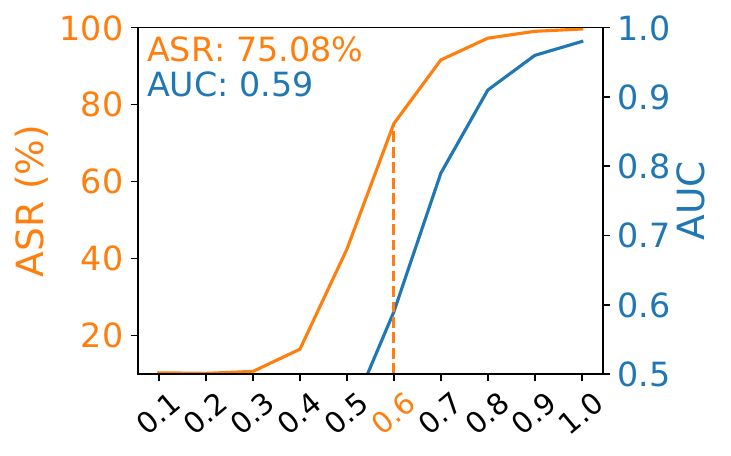}
        \caption{BadNets (Square)}
        \label{figure:defense_inputdet:strip_badnets}  
      \end{subfigure}
      \begin{subfigure}[c]{0.15\linewidth}
        \centering
        \includegraphics[width=\linewidth,trim={60 0 60 0},clip]{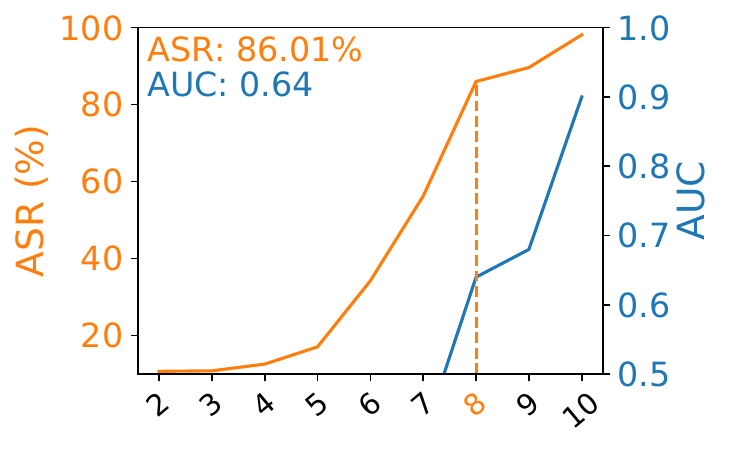}
        \caption{BadNets}
        \label{figure:defense_inputdet:strip_patch}  
      \end{subfigure}
      \begin{subfigure}[c]{0.15\linewidth}
        \centering
        \includegraphics[width=\linewidth,trim={60 0 60 0},clip]{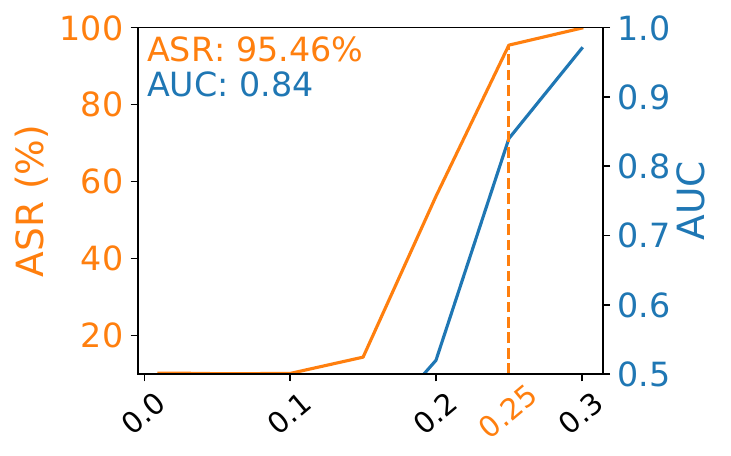}
        \caption{Blended}
        \label{figure:defense_inputdet:strip_blended}
      \end{subfigure}
      \begin{subfigure}[c]{0.15\linewidth}
        \centering
        \includegraphics[width=\linewidth,trim={60 0 60 0},clip]{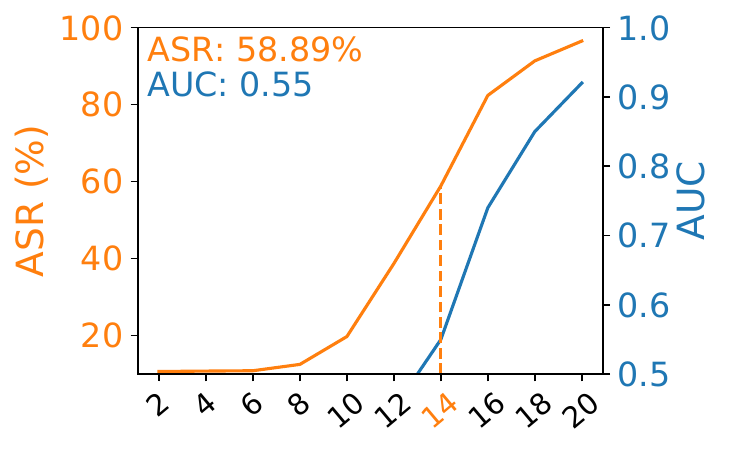}
        \caption{SIG}
        \label{figure:defense_inputdet:strip_sig}
      \end{subfigure}
      \begin{subfigure}[c]{0.15\linewidth}
        \centering
        \includegraphics[width=\linewidth,trim={60 0 60 0},clip]{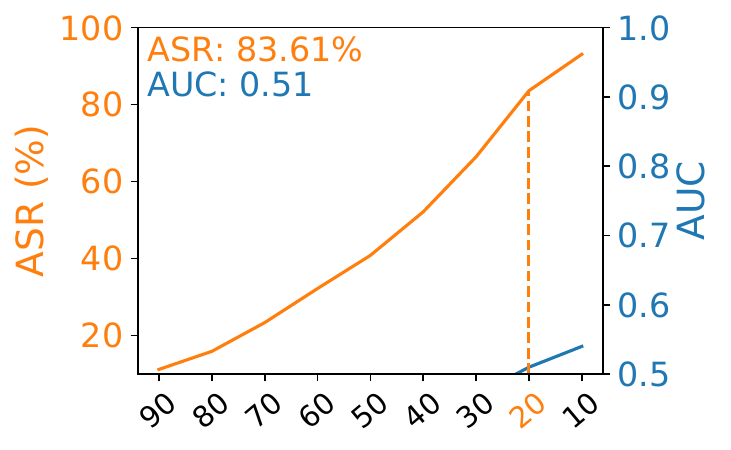}
        \caption{Compress}
        \label{figure:defense_inputdet:strip_compress}  
      \end{subfigure}
      \begin{subfigure}[c]{0.186\linewidth}
        \centering
        \includegraphics[width=\linewidth,trim={60 0 0 0},clip]{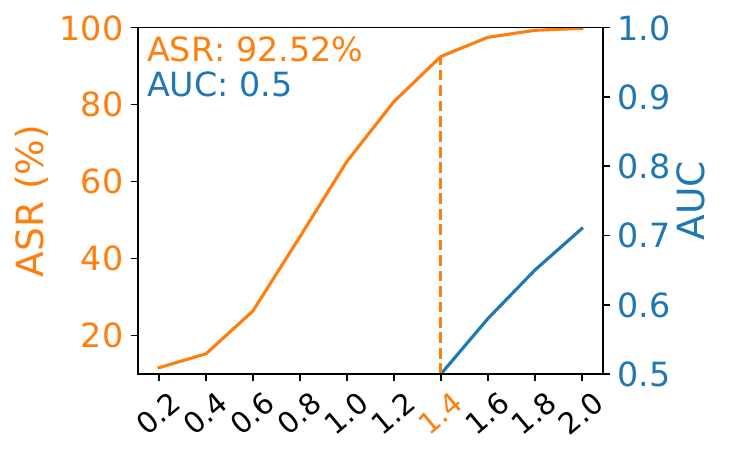}
        \caption{WaNet}
        \label{figure:defense_inputdet:strip_wanet}
      \end{subfigure}
      % \caption{Results of input detection defense against different attacks (STRIP).}
      \label{figure:defense_inputdet:strip}
  \end{subfigure}
  \caption{\highlight{(\textit{Statement}) }{Results of input detection defense against different attacks.
  The upper and lower rows represent Scale-Up and STRIP, respectively.}}
  \label{figure:defense_inputdet}
\end{figure*}

\textbf{Input Detection Defenses.}
Input detection defenses aim to identify attacked samples fed into a deployed model during the inference stage. 
In this context, we first train models using the attack methods listed in Table~\ref{table:experiment_attacks}, setting the intensity to the highest level specified. We then evaluate datasets with different intensities (comprising half-clean and half-poisoned samples) and collected their ASR and AUC for the defense methods.

The results are shown in Figure~\ref{figure:defense_inputdet}. As the intensity of the inference triggers used in the evaluation datasets decreases, the performance of the defenses exhibits a more rapid upward trend compared to the attacks. 
For example, a backdoored model formed with a Square training trigger intensity of 1.0 can still be activated with a high ASR of 91.62\% by using a poisoned dataset with an inference trigger intensity of 0.7. Meanwhile, the AUC of STRIP and Scale-Up can be reduced from 0.99 and 0.96 to 0.80 and 0.62, respectively. 
This indicates that attackers can employ triggers with relatively lower intensity during the inference phase than in the training phase, thereby bypassing input sample detection while maintaining the activatability of the backdoor attack.

\begin{figure}[!t]
  \centering
  \begin{subfigure}[c]{0.56\linewidth}
    \includegraphics[width=\linewidth]{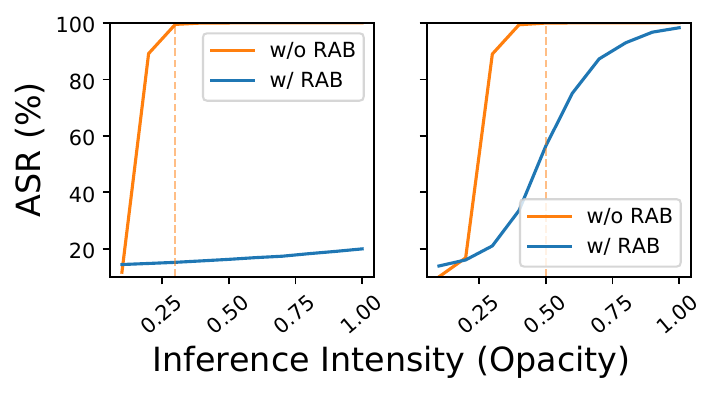}
    \caption{RAB}
    \label{figure:defense_rab}
  \end{subfigure}
  \begin{subfigure}[c]{0.43\linewidth}
    \centering
    \includegraphics[width=\linewidth,trim={5 0 0 0},clip]{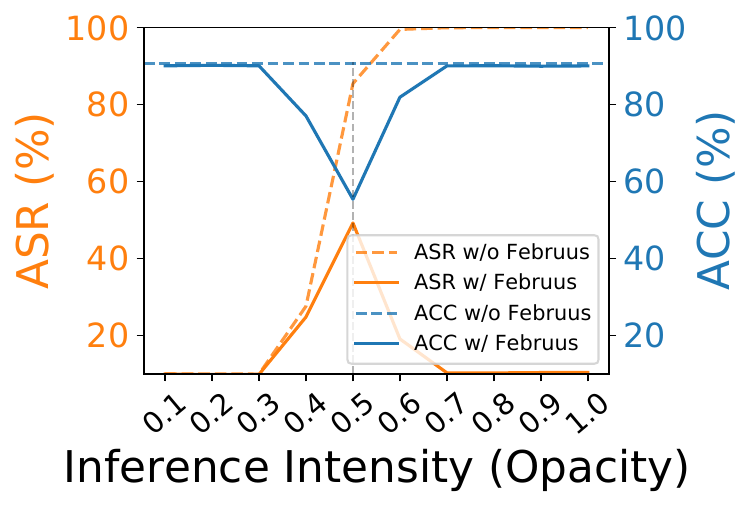}
    \caption{Februus}
    \label{figure:defense_februus}
  \end{subfigure}
  \caption{(a) Results of RAB against BadNets at two training intensities (left: opacity=0.3, right: opacity=0.5) with varying inference-phase trigger intensities. (b) ASR and Acc on sanitized inputs at varied intensities of Februus, which achieves unsatisfactory results around opacity 0.5.
  }
  \label{figure:defense_rab_februus}
\end{figure}

\textbf{Input Preprocessing Defenses.}
\highlight{(\textit{7. More Defenses}) }{
Input preprocessing defenses try to purify poisoned input samples during the inference stage, rendering them incapable of activating the backdoor. 
We test the Februus defense~\cite{DBLP:conf/acsac/DoanAR20} with a model trained with the flower trigger at opacity 1.0.
As illustrated in Figure~\ref{figure:defense_februus}, Februus cannot consistently achieve satisfactory results, with about 50\% of ASR and Acc on sanitized inputs at opacity 0.5.
}

% \begin{figure}[!t]
%   \centering
%   \includegraphics[width=\linewidth]{figure/rab_badnets_mrx.pdf}
%   \caption{\highlight{(\textit{7. More Defenses}) }{Results of RAB against BadNets at two training intensities (left: opacity=0.3, right: opacity=0.5) with varying inference-phase trigger intensities.}}
%   \label{figure:defense_rab}
% \end{figure}

% \begin{figure}[!t]
%   \centering
%   \includegraphics[width=0.9\linewidth]{figure/februus_flower_mrx_v1.1.pdf}
%   \caption{\highlight{(\textit{7. More Defenses}) }{ASR and Acc on sanitized inputs at varied intensities of Februus, which achieves unsatisfactory results around opacity 0.5.}}
%   \label{figure:defense_februus}
% \end{figure}

\textbf{Model Detection Defenses.}
Model detection defenses assess whether a suspicious model is infected through reverse engineering or a meta-classifier. We evaluate backdoored models with varying training trigger intensities using two typical trigger-reversing defenses: Neural Cleanse~\cite{DBLP:conf/sp/WangYSLVZZ19} and FeatureRE~\cite{DBLP:conf/nips/WangMDZM22}. In this context, the defender assumes access to the model and a small dataset containing benign samples of each label. 
Note that the Mixed Value of FeatureRE is a metric from its source code that incorporates ASR, the similarity between poisoned and benign samples, the cosine similarity of their activations, the standard deviation of poisoned sample activations, and the size of the reversed mask.

As shown in Table~\ref{table:experiment_defense_modeldet}, reverse engineering is effective on models trained with weaker triggers, indicating that backdoors formed with weaker training triggers are more easily detected. Both Neural Cleanse and FeatureRE defenses can successfully obtain a reversed trigger even in cases where the original trigger fails to execute the attack, \thatis, when the trigger intensities are lower than 0.4. 
However, backdoors injected with higher intensity training triggers exhibit greater backdoor exclusivity, as they can only be activated within a narrower range of intensities, making them more challenging to be reverse engineerd~\cite{DBLP:journals/corr/abs-2312-04902}. For example, the reverse engineering results of Neural Cleanse show a downward trend as the trigger intensity increases. In the worst cases, FeatureRE consistently reports a backdoored model trained with a trigger intensity of 1.0 as benign with a mixed value of -0.57 (above the threshold of -0.75). We attribute this to the regularizations (\forexample, the $\ell_1$ norm of the trigger mask and the similarity between the poisoned and original images)), which is integral to the optimization process of most trigger reverse-engineering methods.

These findings suggest that attackers may circumvent such defenses by simply employing training triggers with higher intensity, thereby prompting defenders to develop adaptive defenses.

\begin{table}[!t]
    \centering
    \scriptsize
    \caption{Results of model detection defenses against BadNets with different intensities.
    TI and AI denote the trigger intensity and anomaly index, respectively.}
    \label{table:experiment_defense_modeldet}
    \setlength{\tabcolsep}{2pt}{
    \resizebox{\linewidth}{!}{
    \begin{tabular}{ccccccc}
    \toprule \midrule
     &  & \multicolumn{3}{c}{{Neural Cleanse}} & \multicolumn{2}{c}{{FeatureRE}}        \\ \cmidrule(r){3-5} \cmidrule(r){6-7}
     \multirow{-2.5}{*}{TI}                                              & \multirow{-2.5}{*}{ASR (\%)}                     & reASR~$\uparrow$ (\%)       & AI~$\uparrow$ & L1 Norm~$\downarrow$       & reASR~$\uparrow$ (\%)       & Mixed Value~$\downarrow$  \\ \midrule
    % 0.1                                           & 0.11                 & -           & 3.6(±1.3)     & 591.0(±8.4)   & 96.44(±2.89) & -0.91(±0.04) \\ \midrule
    0.2                                           & 9.75                  & 100.0(±1.0) & \phantom{0}2.1(±\phantom{0}1.3)     & 593.4(±\phantom{00}1.9)   & 97.22(±2.33) & -0.89(±0.05) \\[2pt]
    0.3                                           & 11.75                 & 100.0(±1.0) & \phantom{0}1.8(±\phantom{0}0.9)     & 592.6(±\phantom{00}7.4)   & 95.83(±3.61) & -0.89(±0.02) \\[2pt]
    0.4                                           & 67.66                 & 100.0(±1.0) & 25.1(±\phantom{0}5.7)    & 405.4(±\phantom{0}63.5)  & 98.83(±1.17) & -0.93(±0.05) \\[2pt]
    0.5                                           & 95.07                 & 99.99(±1.0) & 70.1(±59.3)   & 260.9(±\phantom{0}86.9)  & 92.83(±1.05) & -0.80(±0.04) \\[2pt]
    0.6                                           & 98.33                 & 100.0(±1.0) & 50.1(±18.5)   & 212.4(±\phantom{0}58.7)  & 92.50(±3.27) & -0.70(±0.16) \\[2pt]
    0.7                                           & 99.24                 & 100.0(±1.0) & 39.6(±21.5)   & 240.8(±\phantom{0}80.8)  & 95.28(±4.72) & -0.53(±0.31) \\[2pt]
    0.8                                           & 99.21                 & 99.99(±1.0) & 41.8(±24.4)   & 279.4(±\phantom{0}86.0)  & 94.83(±5.17) & -0.71(±0.09) \\[2pt]
    0.9                                           & 99.38                 & 100.0(±1.0) & 39.8(±12.1)   & 259.8(±137.3) & 94.11(±5.89) & -0.66(±0.14) \\[2pt]
    1.0                                           & 99.65                 & 100.0(±1.0) & 29.8(±16.5)   & 341.0(±116.8) & 89.77(±0.89) & -0.57(±0.06) \\ \bottomrule
    \end{tabular}}}
\end{table}

\textbf{Robust Training Defenses.}
\highlight{(\textit{7. More Defenses}) }{
Robust training defenses aim to train robust models on untrusted data and obtain a robust bound $R$. They ensure that as long as triggers employed in test instances remain within an $L_p$-ball of radius $R$, the output of robust models is guaranteed to be consistent with benign models.
We evaluate a state-of-the-art defense of this type, namely RAB~\cite{DBLP:conf/sp/WeberXKZL23}, which first adds noise sampled from a smoothing distribution to the original training dataset to create a large number of ``smoothed'' training datasets. It then trains models on these datasets and aggregates their final outputs as the final ``smoothed'' prediction.

Due to its extremely high overhead for training a large number of models, we only evaluate RAB on CIFAR-10 with ResNet-18. We train 1,000 models on each poisoned training dataset and evaluate them on poisoned inference datasets with varying trigger intensities. 
As shown in Figure~\ref{figure:defense_rab}, RAB successfully strengthens backdoored models with lower training intensity (\thatis, opacity 0.3) and reduces the ASR by 80\% across all inference intensities. However, it fails on higher training intensities since the ASR of high-intensity inference triggers remains higher than 98\%.
Moreover, we notice that RAB inevitably sacrifices the model Acc by above 10\%.
}

%!TEX root = main.tex

\section{Discussion}
\label{section:discussion}

\begin{figure}[!t]
  \centering
  \begin{minipage}[t]{\linewidth}
    \centering
    \includegraphics[width=0.365\linewidth,trim={0 0 5 0},clip]{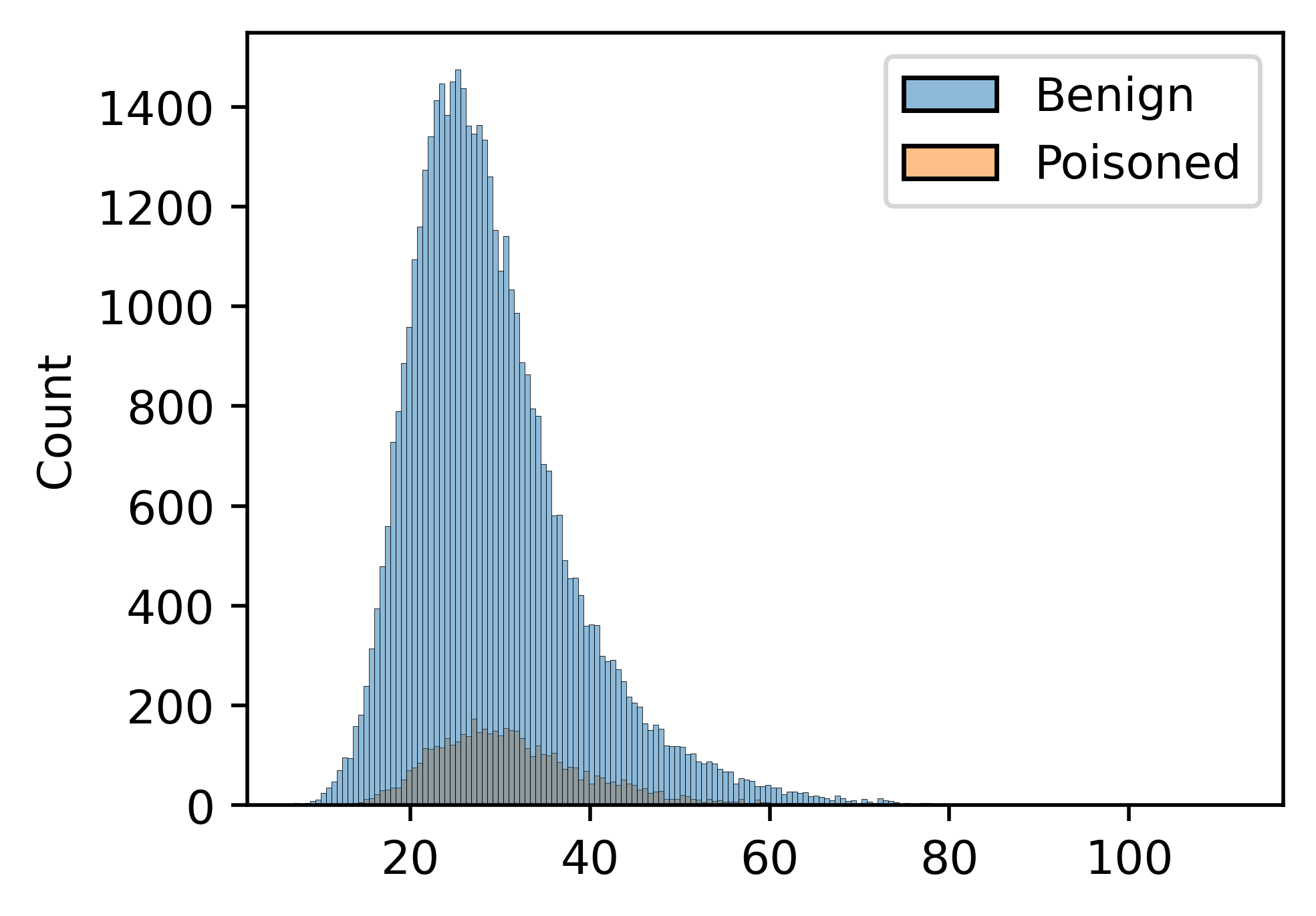}
    \includegraphics[width=0.315\linewidth,trim={47 0 5 0},clip]{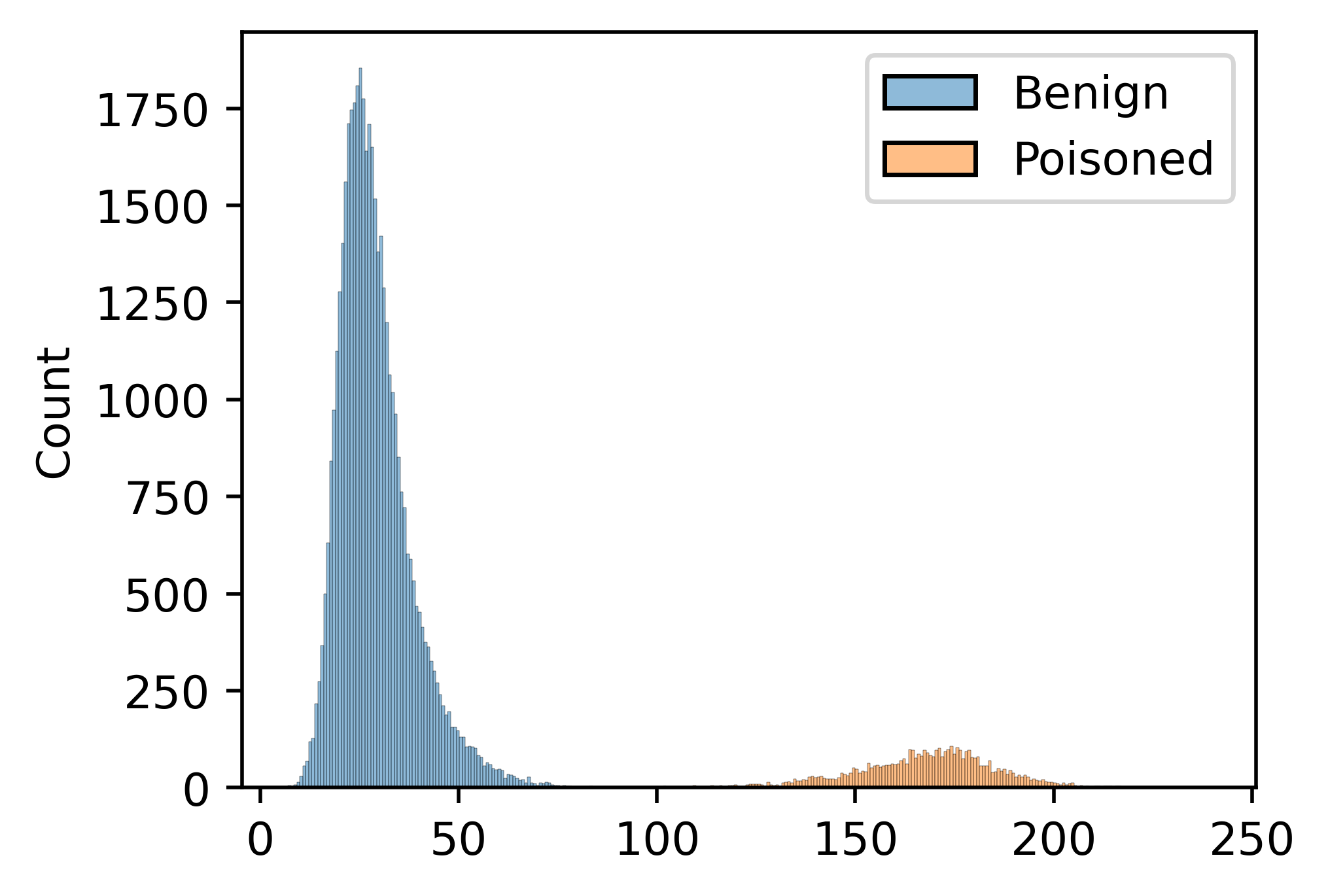}
    \includegraphics[width=0.30\linewidth,trim={50 0 5 0},clip]{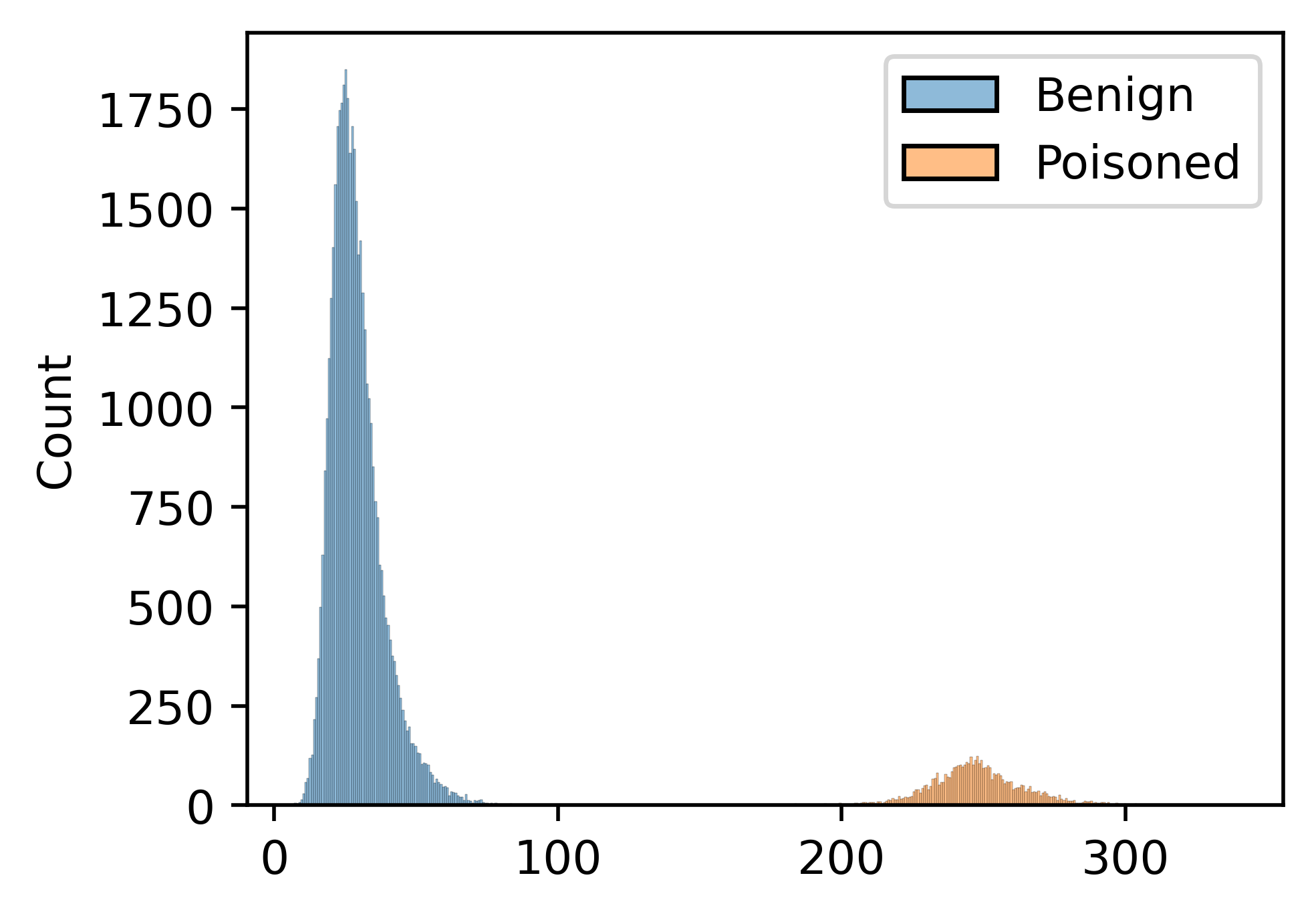}
    \subcaption{Training intensity (opacity)=0.4. }
    \label{figure:explanation:opacity0.4}
  \end{minipage}
  \begin{minipage}[t]{\linewidth}
    \centering
    \includegraphics[width=0.365\linewidth,trim={0 0 5 0},clip]{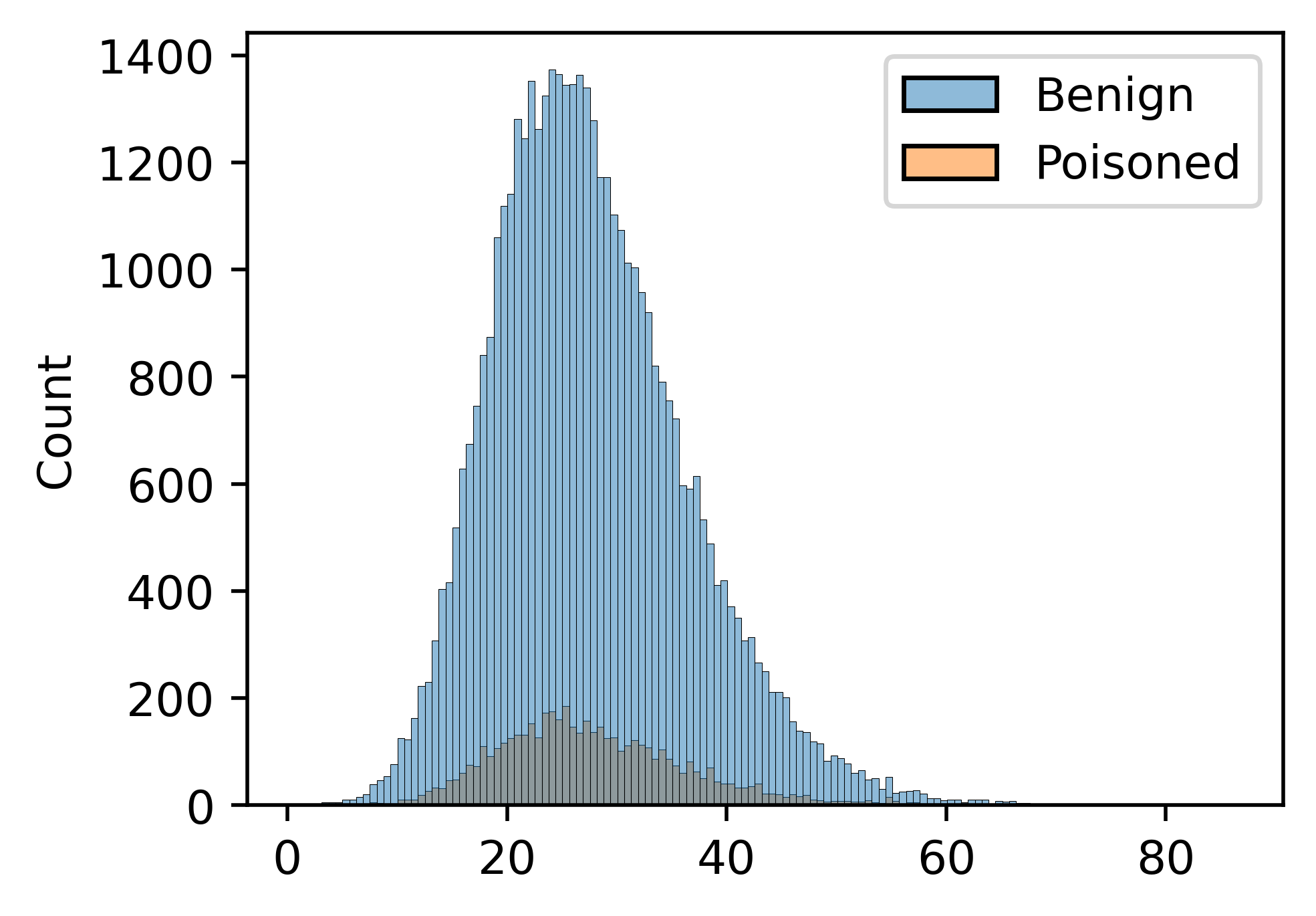}
    \includegraphics[width=0.30\linewidth,trim={50 0 5 0},clip]{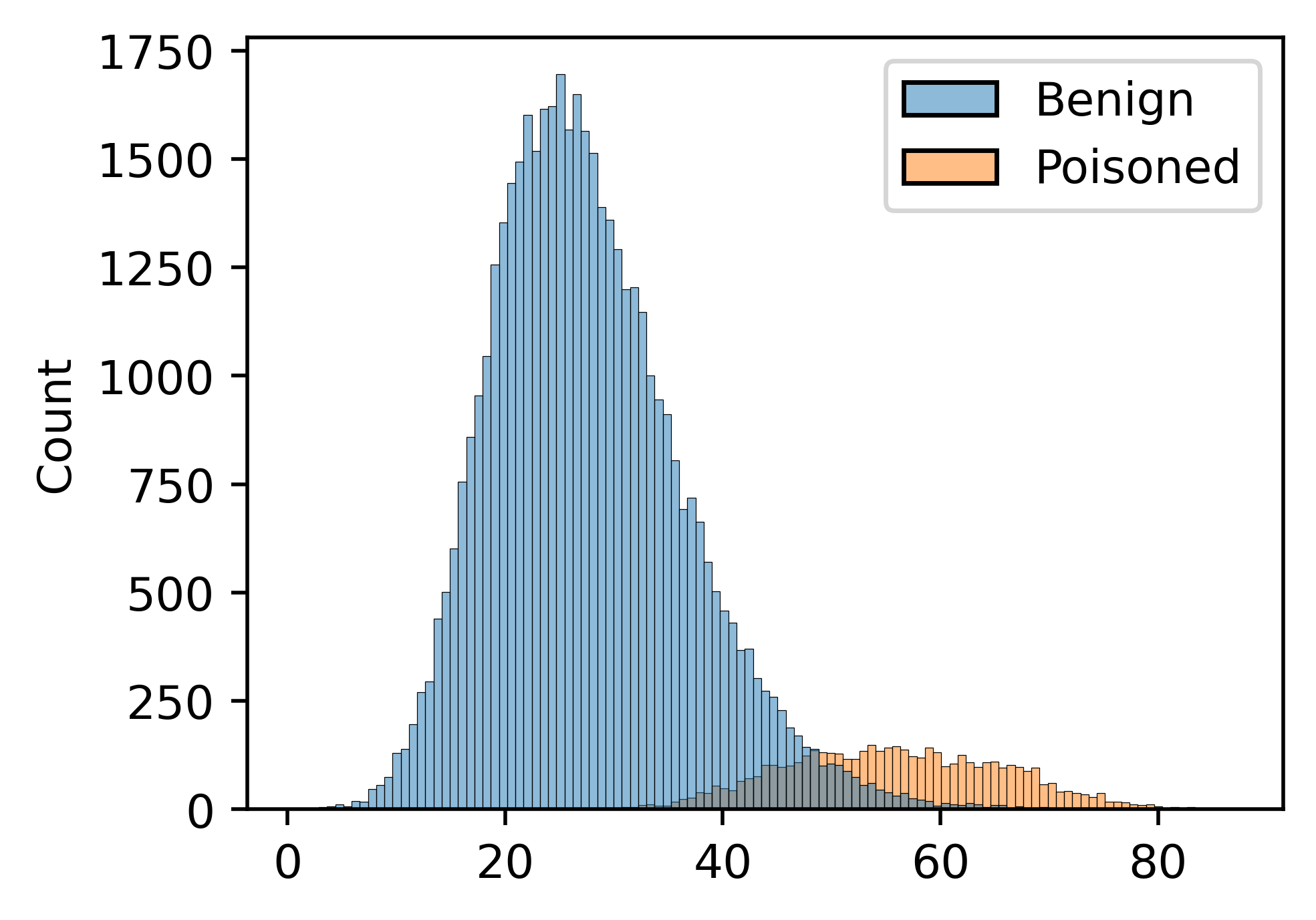}
    \includegraphics[width=0.30\linewidth,trim={50 0 5 0},clip]{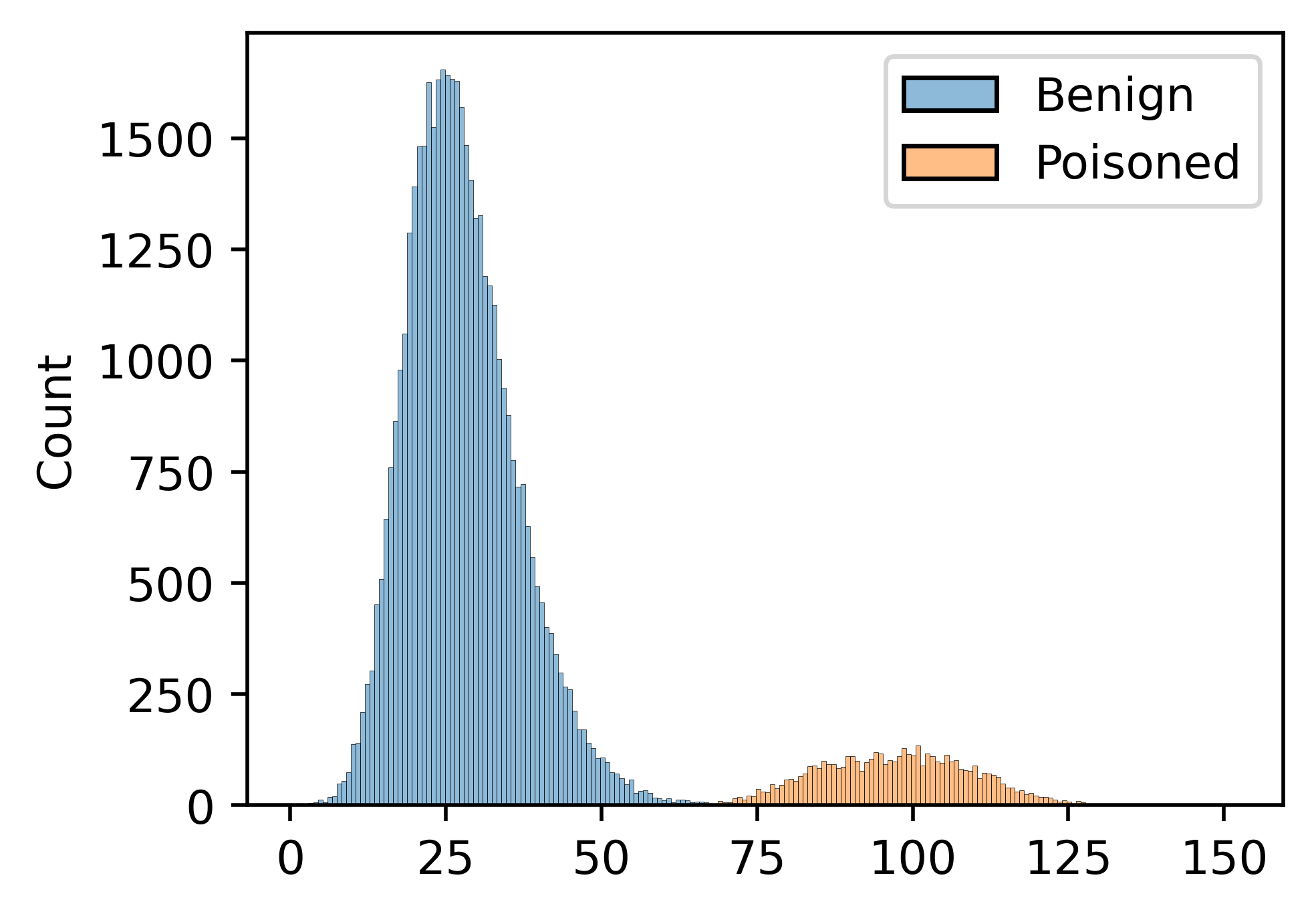}
    \subcaption{Training intensity (opacity)=1.0.}
    \label{figure:explanation:opacity1.0}
  \end{minipage}
  \caption{\highlight{(\textit{6. Deeper Discussion}) }{Neuron activation distribution of two training intensities and three inference intensities (from left to right: opacity=0.1, 0.5, 1.0, respectively; blue: benign, orange: poisoned). }}
  \label{figure:activation}
\end{figure}

\subsection{Explanations of Our Observations}
\label{section:explanation}
In this section, we explain our observations regarding training-inference trigger generalization and overfitting through neuron activations and UMAP visualizations.

\textbf{Neuron Activations.}
\highlight{(\textit{6. Deeper Discussion}) }{
We train a backdoored model with a single training intensity and then identify the compromised neurons employing NONE~\cite{DBLP:conf/nips/WangDZM22}. Subsequently, we input poisoned samples with varying inference intensities into the model to collect the activation values of identified compromised neurons and then compare their distribution with that of benign samples. 

As illustrated in Figure~\ref{figure:activation}, trigger intensity has a monotonic relationship with neuron activation values.
Specifically, as the inference intensity increases, the activation distribution of compromised neurons between poisoned and benign samples becomes more separable as the trigger intensity increases.
This explains the high ASR in the generalization phenomenon.
In contrast, as the training intensity increases, the distribution becomes less separable.
This explains the low ASR in the overfitting phenomenon.
% \todo{Additional results in Appendix~\ref{section:appendix:neuron_activation} further show that when replacing the ReLU function with Sigmoid, the above explanations still hold.}
We find that when replacing the ReLU function with Sigmoid, the above explanations still hold.
}

\begin{figure*}[htbp]
  \centering
  \begin{minipage}[t]{0.44\textwidth}
    \centering
    \includegraphics[width=\linewidth,trim={20 15 20 30},clip]{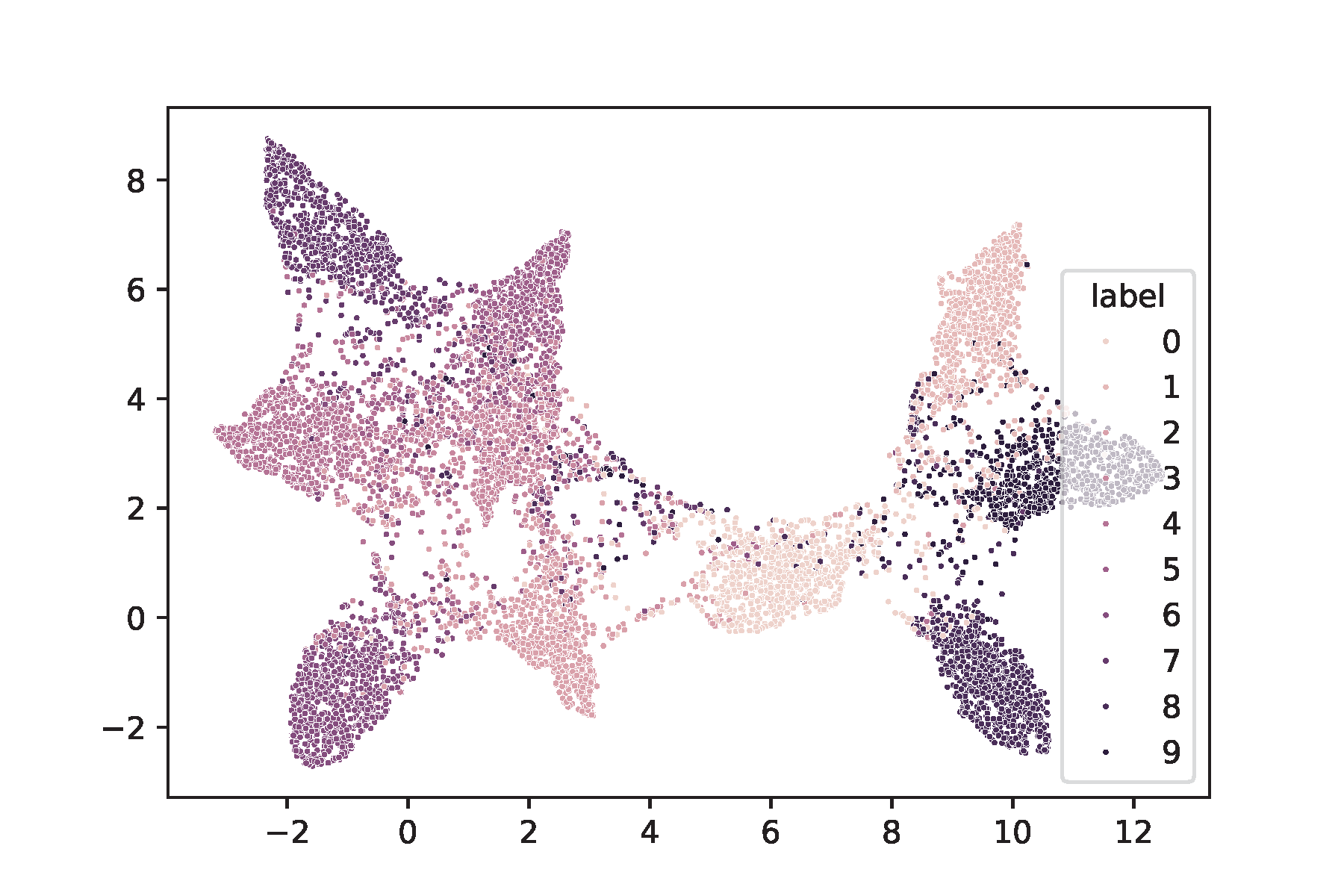}
    % % \fbox{\rule[-.5cm]{0cm}{4cm} \rule[-.5cm]{4cm}{0cm}}
    %\vspace{-20pt}
    \subcaption{Activation visualization of benign images.}
    \label{figure:umap_benign}
  \end{minipage}
  \begin{minipage}[t]{0.44\textwidth}
    \centering
    \includegraphics[width=\linewidth,trim={20 15 20 30},clip]{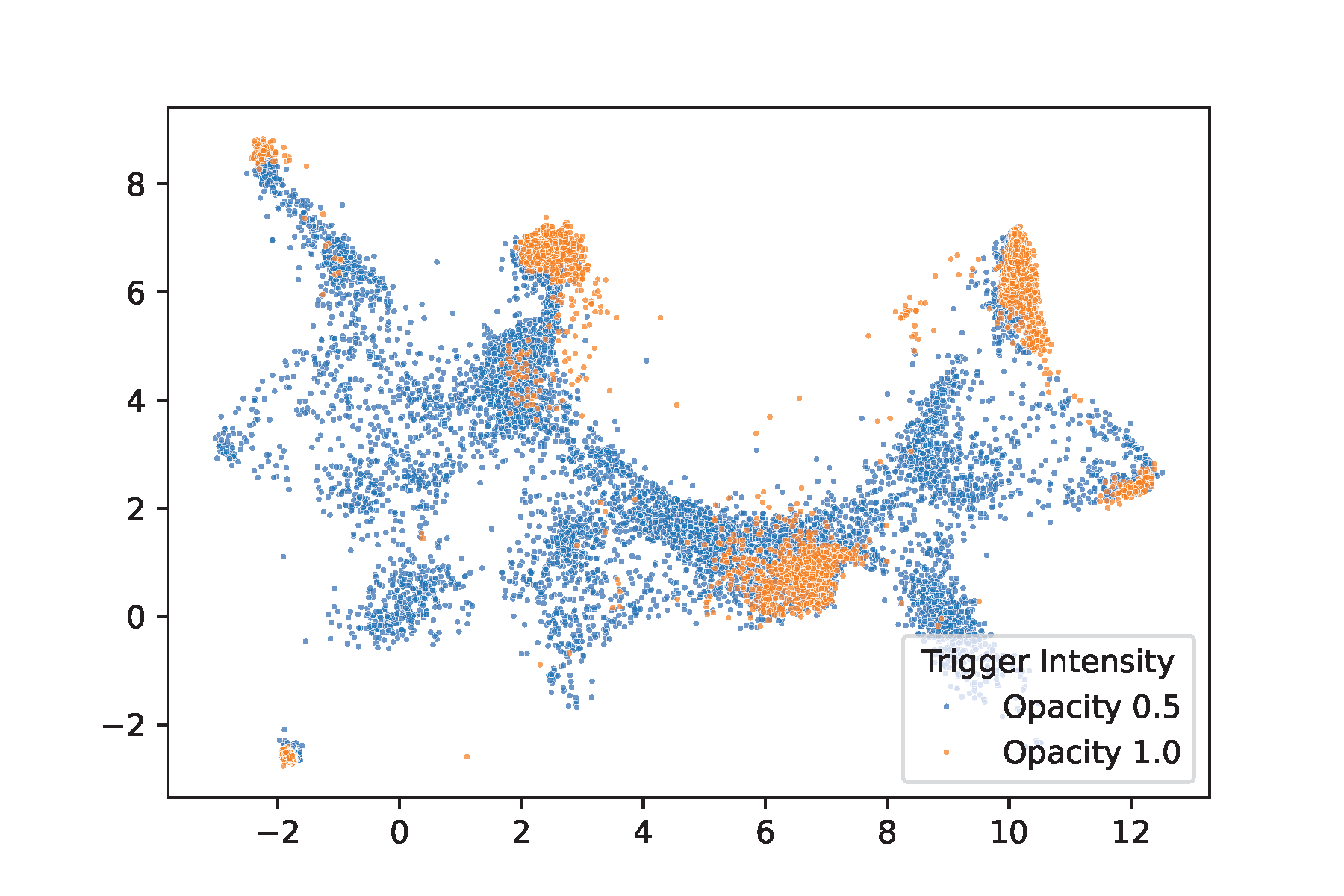}
    % \fbox{\rule[-.5cm]{0cm}{4cm} \rule[-.5cm]{4cm}{0cm}}
    %\vspace{-20pt}
    \subcaption{Activation visualization of poisoned images.}
    \label{figure:umap_poisoned}
  \end{minipage}
  \caption{UMAP visualization for benign and varying intensity samples.}
  \label{figure:umap}
\end{figure*}

\textbf{UMAP Visualizations.}
\highlight{(\textit{6. Deeper Discussion}) }{
We use UMAP~\cite{mcinnes2018umap} to visualize the activations of the fully connected layer inputs of a backdoored model trained with trigger opacity=0.5.
We first construct an embedding space based on the eigenvectors of the benign dataset. Subsequently, we extract the activation values from two poisoned datasets attacked by BadNets with two inference trigger intensities opacity=0.5 and opacity=1.0.
Finally, we embed these poisoned eigenvectors into the previously constructed embedding space. 

As shown in Figure~\ref{figure:umap}, when the training intensity is fixed with opacity$=1.0$, poisoned inference samples with higher trigger intensity (\thatis, opacity$=1.0$) exhibit a more concentrated distribution than opacity$=0.5$ within the embedding space. 
This suggests that activation values with higher inference intensities may reside in a subspace (characterized by a high ASR) within the broader space of low-intensity activation values, explaining the better generalizability of lower-intensity triggers. 
Conversely, the lower inference intensity causes the activation values to form a similar distribution to that of benign samples, \thatis, the trigger is not sufficient to activate the backdoor. 
This explains the low ASR for the overfitting phenomenon of higher training intensities.
}

\subsection{Adaptive Defenses}
% \textbf{(a) Adaptive Defense}
In this section, we discuss two possible adaptive defenses, namely Adaptive Neural Cleanse and Multi-Stage Defenses.

\begin{table}[!t]
    \centering
    \scriptsize
    \caption{Anomaly Index (AI) and $\ell_1$ Norm (L1) of the reversed triggers for the original and adaptive version of Neural Cleanse (NC) against BadNets with different intensities; TI denotes the trigger intensity.}
    \label{table:adap_nc}
        \resizebox{\linewidth}{!}{
    \setlength{\tabcolsep}{2pt}{\begin{tabular}{crrrrrrrrrr}
    \toprule \midrule
    \multicolumn{2}{c}{TI}                                 & 0.2   & \multicolumn{1}{r}{0.3} & \multicolumn{1}{r}{0.4} & \multicolumn{1}{r}{0.5} & \multicolumn{1}{r}{0.6} & \multicolumn{1}{r}{0.7} & \multicolumn{1}{r}{0.8} & \multicolumn{1}{r}{0.9} & \multicolumn{1}{r}{1.0} \\
    \multicolumn{2}{c}{ASR (\%)}                           & 9.75  & 11.75                   & 67.66                   & 95.07                   & 98.33                   & 99.24                   & 99.21                   & 99.38                   & 99.65                   \\ \midrule
    \multirow{2}{*}{NC}                          & AI$\uparrow$      & 2.8   & 3.2                     & 16.4                    & 26.3                    & 43.7                    & 80.0                    & 22.4                    & 25.6                    & 21.7                    \\
                                                 & L1$\downarrow$      & 592.2 & 587.6                   & 392.6                   & 165.8                   & 172.1                   & 175.1                   & 193.5                   & 334.1                   & 274.6                   \\
    \multicolumn{1}{l}{\multirow{2}{*}{Adap-NC}} & AI$\uparrow$      & 5.0   & 4.7                     & 3.5                     & 9.9                     & 0.5                     & 3.9                     & 6.3                     & 4.7                     & 4.7                     \\
    \multicolumn{1}{l}{}                         & L1$\downarrow$      & 489.9 & 465.5                   & 305.3                   & 87.6                    & 144.2                   & 167.4                   & 133.1                   & 327.7                   & 262.9         \\ \bottomrule
    \end{tabular}}}
\end{table}

\textbf{Adaptive Neural Cleanse.}
\highlight{(\textit{7. Adaptive Defenses}) }{
We implement an adaptive version of Neural Cleanse (Adap-NC), which computes the losses for both the reverse-engineered trigger and its lower-opacity version. 
As shown in Table~\ref{table:adap_nc}, the reversing effectiveness of Adap-NC has improved as expected (the $\ell_1$ norm of the reversed pattern is smaller), particularly for lower-intensity triggers. 
However, this improvement comes at the cost of a corresponding drop in the Anomaly Index, as it reverses smaller patterns for all targets, making the overall enhancement modest. Moreover, the necessity of knowing the attacker's choice of intensity renders it impractical.
}

\textbf{Multi-Stage Defense.}
\highlight{(\textit{7. Adaptive Defenses}) }{
As mentioned in Section~\ref{section:application:bypass}, attackers can exploit the intensity mismatch to bypass defenses, potentially leading to severe consequences in real-world applications.
Thus, a single-stage defense mechanism may not be sufficient to protect against such adaptive attacks, as attackers can easily tune the intensity of triggers targeting a single defense at particular stages.

For the design of new defenses, the defenders shall consider integrating mechanisms at different stages to mitigate the impact of intensity mismatch or varying intensity triggers.
Defenders can also incorporate multiple existing defenses at different stages; however, this may introduce challenges such as performance degradation due to interactions between defenses. 
For instance, after applying an incomplete data cleaning defense, the model might learn a weaker backdoor, potentially making it harder to detect by subsequent defenses. More discussion on this topic is left for our future work.
}

%!TEX root = main.tex

\section{Related Work}
\label{section:related_work}

In this section, we introduce some existing works on backdoor attacks and defenses, as well as the backdoor studies related to trigger mismatch.

\textbf{Backdoor Attacks.}
Backdoor attack methods can be roughly divided into poisoning-based, weight-oriented, and structure-modified attacks~\cite{DBLP:journals/corr/abs-2007-08745}.

Data poisoning basically carries out backdoor attacks by manipulating the training data.
BadNets~\cite{DBLP:journals/access/GuLDG19} pioneered the concept of backdoor attacks and exposed the threats in DNN training, i.e., exactly matched trigger between the training and inference phase can cause the victim model to exhibit incorrect behavior. The following methods mainly focus on generating triggers with a higher attack success rate and stealthiness, consisting of visible trigger~\cite{DBLP:journals/access/GuLDG19, DBLP:journals/prl/GuoTB21},
invisible trigger~\cite{DBLP:journals/corr/abs-1712-05526, DBLP:conf/iccv/LiLWLHL21, DBLP:journals/tdsc/LiXZZZ21, DBLP:journals/tip/ZhangCHLZFHY22, DBLP:conf/codaspy/ZhongLSZ020, DBLP:conf/iccv/DoanL0L21, DBLP:conf/nips/DoanLL21, DBLP:conf/eccv/LiuM0020, DBLP:conf/uss/BagdasaryanS21, DBLP:conf/sp/QuiringR20, DBLP:conf/uss/XiaoCS0019, DBLP:conf/iclr/NguyenT21, DBLP:conf/cvpr/Jiang0X023, DBLP:conf/aaai/DuanH0ZZ24,gao2024backdoor,cai2024toward,gao2023not},
style change trigger~\cite{DBLP:conf/ccs/LiuLTMAZ19, DBLP:conf/aaai/0005LMZ21}, frequency domain trigger~\cite{DBLP:conf/bmvc/HammoudG22, DBLP:conf/eccv/WangYXATW22}, clean label attacks~\cite{DBLP:conf/icip/BarniKT19, DBLP:journals/corr/abs-1912-02771, DBLP:conf/cvpr/ZhaoMZ0CJ20, DBLP:conf/aaai/SahaSP20}, and others according to the different types of triggers.
% \swang{merging the references.}

Weight-oriented attacks modify model parameters directly instead of training the whole model on poisoned datasets~\cite{DBLP:conf/icb/DumfordS20, DBLP:conf/cvpr/RakinHF20, DBLP:conf/cikm/Garg0GL20, DBLP:conf/iclr/ZhangLWS022}.
Structure-modified attacks literally inject backdoors by modifying the model structure, \textit{e.g.}, inserting a backdoored module~\cite{DBLP:conf/kdd/TangDLYH20, DBLP:conf/icse/LiH0CL21} or replacing a narrow subnet~\cite{DBLP:journals/corr/abs-2107-07240}.
Since we mainly focus on the impact of trigger intensity mismatch between the training and inference phases, and weight-oriented attacks and structure-modified attacks do not involve trigger intensity during the training phase. Our research does not cover these two types of attack methods.

% In some scenarios, the training process and data are not accessible due to privacy or copyright factors; the attacker may adopt black-box backdoor attacks that utilize substitute training data or models~\cite{DBLP:conf/ndss/LiuMALZW018}.

\textbf{Backdoor Defenses.}
To ensure the security of the model, researchers put forward various methods of backdoor defense. We roughly divide them into data cleaning backdoor defense, model detection backdoor defense, and input detection backdoor defense.

Data cleaning defenses focus on identifying abnormal training samples or apply normalization to them~\cite{DBLP:conf/nips/Tran0M18, DBLP:conf/aaai/ChenCBLELMS19, DBLP:conf/sp/ChouTP20, DBLP:conf/uss/Tang0TZ21, DBLP:journals/corr/abs-2104-11315, DBLP:conf/iccv/ZengPMJ21, tang2023setting}. 
Model detection defenses utilize gradient descent~\cite{DBLP:conf/sp/WangYSLVZZ19, DBLP:conf/icml/ShenLTAX0M021, DBLP:conf/ijcai/ChenFZK19, DBLP:journals/tnn/XiangMK22, DBLP:conf/ccs/LiuLTMAZ19, DBLP:conf/cvpr/SunK23, xu2024towards} or generative adversarial network (GAN)~\cite{DBLP:conf/nips/GoodfellowPMXWOCB14, DBLP:conf/mm/ZhuNWXW20, DBLP:conf/nips/QiaoYL19} to obtain potential trigger patterns and then judge the backdoor by the pattern properties, \textit{e.g.}, size and attack success rate.
Note that there are some implementations in black-box settings in which defenders can reverse trigger patterns even when they only have access to model predictions (logits or labels)~\cite{DBLP:journals/compsec/AikenKWR21, DBLP:conf/iccv/DongYDPX0021, DBLP:conf/iclr/GuoLL22}.
Input detection backdoor defense aims at filtering malicious inputs after deploying the model, and thus preventing the backdoor from being triggered~\cite{DBLP:conf/acsac/GaoXW0RN19, DBLP:journals/corr/abs-2006-14871, DBLP:conf/iclr/DuJS20, DBLP:conf/iccad/JavaheripiSFJK20}.

There are also some methods try to eliminate backdoor through retraining~\cite{DBLP:journals/corr/KirkpatrickPRVD16, DBLP:conf/iccd/LiuXS17, DBLP:conf/iclr/ZengCPM0J22, li2024purifying}, backdoor-related neurons pruning~\cite{DBLP:conf/raid/0017DG18, DBLP:conf/nips/WuW21}, or knowledge distillation~\cite{DBLP:journals/corr/HintonVD15, DBLP:conf/ccs/YoshidaF20, DBLP:conf/iclr/LiLKLLM21, DBLP:conf/sp/GongCYWGHS23}, which are beyond this research and left for our future work.

\textbf{Mismatched Triggers.}
\highlight{(\textit{Statement \& 8. Further Clarification of Difference})}{
Some prior works also studied the generalizability of backdoor triggers. \cite{DBLP:conf/nips/QiaoYL19} attacked a model trained on the CIFAR-10 dataset with a $3\times3$ trigger. By repeating the reverse engineering process with random seeds, they found that the reversed triggers formed a continuous set in the pixel space of all possible $3\times3$ triggers, indicating that the injected backdoor exhibits certain generalizability beyond the original trigger pattern.
\cite{sun2020poisoned, DBLP:conf/cvpr/SunK23} also showed that, given access to a backdoored model, one can reliably generate multiple alternative triggers without accessing the training dataset or the original trigger, suggesting that an injected backdoor trigger pattern usually generalize to other shapes or colors. 
Additionally, other trigger-reversing studies~\cite{DBLP:conf/iclr/GuoLL22,DBLP:conf/iclr/0002LCYJ022,DBLP:conf/iccv/DongYDPX0021} have reported similar phenomena.
\cite{DBLP:journals/corr/abs-2312-04902} defined perturbed triggers while capable of activating backdoor as fuzzy triggers. They also proposed a trigger upper bound that measures how much perturbation a trigger can endure while retaining its ability to activate the backdoor, which implies triggers' generalization. 
}

The above work showed that triggers can vary between training and inference phases, while there are no explicit restrictions on their difference other than $\ell_n$ norms. 
Our works distinguish from existing studies with a semantic restriction, i.e. focusing on the impact of trigger intensity on backdoor attacks, and thoroughly investigated the impact of intensity on the effectiveness and generalizability of backdoor attacks.
% \zcy{the effect and advantage of the restriction}

% and exploring the \theorem~phenomenon in different backdoor attacks, models, and datasets.

% To the best of our knowledge, our work is the first on utilizing triggers' generalization across different intensities to leverage backdoor attack success rate and stealthiness. 

% Firstly, \xxx.
% Secondly, \xxx.
% Additionally, \xxx.
% Above all, our work provides a new perspective on the backdoor attacks and defenses, which is the intensity of the trigger, and we propose a new model called \theorem~to explain the phenomenon of intensity-specific generalization triggers.

%!TEX root = main.tex

\section{Conclusion and Outlook}
\label{section:conclusion}
In this paper, we have systematically investigated the trigger intensity mismatch training and inference.
Based on our new workflow \workflow~(\workflowfull), we reveal that consistent trigger configurations between the training and inference phases may indeed not be optimal for backdoor attacks; Instead, they may be leveraged to improve attacks, even when defenses are present. 
Extensive experiments demonstrate that our findings hold across various attack methods, trigger configurations, model architectures, datasets, etc.
We have also explained the reasons behind our findings and discussed their implications for adapting backdoor defenses.

Future work should explore optimization-based backdoors~\cite{DBLP:conf/ndss/LiuMALZW018, DBLP:conf/iccv/DoanL0L21, DBLP:conf/cvpr/WangZM22}, which generally exploit unpredictable sample-wise triggers, beyond the dynamic triggers (\forexample, WaNet, and Compress) we have considered.
Although we have shown that using the interpolation between the original image and its poisoned version as the intensity is practical, a ``universal intensity metric'' to incorporate different attacks is still worth exploring. 
Additionally, our findings regarding trigger mismatch should be further evaluated in more learning paradigms, such as federated learning and reinforcement learning, which assume distributed and interactive training process, respectively.

\section*{Acknowledgments}
This research is supported by the National Key Research and Development Program of China (2023YFE0209800), the National Natural Science Foundation of China (T2341003, 62376210, 62406240, 62161160337, 62132011, U24B20185, U21B2018, U20A20177, U244120060, 62206217), the Shaanxi Province Key Industry Innovation Program (2023-ZDLGY-38).

\section*{Ethics Considerations}
Our paper proposes a new adaptive backdoor attack strategy under train-inference intensity mismatched triggers, that can bypass several defenses.
It can be used to adversarially manipulate the behavior of a model even in the presence of a specific defense mechanism.

We admit that the proposed attack strategy can be used for malicious purposes, however, it is unlikely to cause severe harm to individuals or society, since selecting the optimal trigger for each setting would require exponentially more resources than training. Moreover, the phenomenon of train-inference intensity mismatch can help form a foundation for developing more robust defenses against backdoor attacks, which can be beneficial to the machine learning community in the long term.

\section*{Open Science}

% In this paper, since the algorithms we use are open source and the datasets include both open-source datasets and those generated using open-source algorithms, we ensure that our work will continue to remain open source.
% \todo{We ensure that our work will be open source and accessible to the public under a permissive license.
% Currently, the source code can be obtained through \url{https://anonymous.4open.science/r/TITIM-5568}.}

% We ensure that our work will be open source and accessible to the public under a permissive license.
The source code and datasets are available on GitHub (\url{https://github.com/cv12ha0/TITIM}) and Zenodo (\url{https://zenodo.org/records/14729436}).

% \swang{compliance with the open science policy}

% \textcolor{blue}{
% This year, USENIX Security introduces a new open science policy, aiming to enhance the reproducibility and replicability of scientific findings:
% Authors are expected to openly share their research artifacts by default. 
% This initiative is part of a broader commitment to foster open science principles, emphasizing the sharing of artifacts such as datasets, scripts, binaries, and source code associated with research papers. 
% If, for some reason (such as licensing restrictions), artifacts cannot be shared, a detailed justification must be provided. 
% Artifacts need to be available for the Artifact Evaluation committee after paper acceptance and before the final papers are due.
% }

%\clearpage

%-------------------------------------------------------------------------------
\bibliographystyle{abbrv}
% \bibliography{\jobname}
\bibliography{main}

% \clearpage
% \newpage
\appendix

\section{Case Study: BadNets on MNIST}
\label{section:appendix:badnets_mnist}

Our initial series of experiments employ the BadNets \cite{gu2019badnets} approach on MNIST~\cite{DBLP:journals/spm/Deng12}, utilizing the trigger opacity as an intuitive metric to quantify the intensity of the trigger.

\begin{figure}[!h]
  \centering
  \includegraphics[width=2cm]{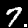}
  % \fbox{\rule[-.5cm]{0cm}{4cm} \rule[-.5cm]{4cm}{0cm}}
  \caption{The backdoor trigger for MNIST.}
  \label{figure:mnist_badnets_sample}
\end{figure}

\textbf{Setup.}
We adopt the configuration employed by \cite{gu2019badnets} to establish our network architecture and training parameters.
The backdoor trigger is set to a 3×3 pixel patch located at the bottom right corner, as shown in Figure~\ref{figure:mnist_badnets_sample}.
The Architecture of the baseline network for this task is shown in Table~\ref{table:mnist_model}.
% \todo{more detailed hyper-parameters}

\begin{figure*}[htbp]
  \centering
  \begin{subfigure}[c]{0.245\linewidth}
    \centering
    \includegraphics[width=\linewidth]{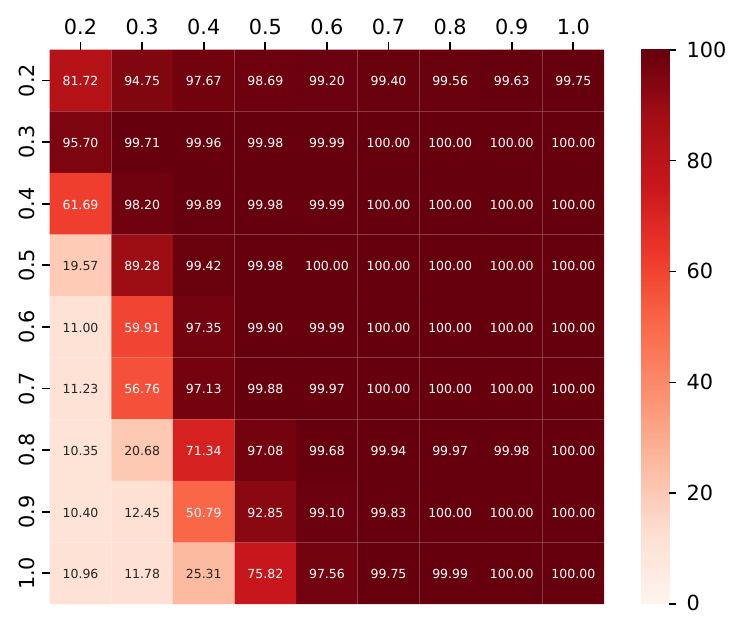}
    \caption{ResNet-18}
    % \label{figure:appendix:model_arch:vgg16}
  \end{subfigure}
  \begin{subfigure}[c]{0.245\linewidth}
    \centering
    \includegraphics[width=\linewidth]{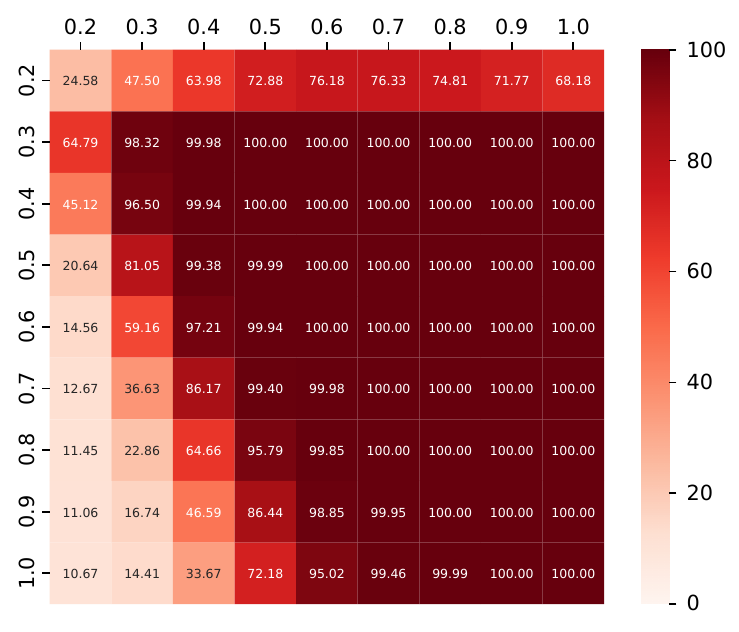}
    \caption{VGG-16}
    % \label{figure:appendix:model_arch:vgg16}
  \end{subfigure}
  \begin{subfigure}[c]{0.245\linewidth}
    \centering
    \includegraphics[width=\linewidth]{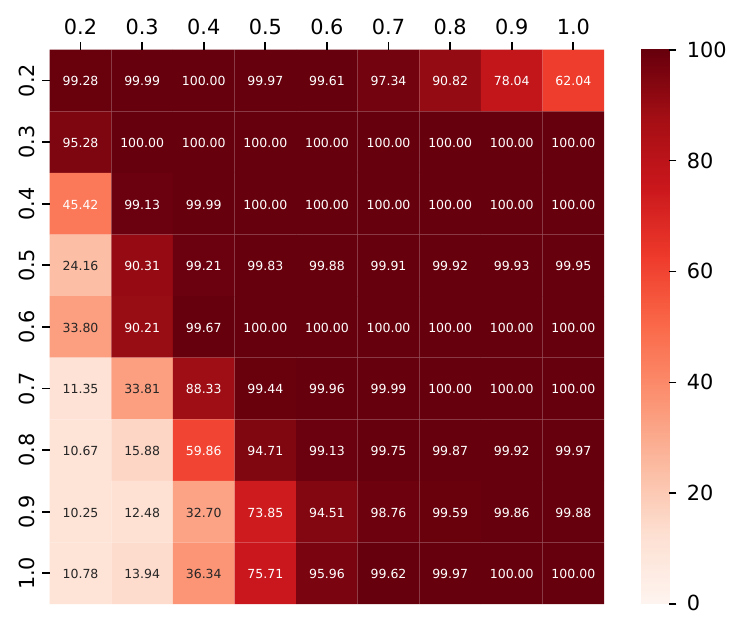}
    \caption{EfficientNet B0}
  \end{subfigure}
  \hfill
  \begin{subfigure}[c]{0.245\linewidth}
    \centering
    \includegraphics[width=\linewidth]{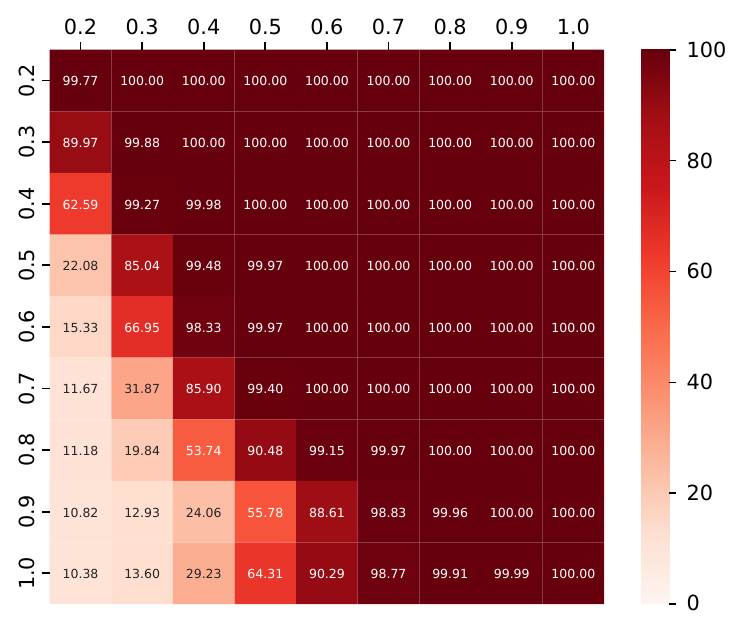}
    \caption{MobileNet V2}
  \end{subfigure}
  \begin{subfigure}[c]{0.245\linewidth}
    \centering
    \includegraphics[width=\linewidth]{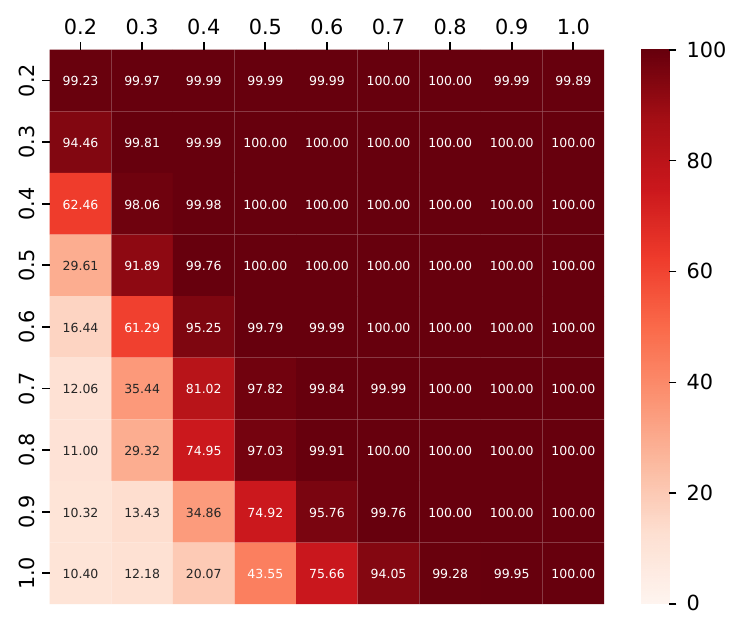}
    \caption{SENet-18}
  \end{subfigure}
  \begin{subfigure}[c]{0.245\linewidth}
    \centering
    \includegraphics[width=\linewidth]{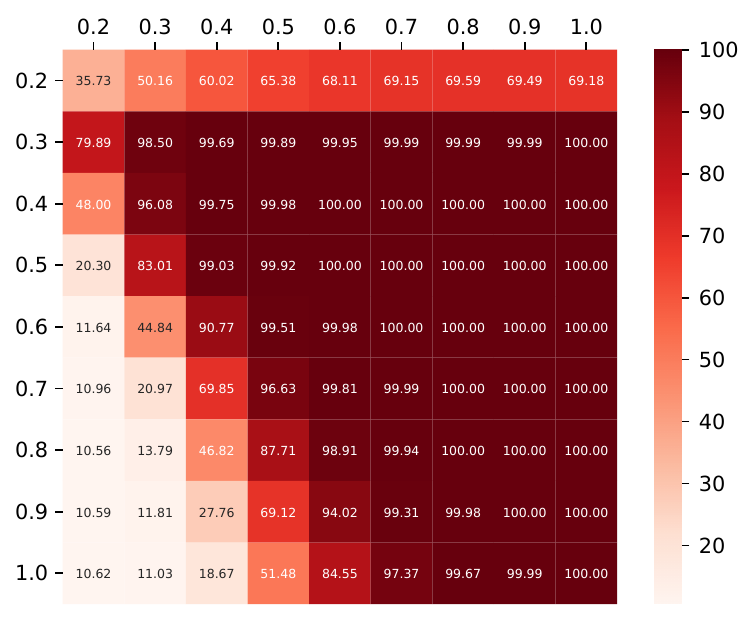}
    \caption{CvT}
  \end{subfigure}
  \begin{subfigure}[c]{0.245\linewidth}
    \centering
    \includegraphics[width=\linewidth]{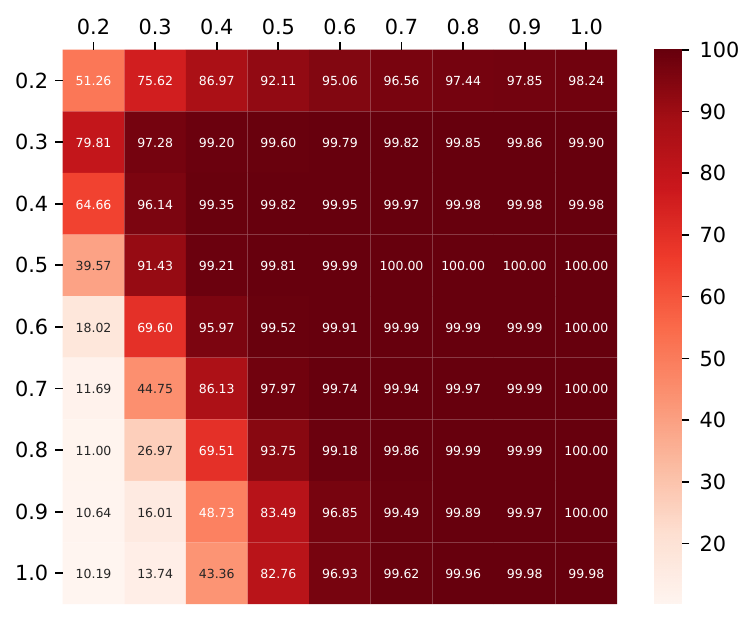}
    \caption{T2T ViT}
  \end{subfigure}
  \begin{subfigure}[c]{0.245\linewidth}
    \centering
    \includegraphics[width=\linewidth]{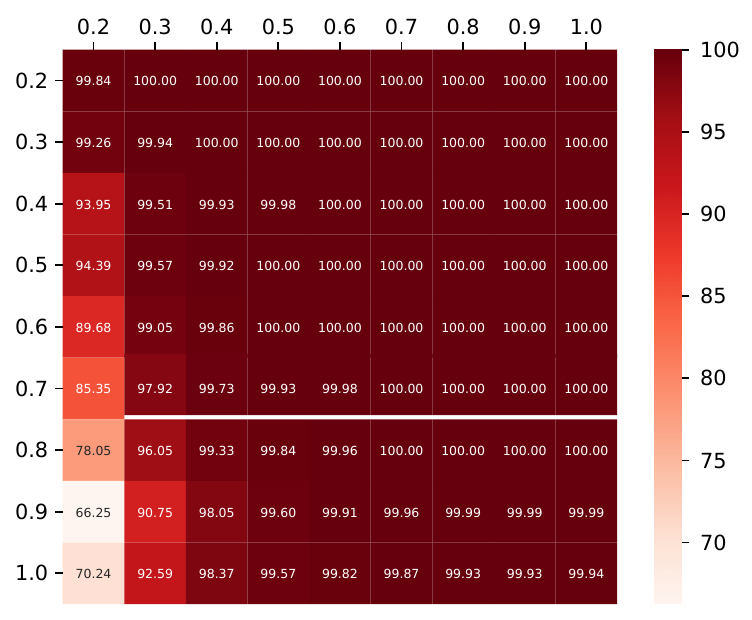}
    \caption{SimpleViT}
  \end{subfigure}
  \caption{Results of BadNets with Square patch on different model architectures. }
  \label{figure:appendix:model_arch}
\end{figure*}

\begin{figure*}[!t]
  \centering
  \begin{minipage}[t]{0.8\textwidth}
    \begin{subfigure}[c]{0.32\linewidth}
      \centering
      \includegraphics[width=\linewidth]{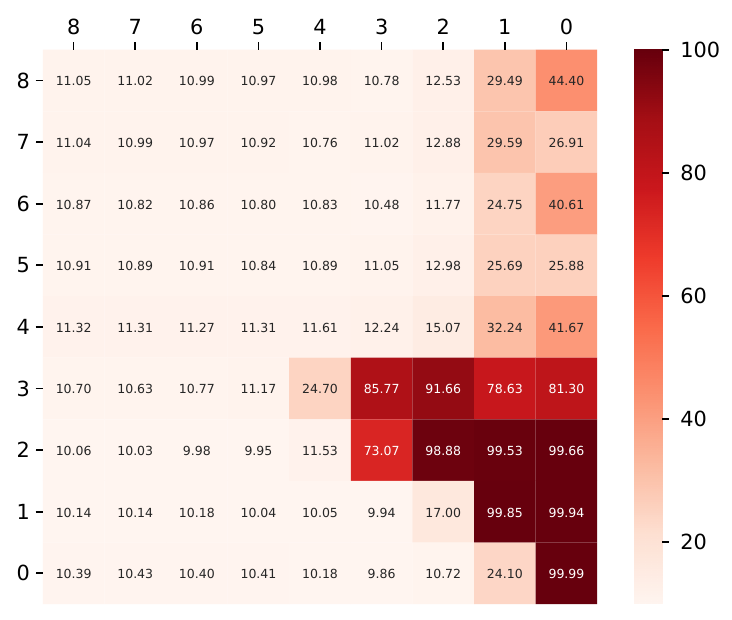}
      \caption{CIFAR-10}
      % \label{figure:appendix:model_arch:vgg16}
    \end{subfigure}
    \begin{subfigure}[c]{0.32\linewidth}
      \centering
      \includegraphics[width=\linewidth]{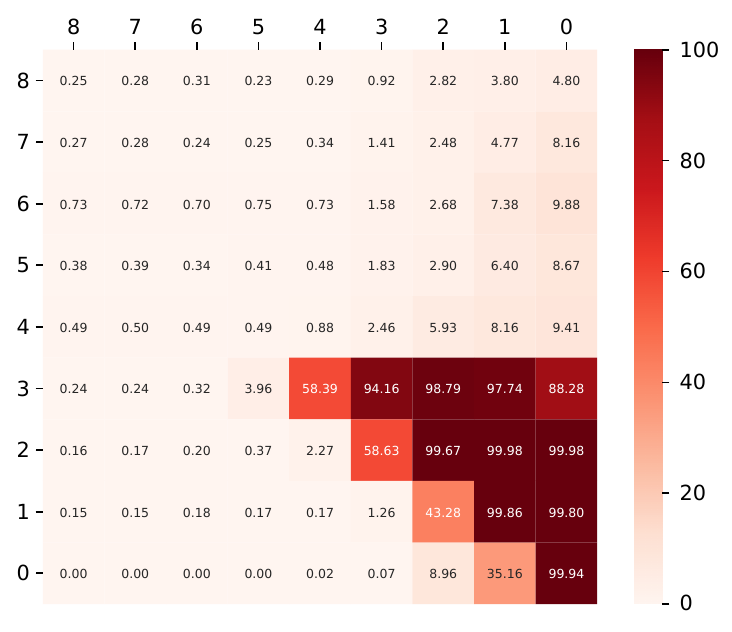}
      \caption{GTSRB}
    \end{subfigure}
    \begin{subfigure}[c]{0.32\linewidth}
      \centering
      \includegraphics[width=\linewidth]{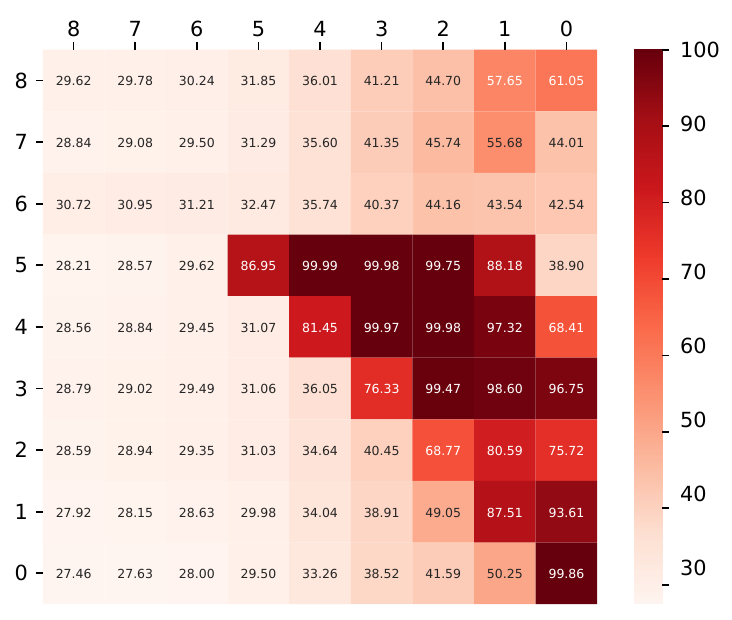}
      \caption{CelebA}
    \end{subfigure}
  \end{minipage}
  \caption{Results of BppAttack. The x- and y-axis indicate inference- and training-phase trigger intensity, respectively.}
  \label{figure:appendix:full_res:bpp}
\end{figure*}

\textbf{Results. }
We conduct two sets of experiments on MNIST, where the opacity of the trigger is set to 0.5 during both the training and inference phases, respectively. Then we examine the impact on the attack success rate when the trigger's opacity is adjusted during the other stage.
Here, we measure the trigger intensity with its opacity.

We train three models on a poisoned dataset with the trigger opacity set to 0.25, 0.5, and 0.75 and then evaluate its performance across 20 distinct poisoned datasets, each characterized by varying levels of trigger opacity. The results are shown in Figure~\ref{figure:mnist_square} (left).
The result is in line with our experience: the backdoor is successfully activated when the trigger intensity of the test reaches that of the training set, and the attack success rate increases as the trigger intensity is augmented.
In Figure~\ref{figure:mnist_square} (right), we train 20 models attacked by different intensities of the backdoor and evaluate them on three poisoned test sets.
The attack success rate firstly goes up as the trigger is injected successfully but then drops significantly.
In similar repeated experiments, the ASR dropping consistently occurs after a certain point, which means injecting a more potent backdoor does not always yield a better attack result. 

\textbf{Additional Analysis of ASR Dropping. }
To gain further insights into the ASR dropping phenomenon, we conduct cross-testing on each pair of trigger configurations for the training/inference stages mentioned above.
Figure~\ref{figure:mnist_heatmap} illustrates the attack success rate across each pair of configurations in two phases, wherein each set employs an identical white square as the trigger pattern and varies only in the trigger opacity.
The ASR always appears to drop when the attack intensity of the training set exceeds that of the test set, this observation implies that models may be overfitting to the trigger pattern with a higher intensity.

\section{Results on More Model Architectures}
\label{section:appendix:more_model}
% \textbf{Generalization to more architectures.}
\highlight{(\textit{2. Black-Box Scenarios}) }{
In this section, we additionally verify the generalization of our work to various architectures on 4 CNNs and 3 ViTs: for CNN, we tested VGG-16~\cite{DBLP:journals/corr/SimonyanZ14a}, EfficientNet B0~\cite{DBLP:conf/icml/TanL19}, MobileNet V2~\cite{DBLP:conf/cvpr/SandlerHZZC18}, and SENet-18~\cite{DBLP:journals/corr/abs-1709-01507}; and for ViT, we tested SimpleViT~\cite{DBLP:conf/iclr/DosovitskiyB0WZ21}, CvT~\cite{DBLP:conf/iccv/WuXCLDY021}, and T2T ViT~\cite{DBLP:conf/iccv/0007CWYSJTFY21}
We adopt the BadNets with square triggers, as noted in Table~\ref{table:experiment_attacks}, and set the poisoning rate to 10\% for the following experiments. 

As shown in Figure~\ref{figure:appendix:model_arch}, the ASR distribution across varying training- and inference-phase trigger intensity settings is mostly consistent, except for SimpleViT, which exhibits a consistently higher ASR than other models. While the overall pattern aligns with our previous findings, this is reasonable for its significantly larger size, confirming that our observed phenomena are model architecture agnostic. 
Moreover, the similar ASR distribution indicates that it is feasible for attackers to obtain desired training or inference intensities on surrogate models, enabling attacks in black-box scenarios.
% our findings still hold on various model architectures.
}

\begin{figure*}[!t]
  \centering
  \begin{minipage}[t]{0.8\textwidth}
    \begin{subfigure}[c]{0.28\linewidth}
      \centering
      \includegraphics[width=\linewidth,trim={0 20 60pt 0},clip]{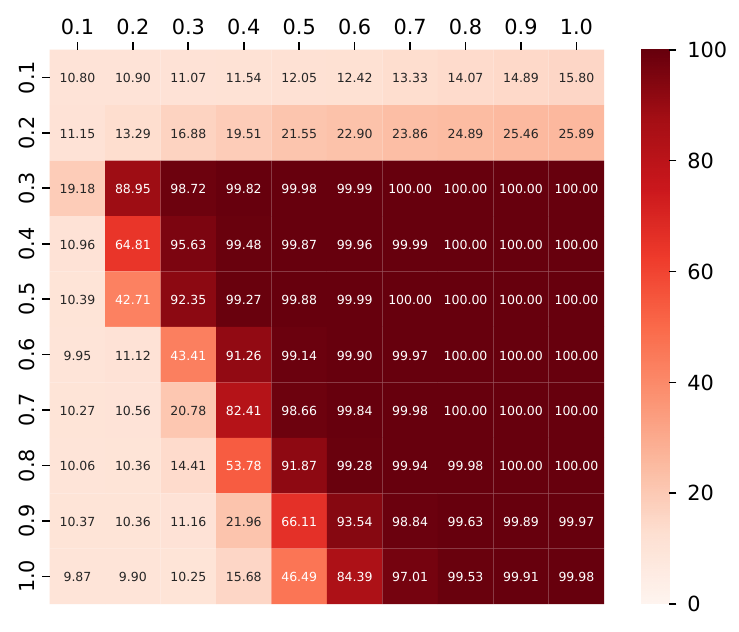}
      % \caption{non-mix 5\% poisoning rate}
      \label{figure:mix_0.05mrx}
    \end{subfigure}
    \begin{subfigure}[c]{0.28\linewidth}
      \centering
      \includegraphics[width=\linewidth,trim={0 20 60pt 0},clip]{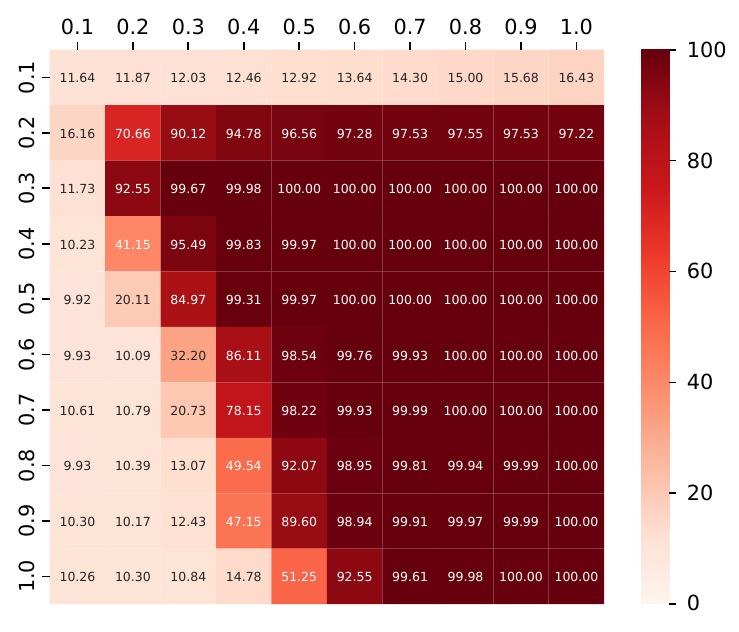}
      % \caption{non-mix 10\% poisoning rate}
      \label{figure:mix_0.1mrx}
    \end{subfigure}
    \begin{subfigure}[c]{0.38\linewidth}
      \centering
      \includegraphics[width=\linewidth,trim={0 20 0 0},clip]{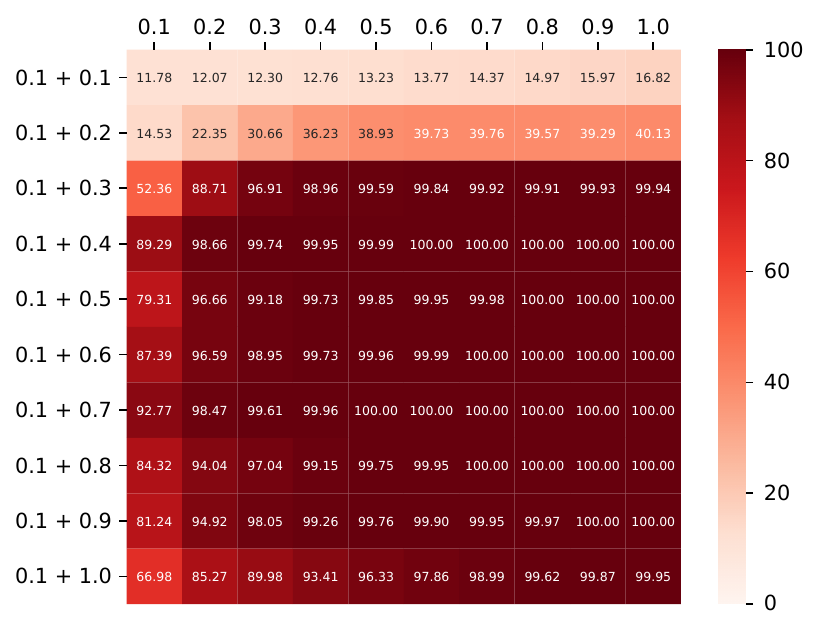}
       % \caption{mix 10\% poisoning rate}
      \label{figure:mix_0.05mr0.1}  
    \end{subfigure}
    % \subcaption{The detailed ASR of the BadNets with mixed training trigger intensities (opacity).}
    \label{figure:appendix:mixtures_mr}
  \end{minipage}
  \begin{minipage}[t]{0.8\textwidth}
    \begin{subfigure}[c]{0.28\linewidth}
      \centering
      \includegraphics[width=\linewidth,trim={0 0 60pt 0},clip]{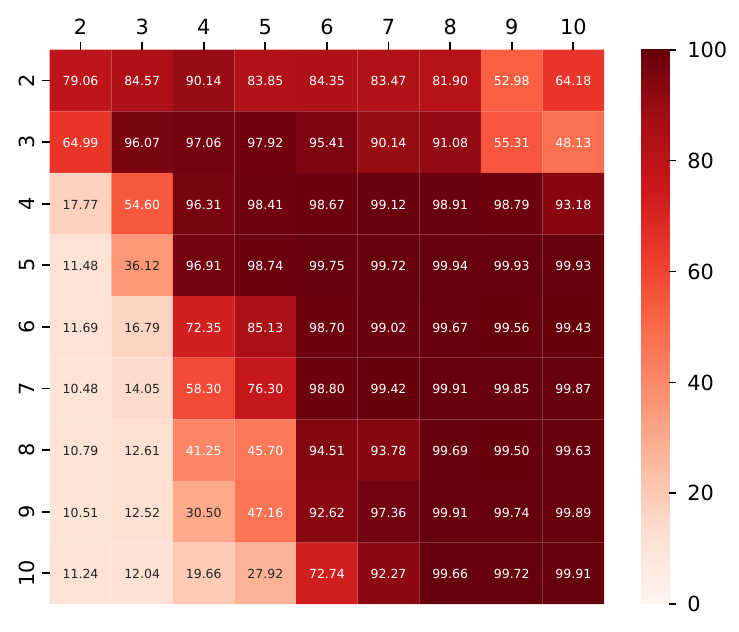}
      \caption{non-mix 5\% poisoning rate}
      \label{figure:mix_0.05szx}
    \end{subfigure}
    \begin{subfigure}[c]{0.28\linewidth}
      \centering
      \includegraphics[width=\linewidth,trim={0 0 60pt 0},clip]{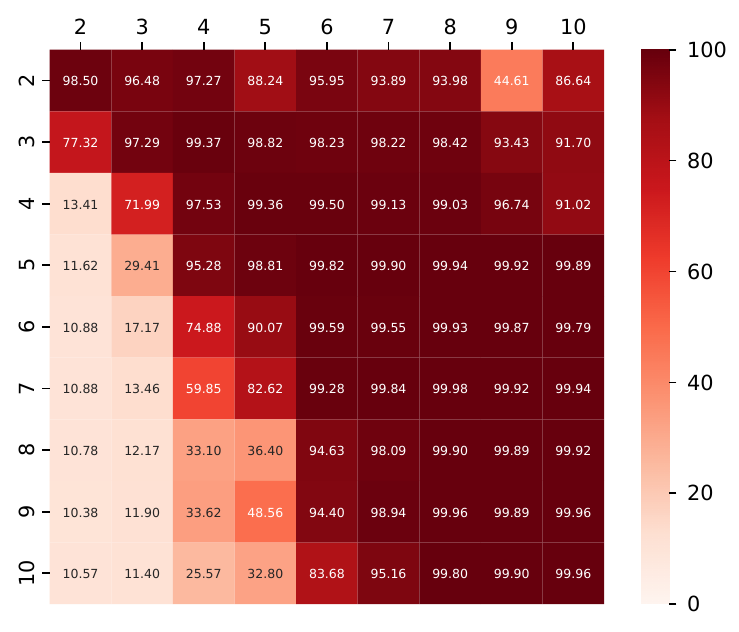}
      \caption{non-mix 10\% poisoning rate}
      \label{figure:mix_0.1szx}
    \end{subfigure}
    \begin{subfigure}[c]{0.36\linewidth}
      \centering
      \includegraphics[width=\linewidth,trim={-20 0 0 0},clip]{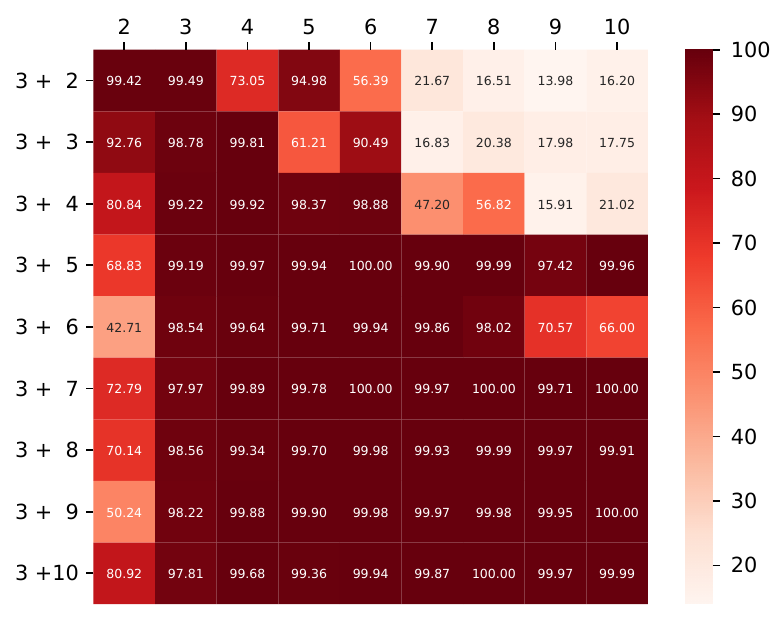}
       \caption{mix 10\% poisoning rate}
      \label{figure:mix_0.05sz3}  
    \end{subfigure}
    % \subcaption{The detailed ASR of the BadNets with mixed training trigger intensities (upper row: opacity mixing, lower row: size mixing).}
    \label{figure:appendix:mixtures_sz}
  \end{minipage}
  \caption{The detailed ASR of the BadNets with mixed training trigger intensities (upper row: opacity mixing, lower row: size mixing).}
  \label{figure:appendix:mixtures}
\end{figure*}

\begin{figure}[!t]
  \centering
  \begin{subfigure}[c]{0.44\linewidth}
    \centering
    \includegraphics[width=\linewidth,trim={0 0 60 0},clip]{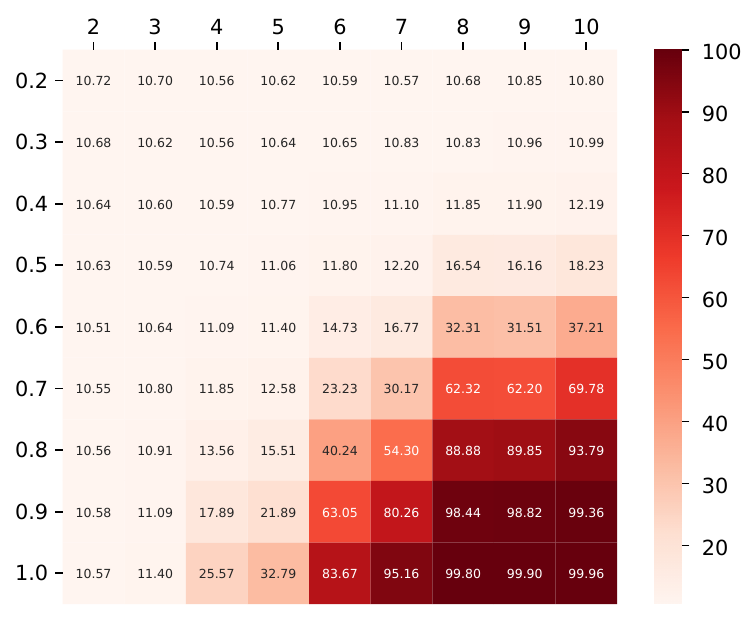}
    \caption{w/o mixing}
  \end{subfigure}
  \begin{subfigure}[c]{0.525\linewidth}
    \centering
    \includegraphics[width=\linewidth,trim={0 0 0 0},clip]{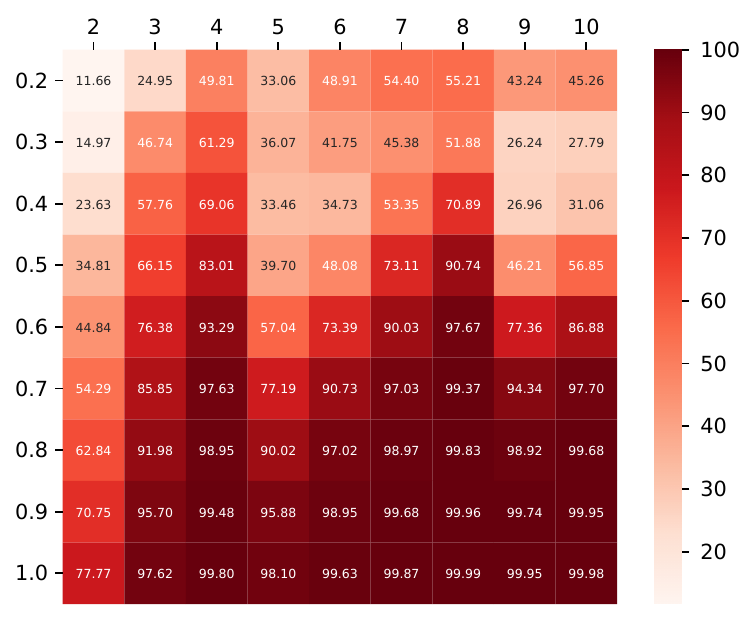}
    \caption{w/ mixing}
  \end{subfigure}
  \caption{Results of simultaneously mixing triggers of varying opacities and sizes. }
  \label{figure:appendix:mixtures_mrsz}
\end{figure}

\begin{figure}[!t]
  \centering
  \begin{minipage}[t]{0.98\linewidth}
      \includegraphics[width=0.3\linewidth,trim={0 0 60 0},clip]{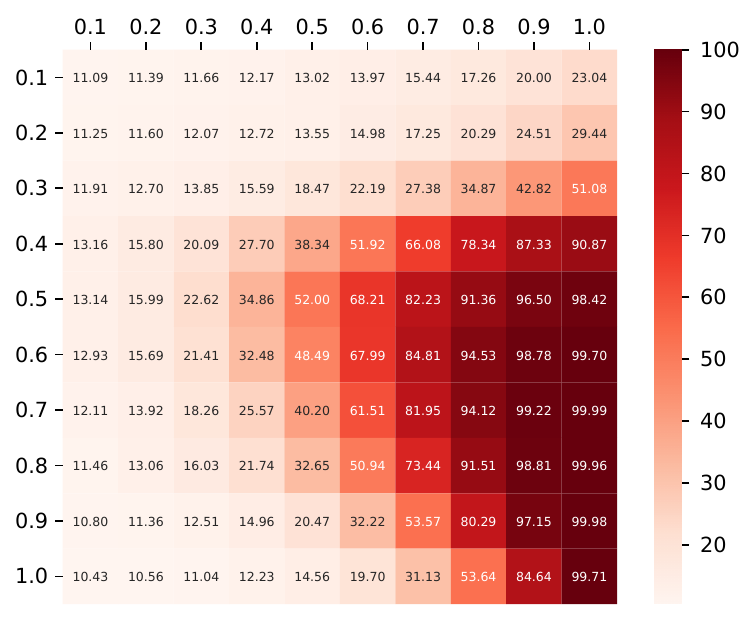}
      \includegraphics[width=0.3\linewidth,trim={0 0 60 0},clip]{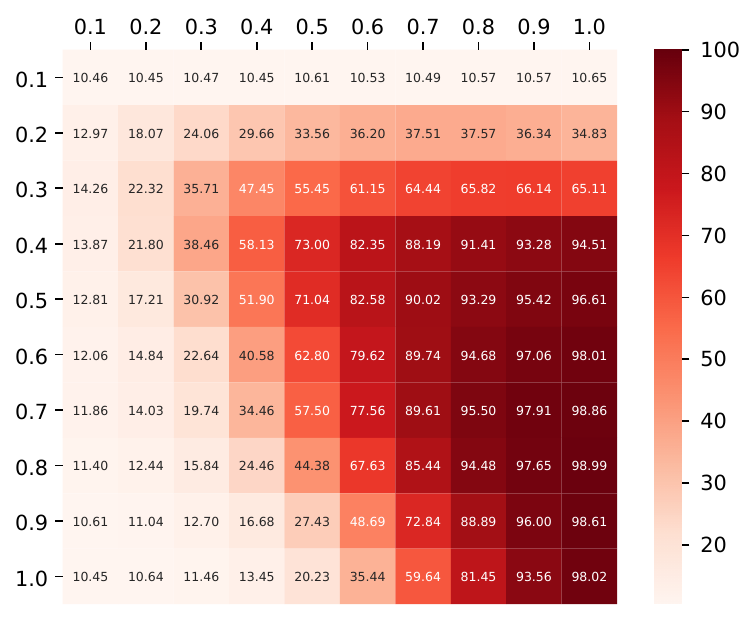}
      \includegraphics[width=0.362\linewidth,trim={0 0 0 0},clip]{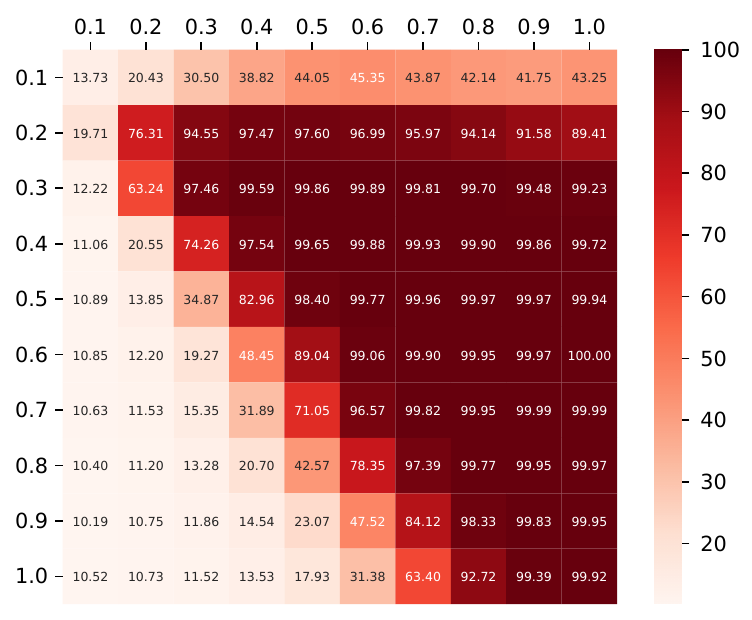}
  \end{minipage}

  \begin{minipage}[t]{0.98\linewidth}
      \includegraphics[width=0.3\linewidth,trim={0 0 60 0},clip]{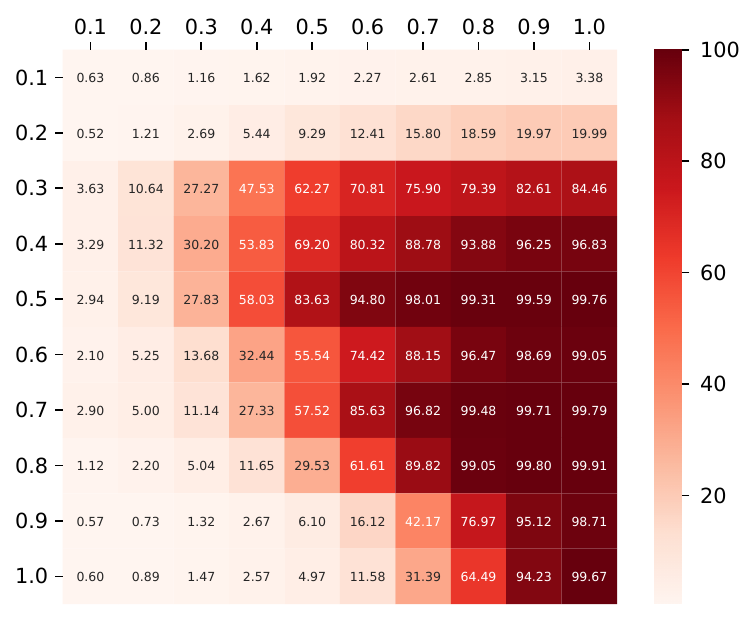}
      \includegraphics[width=0.3\linewidth,trim={0 0 60 0},clip]{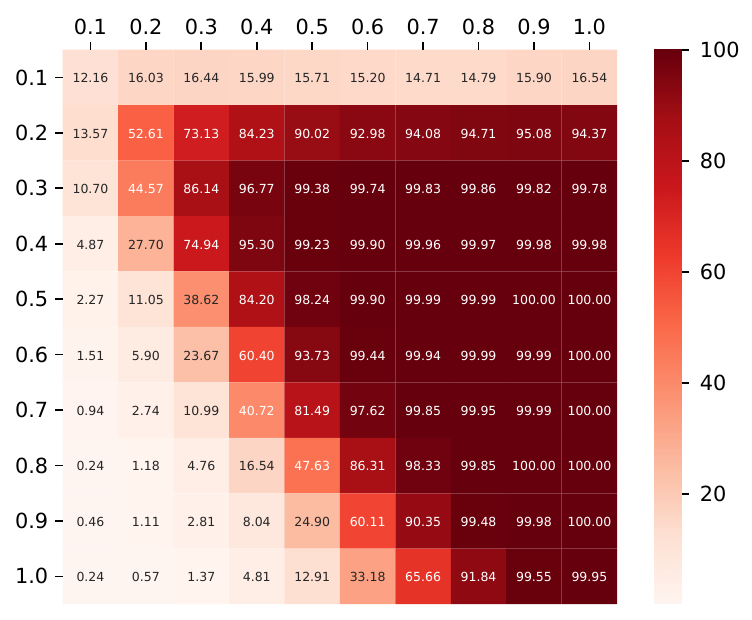}
      \includegraphics[width=0.362\linewidth,trim={0 0 0 0},clip]{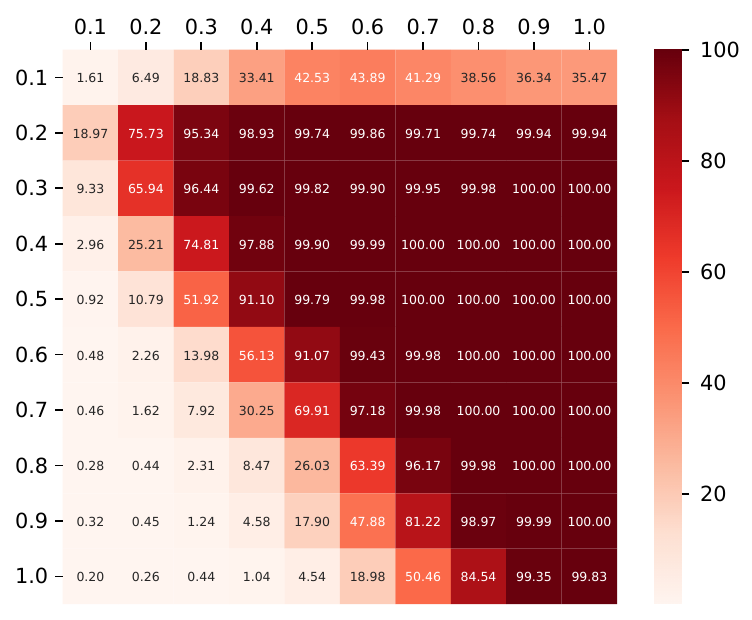}
  \end{minipage}

  \begin{minipage}[t]{0.98\linewidth}
    \begin{subfigure}[c]{0.3\linewidth}
      \centering
      \includegraphics[width=\linewidth,trim={0 0 60 0},clip]{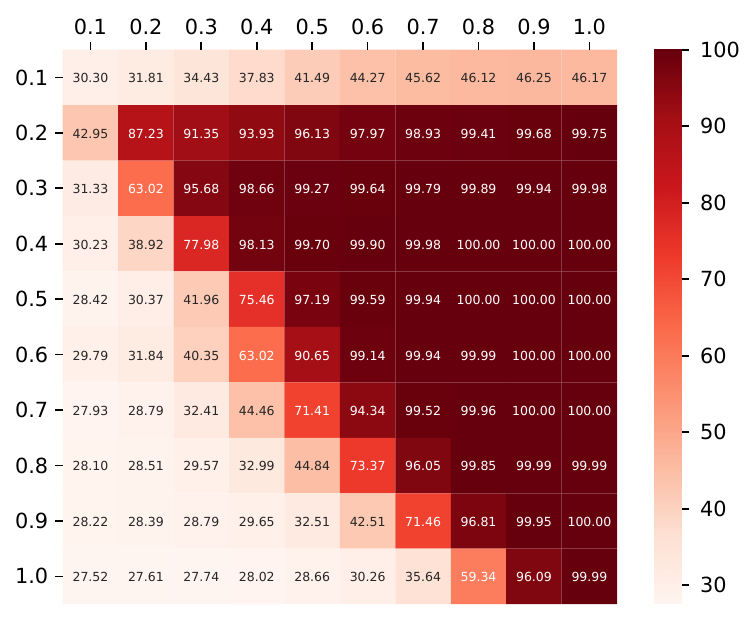}
      \caption{Gotham}
    \end{subfigure}
    \begin{subfigure}[c]{0.3\linewidth}
      \centering
      \includegraphics[width=\linewidth,trim={0 0 60 0},clip]{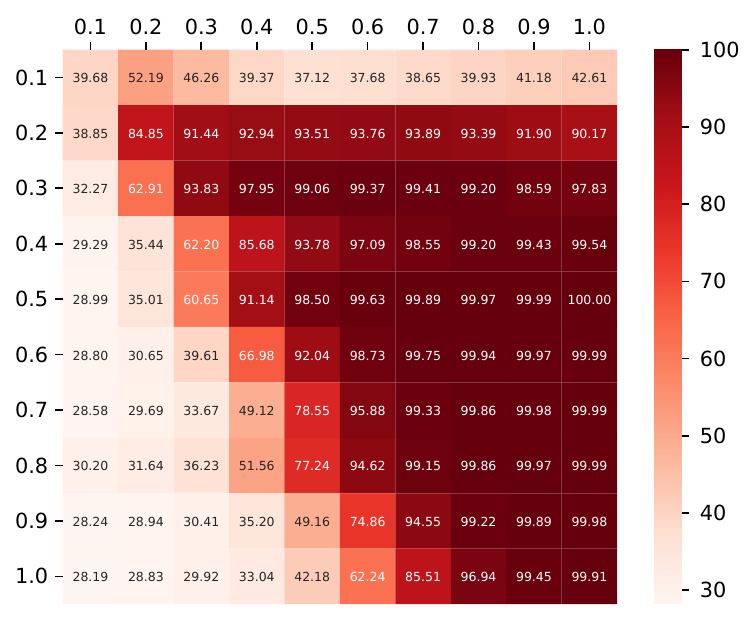}
      \caption{Kelvin}
    \end{subfigure}
    \begin{subfigure}[c]{0.362\linewidth}
      \centering
      \includegraphics[width=\linewidth,trim={0 0 0 0},clip]{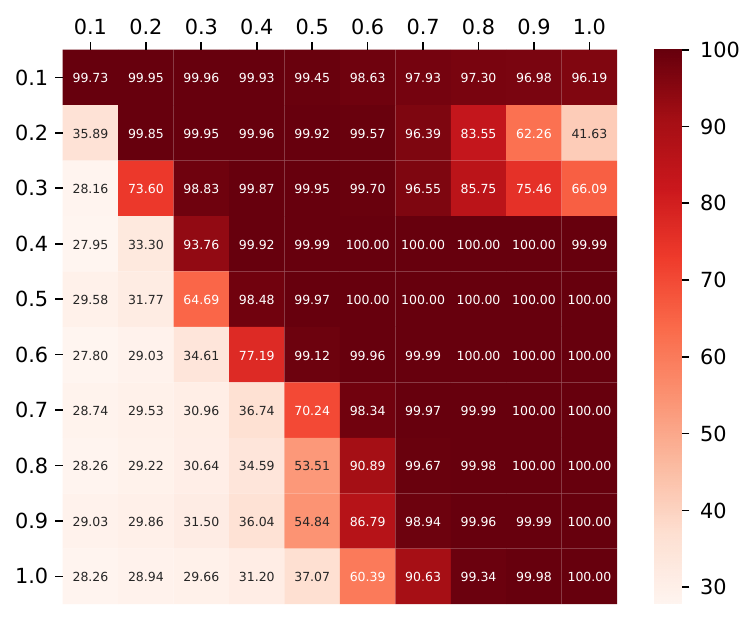}
      \caption{Lomo\phantom{0}}
    \end{subfigure}
  \end{minipage}

  \caption{Results of Styled (the dataset for each row being CIFAR-10, GTSRB, and CelebA from top to bottom, respectively).}
  \label{figure:appendix:full_res:styled}
  \vspace{-4pt}
\end{figure}

\section{Detailed Results of Intensity Mixture}
\label{section:appendix:intensity_mixture}
The detailed results of the intensity mixture are shown in Figure~\ref{figure:appendix:mixtures} (upper row). 
% The x-axis and y-axis represent the trigger intensity employed during the inference and training phases, respectively.
Triggers with intensity mixtures consistently outperform those utilizing a single intensity, except for a trigger intensity of 0.2 at a poisoning rate of 0.1, where the mixed triggers are too weak to execute the attack effectively.
As shown in Figure~\ref{figure:appendix:mixtures} (lower row), We further confirmed that mixing other intensities can yield similar outcomes by mixing triggers of different sizes. Although the localized nature of BadNets triggers restricts the effect of low-intensity mixing (\forexample, mixing size 3 with 2 to 4) on larger triggers, the mixture of size 3 with higher intensities continues to demonstrate promising results.

\highlight{(1. Simultaneously Mixing Multiple Intensities)}{
Additionally, attackers can simultaneously mix varying intensities. For instance, we equally mix triggers with three configurations, resulting in a total poisoning rate of 10\%: opacity=1.0\&size=10, opacity=0.1\&size=10, and opacity=1.0\&size=3.
Subsequently, we collect the backdoored model's ASR on triggers with varying opacities and sizes and then compare wit ith the ASR of another victim model trained with single intensity (opacity=1.0\&size=10, with a poisoning rate at 10\%, too). As shown in Figure~\ref{figure:appendix:mixtures_mrsz}, this mixing strategy enhances the generalizability across varying trigger intensities. This approach is similar to trigger-augmentation, which is utilized to enhance the robustness of backdoors in the physical world~\cite{DBLP:journals/corr/abs-2104-02361}, \etc.
}

% Next, we collect the backdoored model's Attack Success Rate (ASR) on triggers with varying opacities and sizes, and we perform the same for another victim model trained with a single intensity (opacity=1.0&size=10, also with a poisoning rate of 10%). As shown in Figure~\ref{figure:appendix:mixtures_mrsz}, this mixing strategy enhances generalizability across varying trigger intensities. This approach is similar to trigger-augmentation, which is utilized to enhance the robustness of backdoors in the physical world, etc."

% \todo{mixing both opacity and size}
%and Figure~\ref{figure:mixtures_0.01}. 

\section{Additional Overall Attack Results}
\label{section:appendix:full_res}
\highlight{(5. More Triggers)}{
% In this section, we first present results on additional trigger patterns for BadNets and Blended. Specifically, we use a bomb and a flower as BadNets trigger patches and noise as the Blended pattern.
Here we demonstrate the general applicability of our findings to two more attacks: BppAttack~\cite{DBLP:conf/cvpr/WangZM22} and Styled~\cite{DBLP:conf/ccs/LiuLTMAZ19}. 

In BppAttack, although the ASR is lower at low intensities (\thatis, large color depth) as we omit the contrastive learning that modifies the training process, the overall pattern of ASR in Figure~\ref{figure:appendix:full_res:bpp} is consistent with our previous findings. 

In Styled, we define an artificial intensity by the interpolation between the original images and their poisoned versions, as this attack lacks tunable parameters other than the image filter. As shown in Figure~\ref{figure:appendix:full_res:styled}, the ASR pattern of this artificial intensity aligns well with our previous findings, demonstrating a practical approach to adjusting the intensity of triggers in backdoor attacks without requiring tunable parameters. 
}

\end{document}